\documentclass[useAMS]{mn2e}

\usepackage[fleqn]{amsmath}
\usepackage{color}
\usepackage{graphicx}
\usepackage{url}
\usepackage{float}
\DeclareGraphicsExtensions{.eps,.ps,.eps.gz,.ps.gz,.eps.Z}
\definecolor{AliceBlue}{rgb}{0.94,0.97,1.00}
\definecolor{AntiqueWhite1}{rgb}{1.00,0.94,0.86}
\definecolor{AntiqueWhite2}{rgb}{0.93,0.87,0.80}
\definecolor{AntiqueWhite3}{rgb}{0.80,0.75,0.69}
\definecolor{AntiqueWhite4}{rgb}{0.55,0.51,0.47}
\definecolor{AntiqueWhite}{rgb}{0.98,0.92,0.84}
\definecolor{BlanchedAlmond}{rgb}{1.00,0.92,0.80}
\definecolor{BlueViolet}{rgb}{0.54,0.17,0.89}
\definecolor{CadetBlue1}{rgb}{0.60,0.96,1.00}
\definecolor{CadetBlue2}{rgb}{0.56,0.90,0.93}
\definecolor{CadetBlue3}{rgb}{0.48,0.77,0.80}
\definecolor{CadetBlue4}{rgb}{0.33,0.53,0.55}
\definecolor{CadetBlue}{rgb}{0.37,0.62,0.63}
\definecolor{CornflowerBlue}{rgb}{0.39,0.58,0.93}
\definecolor{DarkBlue}{rgb}{0.00,0.00,0.55}
\definecolor{DarkCyan}{rgb}{0.00,0.55,0.55}
\definecolor{DarkGoldenrod1}{rgb}{1.00,0.73,0.06}
\definecolor{DarkGoldenrod2}{rgb}{0.93,0.68,0.05}
\definecolor{DarkGoldenrod3}{rgb}{0.80,0.58,0.05}
\definecolor{DarkGoldenrod4}{rgb}{0.55,0.40,0.03}
\definecolor{DarkGoldenrod}{rgb}{0.72,0.53,0.04}
\definecolor{DarkGray}{rgb}{0.66,0.66,0.66}
\definecolor{DarkGreen}{rgb}{0.00,0.39,0.00}
\definecolor{DarkGrey}{rgb}{0.66,0.66,0.66}
\definecolor{DarkKhaki}{rgb}{0.74,0.72,0.42}
\definecolor{DarkMagenta}{rgb}{0.55,0.00,0.55}
\definecolor{DarkOliveGreen1}{rgb}{0.79,1.00,0.44}
\definecolor{DarkOliveGreen2}{rgb}{0.74,0.93,0.41}
\definecolor{DarkOliveGreen3}{rgb}{0.64,0.80,0.35}
\definecolor{DarkOliveGreen4}{rgb}{0.43,0.55,0.24}
\definecolor{DarkOliveGreen}{rgb}{0.33,0.42,0.18}
\definecolor{DarkOrange1}{rgb}{1.00,0.50,0.00}
\definecolor{DarkOrange2}{rgb}{0.93,0.46,0.00}
\definecolor{DarkOrange3}{rgb}{0.80,0.40,0.00}
\definecolor{DarkOrange4}{rgb}{0.55,0.27,0.00}
\definecolor{DarkOrange}{rgb}{1.00,0.55,0.00}
\definecolor{DarkOrchid1}{rgb}{0.75,0.24,1.00}
\definecolor{DarkOrchid2}{rgb}{0.70,0.23,0.93}
\definecolor{DarkOrchid3}{rgb}{0.60,0.20,0.80}
\definecolor{DarkOrchid4}{rgb}{0.41,0.13,0.55}
\definecolor{DarkOrchid}{rgb}{0.60,0.20,0.80}
\definecolor{DarkRed}{rgb}{0.55,0.00,0.00}
\definecolor{DarkSalmon}{rgb}{0.91,0.59,0.48}
\definecolor{DarkSeaGreen1}{rgb}{0.76,1.00,0.76}
\definecolor{DarkSeaGreen2}{rgb}{0.71,0.93,0.71}
\definecolor{DarkSeaGreen3}{rgb}{0.61,0.80,0.61}
\definecolor{DarkSeaGreen4}{rgb}{0.41,0.55,0.41}
\definecolor{DarkSeaGreen}{rgb}{0.56,0.74,0.56}
\definecolor{DarkSlateBlue}{rgb}{0.28,0.24,0.55}
\definecolor{DarkSlateGray1}{rgb}{0.59,1.00,1.00}
\definecolor{DarkSlateGray2}{rgb}{0.55,0.93,0.93}
\definecolor{DarkSlateGray3}{rgb}{0.47,0.80,0.80}
\definecolor{DarkSlateGray4}{rgb}{0.32,0.55,0.55}
\definecolor{DarkSlateGray}{rgb}{0.18,0.31,0.31}
\definecolor{DarkSlateGrey}{rgb}{0.18,0.31,0.31}
\definecolor{DarkTurquoise}{rgb}{0.00,0.81,0.82}
\definecolor{DarkViolet}{rgb}{0.58,0.00,0.83}
\definecolor{DeepPink1}{rgb}{1.00,0.08,0.58}
\definecolor{DeepPink2}{rgb}{0.93,0.07,0.54}
\definecolor{DeepPink3}{rgb}{0.80,0.06,0.46}
\definecolor{DeepPink4}{rgb}{0.55,0.04,0.31}
\definecolor{DeepPink}{rgb}{1.00,0.08,0.58}
\definecolor{DeepSkyBlue1}{rgb}{0.00,0.75,1.00}
\definecolor{DeepSkyBlue2}{rgb}{0.00,0.70,0.93}
\definecolor{DeepSkyBlue3}{rgb}{0.00,0.60,0.80}
\definecolor{DeepSkyBlue4}{rgb}{0.00,0.41,0.55}
\definecolor{DeepSkyBlue}{rgb}{0.00,0.75,1.00}
\definecolor{DimGray}{rgb}{0.41,0.41,0.41}
\definecolor{DimGrey}{rgb}{0.41,0.41,0.41}
\definecolor{DodgerBlue1}{rgb}{0.12,0.56,1.00}
\definecolor{DodgerBlue2}{rgb}{0.11,0.53,0.93}
\definecolor{DodgerBlue3}{rgb}{0.09,0.45,0.80}
\definecolor{DodgerBlue4}{rgb}{0.06,0.31,0.55}
\definecolor{DodgerBlue}{rgb}{0.12,0.56,1.00}
\definecolor{FloralWhite}{rgb}{1.00,0.98,0.94}
\definecolor{ForestGreen}{rgb}{0.13,0.55,0.13}
\definecolor{GhostWhite}{rgb}{0.97,0.97,1.00}
\definecolor{GreenYellow}{rgb}{0.68,1.00,0.18}
\definecolor{HotPink1}{rgb}{1.00,0.43,0.71}
\definecolor{HotPink2}{rgb}{0.93,0.42,0.65}
\definecolor{HotPink3}{rgb}{0.80,0.38,0.56}
\definecolor{HotPink4}{rgb}{0.55,0.23,0.38}
\definecolor{HotPink}{rgb}{1.00,0.41,0.71}
\definecolor{IndianRed1}{rgb}{1.00,0.42,0.42}
\definecolor{IndianRed2}{rgb}{0.93,0.39,0.39}
\definecolor{IndianRed3}{rgb}{0.80,0.33,0.33}
\definecolor{IndianRed4}{rgb}{0.55,0.23,0.23}
\definecolor{IndianRed}{rgb}{0.80,0.36,0.36}
\definecolor{LavenderBlush1}{rgb}{1.00,0.94,0.96}
\definecolor{LavenderBlush2}{rgb}{0.93,0.88,0.90}
\definecolor{LavenderBlush3}{rgb}{0.80,0.76,0.77}
\definecolor{LavenderBlush4}{rgb}{0.55,0.51,0.53}
\definecolor{LavenderBlush}{rgb}{1.00,0.94,0.96}
\definecolor{LawnGreen}{rgb}{0.49,0.99,0.00}
\definecolor{LemonChiffon1}{rgb}{1.00,0.98,0.80}
\definecolor{LemonChiffon2}{rgb}{0.93,0.91,0.75}
\definecolor{LemonChiffon3}{rgb}{0.80,0.79,0.65}
\definecolor{LemonChiffon4}{rgb}{0.55,0.54,0.44}
\definecolor{LemonChiffon}{rgb}{1.00,0.98,0.80}
\definecolor{LightBlue1}{rgb}{0.75,0.94,1.00}
\definecolor{LightBlue2}{rgb}{0.70,0.87,0.93}
\definecolor{LightBlue3}{rgb}{0.60,0.75,0.80}
\definecolor{LightBlue4}{rgb}{0.41,0.51,0.55}
\definecolor{LightBlue}{rgb}{0.68,0.85,0.90}
\definecolor{LightCoral}{rgb}{0.94,0.50,0.50}
\definecolor{LightCyan1}{rgb}{0.88,1.00,1.00}
\definecolor{LightCyan2}{rgb}{0.82,0.93,0.93}
\definecolor{LightCyan3}{rgb}{0.71,0.80,0.80}
\definecolor{LightCyan4}{rgb}{0.48,0.55,0.55}
\definecolor{LightCyan}{rgb}{0.88,1.00,1.00}
\definecolor{LightGoldenrod1}{rgb}{1.00,0.93,0.55}
\definecolor{LightGoldenrod2}{rgb}{0.93,0.86,0.51}
\definecolor{LightGoldenrod3}{rgb}{0.80,0.75,0.44}
\definecolor{LightGoldenrod4}{rgb}{0.55,0.51,0.30}
\definecolor{LightGoldenrodYellow}{rgb}{0.98,0.98,0.82}
\definecolor{LightGoldenrod}{rgb}{0.93,0.87,0.51}
\definecolor{LightGray}{rgb}{0.83,0.83,0.83}
\definecolor{LightGreen}{rgb}{0.56,0.93,0.56}
\definecolor{LightGrey}{rgb}{0.83,0.83,0.83}
\definecolor{LightPink1}{rgb}{1.00,0.68,0.73}
\definecolor{LightPink2}{rgb}{0.93,0.64,0.68}
\definecolor{LightPink3}{rgb}{0.80,0.55,0.58}
\definecolor{LightPink4}{rgb}{0.55,0.37,0.40}
\definecolor{LightPink}{rgb}{1.00,0.71,0.76}
\definecolor{LightSalmon1}{rgb}{1.00,0.63,0.48}
\definecolor{LightSalmon2}{rgb}{0.93,0.58,0.45}
\definecolor{LightSalmon3}{rgb}{0.80,0.51,0.38}
\definecolor{LightSalmon4}{rgb}{0.55,0.34,0.26}
\definecolor{LightSalmon}{rgb}{1.00,0.63,0.48}
\definecolor{LightSeaGreen}{rgb}{0.13,0.70,0.67}
\definecolor{LightSkyBlue1}{rgb}{0.69,0.89,1.00}
\definecolor{LightSkyBlue2}{rgb}{0.64,0.83,0.93}
\definecolor{LightSkyBlue3}{rgb}{0.55,0.71,0.80}
\definecolor{LightSkyBlue4}{rgb}{0.38,0.48,0.55}
\definecolor{LightSkyBlue}{rgb}{0.53,0.81,0.98}
\definecolor{LightSlateBlue}{rgb}{0.52,0.44,1.00}
\definecolor{LightSlateGray}{rgb}{0.47,0.53,0.60}
\definecolor{LightSlateGrey}{rgb}{0.47,0.53,0.60}
\definecolor{LightSteelBlue1}{rgb}{0.79,0.88,1.00}
\definecolor{LightSteelBlue2}{rgb}{0.74,0.82,0.93}
\definecolor{LightSteelBlue3}{rgb}{0.64,0.71,0.80}
\definecolor{LightSteelBlue4}{rgb}{0.43,0.48,0.55}
\definecolor{LightSteelBlue}{rgb}{0.69,0.77,0.87}
\definecolor{LightYellow1}{rgb}{1.00,1.00,0.88}
\definecolor{LightYellow2}{rgb}{0.93,0.93,0.82}
\definecolor{LightYellow3}{rgb}{0.80,0.80,0.71}
\definecolor{LightYellow4}{rgb}{0.55,0.55,0.48}
\definecolor{LightYellow}{rgb}{1.00,1.00,0.88}
\definecolor{LimeGreen}{rgb}{0.20,0.80,0.20}
\definecolor{MediumAquamarine}{rgb}{0.40,0.80,0.67}
\definecolor{MediumBlue}{rgb}{0.00,0.00,0.80}
\definecolor{MediumOrchid1}{rgb}{0.88,0.40,1.00}
\definecolor{MediumOrchid2}{rgb}{0.82,0.37,0.93}
\definecolor{MediumOrchid3}{rgb}{0.71,0.32,0.80}
\definecolor{MediumOrchid4}{rgb}{0.48,0.22,0.55}
\definecolor{MediumOrchid}{rgb}{0.73,0.33,0.83}
\definecolor{MediumPurple1}{rgb}{0.67,0.51,1.00}
\definecolor{MediumPurple2}{rgb}{0.62,0.47,0.93}
\definecolor{MediumPurple3}{rgb}{0.54,0.41,0.80}
\definecolor{MediumPurple4}{rgb}{0.36,0.28,0.55}
\definecolor{MediumPurple}{rgb}{0.58,0.44,0.86}
\definecolor{MediumSeaGreen}{rgb}{0.24,0.70,0.44}
\definecolor{MediumSlateBlue}{rgb}{0.48,0.41,0.93}
\definecolor{MediumSpringGreen}{rgb}{0.00,0.98,0.60}
\definecolor{MediumTurquoise}{rgb}{0.28,0.82,0.80}
\definecolor{MediumVioletRed}{rgb}{0.78,0.08,0.52}
\definecolor{MidnightBlue}{rgb}{0.10,0.10,0.44}
\definecolor{MintCream}{rgb}{0.96,1.00,0.98}
\definecolor{MistyRose1}{rgb}{1.00,0.89,0.88}
\definecolor{MistyRose2}{rgb}{0.93,0.84,0.82}
\definecolor{MistyRose3}{rgb}{0.80,0.72,0.71}
\definecolor{MistyRose4}{rgb}{0.55,0.49,0.48}
\definecolor{MistyRose}{rgb}{1.00,0.89,0.88}
\definecolor{NavajoWhite1}{rgb}{1.00,0.87,0.68}
\definecolor{NavajoWhite2}{rgb}{0.93,0.81,0.63}
\definecolor{NavajoWhite3}{rgb}{0.80,0.70,0.55}
\definecolor{NavajoWhite4}{rgb}{0.55,0.47,0.37}
\definecolor{NavajoWhite}{rgb}{1.00,0.87,0.68}
\definecolor{NavyBlue}{rgb}{0.00,0.00,0.50}
\definecolor{OldLace}{rgb}{0.99,0.96,0.90}
\definecolor{OliveDrab1}{rgb}{0.75,1.00,0.24}
\definecolor{OliveDrab2}{rgb}{0.70,0.93,0.23}
\definecolor{OliveDrab3}{rgb}{0.60,0.80,0.20}
\definecolor{OliveDrab4}{rgb}{0.41,0.55,0.13}
\definecolor{OliveDrab}{rgb}{0.42,0.56,0.14}
\definecolor{OrangeRed1}{rgb}{1.00,0.27,0.00}
\definecolor{OrangeRed2}{rgb}{0.93,0.25,0.00}
\definecolor{OrangeRed3}{rgb}{0.80,0.22,0.00}
\definecolor{OrangeRed4}{rgb}{0.55,0.15,0.00}
\definecolor{OrangeRed}{rgb}{1.00,0.27,0.00}
\definecolor{PaleGoldenrod}{rgb}{0.93,0.91,0.67}
\definecolor{PaleGreen1}{rgb}{0.60,1.00,0.60}
\definecolor{PaleGreen2}{rgb}{0.56,0.93,0.56}
\definecolor{PaleGreen3}{rgb}{0.49,0.80,0.49}
\definecolor{PaleGreen4}{rgb}{0.33,0.55,0.33}
\definecolor{PaleGreen}{rgb}{0.60,0.98,0.60}
\definecolor{PaleTurquoise1}{rgb}{0.73,1.00,1.00}
\definecolor{PaleTurquoise2}{rgb}{0.68,0.93,0.93}
\definecolor{PaleTurquoise3}{rgb}{0.59,0.80,0.80}
\definecolor{PaleTurquoise4}{rgb}{0.40,0.55,0.55}
\definecolor{PaleTurquoise}{rgb}{0.69,0.93,0.93}
\definecolor{PaleVioletRed1}{rgb}{1.00,0.51,0.67}
\definecolor{PaleVioletRed2}{rgb}{0.93,0.47,0.62}
\definecolor{PaleVioletRed3}{rgb}{0.80,0.41,0.54}
\definecolor{PaleVioletRed4}{rgb}{0.55,0.28,0.36}
\definecolor{PaleVioletRed}{rgb}{0.86,0.44,0.58}
\definecolor{PapayaWhip}{rgb}{1.00,0.94,0.84}
\definecolor{PeachPuff1}{rgb}{1.00,0.85,0.73}
\definecolor{PeachPuff2}{rgb}{0.93,0.80,0.68}
\definecolor{PeachPuff3}{rgb}{0.80,0.69,0.58}
\definecolor{PeachPuff4}{rgb}{0.55,0.47,0.40}
\definecolor{PeachPuff}{rgb}{1.00,0.85,0.73}
\definecolor{PowderBlue}{rgb}{0.69,0.88,0.90}
\definecolor{RosyBrown1}{rgb}{1.00,0.76,0.76}
\definecolor{RosyBrown2}{rgb}{0.93,0.71,0.71}
\definecolor{RosyBrown3}{rgb}{0.80,0.61,0.61}
\definecolor{RosyBrown4}{rgb}{0.55,0.41,0.41}
\definecolor{RosyBrown}{rgb}{0.74,0.56,0.56}
\definecolor{RoyalBlue1}{rgb}{0.28,0.46,1.00}
\definecolor{RoyalBlue2}{rgb}{0.26,0.43,0.93}
\definecolor{RoyalBlue3}{rgb}{0.23,0.37,0.80}
\definecolor{RoyalBlue4}{rgb}{0.15,0.25,0.55}
\definecolor{RoyalBlue}{rgb}{0.25,0.41,0.88}
\definecolor{SaddleBrown}{rgb}{0.55,0.27,0.07}
\definecolor{SandyBrown}{rgb}{0.96,0.64,0.38}
\definecolor{SeaGreen1}{rgb}{0.33,1.00,0.62}
\definecolor{SeaGreen2}{rgb}{0.31,0.93,0.58}
\definecolor{SeaGreen3}{rgb}{0.26,0.80,0.50}
\definecolor{SeaGreen4}{rgb}{0.18,0.55,0.34}
\definecolor{SeaGreen}{rgb}{0.18,0.55,0.34}
\definecolor{SkyBlue1}{rgb}{0.53,0.81,1.00}
\definecolor{SkyBlue2}{rgb}{0.49,0.75,0.93}
\definecolor{SkyBlue3}{rgb}{0.42,0.65,0.80}
\definecolor{SkyBlue4}{rgb}{0.29,0.44,0.55}
\definecolor{SkyBlue}{rgb}{0.53,0.81,0.92}
\definecolor{SlateBlue1}{rgb}{0.51,0.44,1.00}
\definecolor{SlateBlue2}{rgb}{0.48,0.40,0.93}
\definecolor{SlateBlue3}{rgb}{0.41,0.35,0.80}
\definecolor{SlateBlue4}{rgb}{0.28,0.24,0.55}
\definecolor{SlateBlue}{rgb}{0.42,0.35,0.80}
\definecolor{SlateGray1}{rgb}{0.78,0.89,1.00}
\definecolor{SlateGray2}{rgb}{0.73,0.83,0.93}
\definecolor{SlateGray3}{rgb}{0.62,0.71,0.80}
\definecolor{SlateGray4}{rgb}{0.42,0.48,0.55}
\definecolor{SlateGray}{rgb}{0.44,0.50,0.56}
\definecolor{SlateGrey}{rgb}{0.44,0.50,0.56}
\definecolor{SpringGreen1}{rgb}{0.00,1.00,0.50}
\definecolor{SpringGreen2}{rgb}{0.00,0.93,0.46}
\definecolor{SpringGreen3}{rgb}{0.00,0.80,0.40}
\definecolor{SpringGreen4}{rgb}{0.00,0.55,0.27}
\definecolor{SpringGreen}{rgb}{0.00,1.00,0.50}
\definecolor{SteelBlue1}{rgb}{0.39,0.72,1.00}
\definecolor{SteelBlue2}{rgb}{0.36,0.67,0.93}
\definecolor{SteelBlue3}{rgb}{0.31,0.58,0.80}
\definecolor{SteelBlue4}{rgb}{0.21,0.39,0.55}
\definecolor{SteelBlue}{rgb}{0.27,0.51,0.71}
\definecolor{VioletRed1}{rgb}{1.00,0.24,0.59}
\definecolor{VioletRed2}{rgb}{0.93,0.23,0.55}
\definecolor{VioletRed3}{rgb}{0.80,0.20,0.47}
\definecolor{VioletRed4}{rgb}{0.55,0.13,0.32}
\definecolor{VioletRed}{rgb}{0.82,0.13,0.56}
\definecolor{WhiteSmoke}{rgb}{0.96,0.96,0.96}
\definecolor{YellowGreen}{rgb}{0.60,0.80,0.20}
\definecolor{aliceblue}{rgb}{0.94,0.97,1.00}
\definecolor{antiquewhite}{rgb}{0.98,0.92,0.84}
\definecolor{aquamarine1}{rgb}{0.50,1.00,0.83}
\definecolor{aquamarine2}{rgb}{0.46,0.93,0.78}
\definecolor{aquamarine3}{rgb}{0.40,0.80,0.67}
\definecolor{aquamarine4}{rgb}{0.27,0.55,0.45}
\definecolor{aquamarine}{rgb}{0.50,1.00,0.83}
\definecolor{azure1}{rgb}{0.94,1.00,1.00}
\definecolor{azure2}{rgb}{0.88,0.93,0.93}
\definecolor{azure3}{rgb}{0.76,0.80,0.80}
\definecolor{azure4}{rgb}{0.51,0.55,0.55}
\definecolor{azure}{rgb}{0.94,1.00,1.00}
\definecolor{beige}{rgb}{0.96,0.96,0.86}
\definecolor{bisque1}{rgb}{1.00,0.89,0.77}
\definecolor{bisque2}{rgb}{0.93,0.84,0.72}
\definecolor{bisque3}{rgb}{0.80,0.72,0.62}
\definecolor{bisque4}{rgb}{0.55,0.49,0.42}
\definecolor{bisque}{rgb}{1.00,0.89,0.77}
\definecolor{black}{rgb}{0.00,0.00,0.00}
\definecolor{blanchedalmond}{rgb}{1.00,0.92,0.80}
\definecolor{blue1}{rgb}{0.00,0.00,1.00}
\definecolor{blue2}{rgb}{0.00,0.00,0.93}
\definecolor{blue3}{rgb}{0.00,0.00,0.80}
\definecolor{blue4}{rgb}{0.00,0.00,0.55}
\definecolor{blueviolet}{rgb}{0.54,0.17,0.89}
\definecolor{blue}{rgb}{0.00,0.00,1.00}
\definecolor{brown1}{rgb}{1.00,0.25,0.25}
\definecolor{brown2}{rgb}{0.93,0.23,0.23}
\definecolor{brown3}{rgb}{0.80,0.20,0.20}
\definecolor{brown4}{rgb}{0.55,0.14,0.14}
\definecolor{brown}{rgb}{0.65,0.16,0.16}
\definecolor{burlywood1}{rgb}{1.00,0.83,0.61}
\definecolor{burlywood2}{rgb}{0.93,0.77,0.57}
\definecolor{burlywood3}{rgb}{0.80,0.67,0.49}
\definecolor{burlywood4}{rgb}{0.55,0.45,0.33}
\definecolor{burlywood}{rgb}{0.87,0.72,0.53}
\definecolor{cadetblue}{rgb}{0.37,0.62,0.63}
\definecolor{chartreuse1}{rgb}{0.50,1.00,0.00}
\definecolor{chartreuse2}{rgb}{0.46,0.93,0.00}
\definecolor{chartreuse3}{rgb}{0.40,0.80,0.00}
\definecolor{chartreuse4}{rgb}{0.27,0.55,0.00}
\definecolor{chartreuse}{rgb}{0.50,1.00,0.00}
\definecolor{chocolate1}{rgb}{1.00,0.50,0.14}
\definecolor{chocolate2}{rgb}{0.93,0.46,0.13}
\definecolor{chocolate3}{rgb}{0.80,0.40,0.11}
\definecolor{chocolate4}{rgb}{0.55,0.27,0.07}
\definecolor{chocolate}{rgb}{0.82,0.41,0.12}
\definecolor{coral1}{rgb}{1.00,0.45,0.34}
\definecolor{coral2}{rgb}{0.93,0.42,0.31}
\definecolor{coral3}{rgb}{0.80,0.36,0.27}
\definecolor{coral4}{rgb}{0.55,0.24,0.18}
\definecolor{coral}{rgb}{1.00,0.50,0.31}
\definecolor{cornflowerblue}{rgb}{0.39,0.58,0.93}
\definecolor{cornsilk1}{rgb}{1.00,0.97,0.86}
\definecolor{cornsilk2}{rgb}{0.93,0.91,0.80}
\definecolor{cornsilk3}{rgb}{0.80,0.78,0.69}
\definecolor{cornsilk4}{rgb}{0.55,0.53,0.47}
\definecolor{cornsilk}{rgb}{1.00,0.97,0.86}
\definecolor{cyan1}{rgb}{0.00,1.00,1.00}
\definecolor{cyan2}{rgb}{0.00,0.93,0.93}
\definecolor{cyan3}{rgb}{0.00,0.80,0.80}
\definecolor{cyan4}{rgb}{0.00,0.55,0.55}
\definecolor{cyan}{rgb}{0.00,1.00,1.00}
\definecolor{darkblue}{rgb}{0.00,0.00,0.55}
\definecolor{darkcyan}{rgb}{0.00,0.55,0.55}
\definecolor{darkgoldenrod}{rgb}{0.72,0.53,0.04}
\definecolor{darkgray}{rgb}{0.66,0.66,0.66}
\definecolor{darkgreen}{rgb}{0.00,0.39,0.00}
\definecolor{darkgrey}{rgb}{0.66,0.66,0.66}
\definecolor{darkkhaki}{rgb}{0.74,0.72,0.42}
\definecolor{darkmagenta}{rgb}{0.55,0.00,0.55}
\definecolor{darkolive}{rgb}{0.33,0.42,0.18}
\definecolor{darkorange}{rgb}{1.00,0.55,0.00}
\definecolor{darkorchid}{rgb}{0.60,0.20,0.80}
\definecolor{darkred}{rgb}{0.55,0.00,0.00}
\definecolor{darksalmon}{rgb}{0.91,0.59,0.48}
\definecolor{darksea}{rgb}{0.56,0.74,0.56}
\definecolor{darkslate}{rgb}{0.18,0.31,0.31}
\definecolor{darkslate}{rgb}{0.18,0.31,0.31}
\definecolor{darkslate}{rgb}{0.28,0.24,0.55}
\definecolor{darkturquoise}{rgb}{0.00,0.81,0.82}
\definecolor{darkviolet}{rgb}{0.58,0.00,0.83}
\definecolor{deeppink}{rgb}{1.00,0.08,0.58}
\definecolor{deepsky}{rgb}{0.00,0.75,1.00}
\definecolor{dimgray}{rgb}{0.41,0.41,0.41}
\definecolor{dimgrey}{rgb}{0.41,0.41,0.41}
\definecolor{dodgerblue}{rgb}{0.12,0.56,1.00}
\definecolor{firebrick1}{rgb}{1.00,0.19,0.19}
\definecolor{firebrick2}{rgb}{0.93,0.17,0.17}
\definecolor{firebrick3}{rgb}{0.80,0.15,0.15}
\definecolor{firebrick4}{rgb}{0.55,0.10,0.10}
\definecolor{firebrick}{rgb}{0.70,0.13,0.13}
\definecolor{floralwhite}{rgb}{1.00,0.98,0.94}
\definecolor{forestgreen}{rgb}{0.13,0.55,0.13}
\definecolor{gainsboro}{rgb}{0.86,0.86,0.86}
\definecolor{ghostwhite}{rgb}{0.97,0.97,1.00}
\definecolor{gold1}{rgb}{1.00,0.84,0.00}
\definecolor{gold2}{rgb}{0.93,0.79,0.00}
\definecolor{gold3}{rgb}{0.80,0.68,0.00}
\definecolor{gold4}{rgb}{0.55,0.46,0.00}
\definecolor{goldenrod1}{rgb}{1.00,0.76,0.15}
\definecolor{goldenrod2}{rgb}{0.93,0.71,0.13}
\definecolor{goldenrod3}{rgb}{0.80,0.61,0.11}
\definecolor{goldenrod4}{rgb}{0.55,0.41,0.08}
\definecolor{goldenrod}{rgb}{0.85,0.65,0.13}
\definecolor{gold}{rgb}{1.00,0.84,0.00}
\definecolor{gray0}{rgb}{0.00,0.00,0.00}
\definecolor{gray100}{rgb}{1.00,1.00,1.00}
\definecolor{gray10}{rgb}{0.10,0.10,0.10}
\definecolor{gray11}{rgb}{0.11,0.11,0.11}
\definecolor{gray12}{rgb}{0.12,0.12,0.12}
\definecolor{gray13}{rgb}{0.13,0.13,0.13}
\definecolor{gray14}{rgb}{0.14,0.14,0.14}
\definecolor{gray15}{rgb}{0.15,0.15,0.15}
\definecolor{gray16}{rgb}{0.16,0.16,0.16}
\definecolor{gray17}{rgb}{0.17,0.17,0.17}
\definecolor{gray18}{rgb}{0.18,0.18,0.18}
\definecolor{gray19}{rgb}{0.19,0.19,0.19}
\definecolor{gray1}{rgb}{0.01,0.01,0.01}
\definecolor{gray20}{rgb}{0.20,0.20,0.20}
\definecolor{gray21}{rgb}{0.21,0.21,0.21}
\definecolor{gray22}{rgb}{0.22,0.22,0.22}
\definecolor{gray23}{rgb}{0.23,0.23,0.23}
\definecolor{gray24}{rgb}{0.24,0.24,0.24}
\definecolor{gray25}{rgb}{0.25,0.25,0.25}
\definecolor{gray26}{rgb}{0.26,0.26,0.26}
\definecolor{gray27}{rgb}{0.27,0.27,0.27}
\definecolor{gray28}{rgb}{0.28,0.28,0.28}
\definecolor{gray29}{rgb}{0.29,0.29,0.29}
\definecolor{gray2}{rgb}{0.02,0.02,0.02}
\definecolor{gray30}{rgb}{0.30,0.30,0.30}
\definecolor{gray31}{rgb}{0.31,0.31,0.31}
\definecolor{gray32}{rgb}{0.32,0.32,0.32}
\definecolor{gray33}{rgb}{0.33,0.33,0.33}
\definecolor{gray34}{rgb}{0.34,0.34,0.34}
\definecolor{gray35}{rgb}{0.35,0.35,0.35}
\definecolor{gray36}{rgb}{0.36,0.36,0.36}
\definecolor{gray37}{rgb}{0.37,0.37,0.37}
\definecolor{gray38}{rgb}{0.38,0.38,0.38}
\definecolor{gray39}{rgb}{0.39,0.39,0.39}
\definecolor{gray3}{rgb}{0.03,0.03,0.03}
\definecolor{gray40}{rgb}{0.40,0.40,0.40}
\definecolor{gray41}{rgb}{0.41,0.41,0.41}
\definecolor{gray42}{rgb}{0.42,0.42,0.42}
\definecolor{gray43}{rgb}{0.43,0.43,0.43}
\definecolor{gray44}{rgb}{0.44,0.44,0.44}
\definecolor{gray45}{rgb}{0.45,0.45,0.45}
\definecolor{gray46}{rgb}{0.46,0.46,0.46}
\definecolor{gray47}{rgb}{0.47,0.47,0.47}
\definecolor{gray48}{rgb}{0.48,0.48,0.48}
\definecolor{gray49}{rgb}{0.49,0.49,0.49}
\definecolor{gray4}{rgb}{0.04,0.04,0.04}
\definecolor{gray50}{rgb}{0.50,0.50,0.50}
\definecolor{gray51}{rgb}{0.51,0.51,0.51}
\definecolor{gray52}{rgb}{0.52,0.52,0.52}
\definecolor{gray53}{rgb}{0.53,0.53,0.53}
\definecolor{gray54}{rgb}{0.54,0.54,0.54}
\definecolor{gray55}{rgb}{0.55,0.55,0.55}
\definecolor{gray56}{rgb}{0.56,0.56,0.56}
\definecolor{gray57}{rgb}{0.57,0.57,0.57}
\definecolor{gray58}{rgb}{0.58,0.58,0.58}
\definecolor{gray59}{rgb}{0.59,0.59,0.59}
\definecolor{gray5}{rgb}{0.05,0.05,0.05}
\definecolor{gray60}{rgb}{0.60,0.60,0.60}
\definecolor{gray61}{rgb}{0.61,0.61,0.61}
\definecolor{gray62}{rgb}{0.62,0.62,0.62}
\definecolor{gray63}{rgb}{0.63,0.63,0.63}
\definecolor{gray64}{rgb}{0.64,0.64,0.64}
\definecolor{gray65}{rgb}{0.65,0.65,0.65}
\definecolor{gray66}{rgb}{0.66,0.66,0.66}
\definecolor{gray67}{rgb}{0.67,0.67,0.67}
\definecolor{gray68}{rgb}{0.68,0.68,0.68}
\definecolor{gray69}{rgb}{0.69,0.69,0.69}
\definecolor{gray6}{rgb}{0.06,0.06,0.06}
\definecolor{gray70}{rgb}{0.70,0.70,0.70}
\definecolor{gray71}{rgb}{0.71,0.71,0.71}
\definecolor{gray72}{rgb}{0.72,0.72,0.72}
\definecolor{gray73}{rgb}{0.73,0.73,0.73}
\definecolor{gray74}{rgb}{0.74,0.74,0.74}
\definecolor{gray75}{rgb}{0.75,0.75,0.75}
\definecolor{gray76}{rgb}{0.76,0.76,0.76}
\definecolor{gray77}{rgb}{0.77,0.77,0.77}
\definecolor{gray78}{rgb}{0.78,0.78,0.78}
\definecolor{gray79}{rgb}{0.79,0.79,0.79}
\definecolor{gray7}{rgb}{0.07,0.07,0.07}
\definecolor{gray80}{rgb}{0.80,0.80,0.80}
\definecolor{gray81}{rgb}{0.81,0.81,0.81}
\definecolor{gray82}{rgb}{0.82,0.82,0.82}
\definecolor{gray83}{rgb}{0.83,0.83,0.83}
\definecolor{gray84}{rgb}{0.84,0.84,0.84}
\definecolor{gray85}{rgb}{0.85,0.85,0.85}
\definecolor{gray86}{rgb}{0.86,0.86,0.86}
\definecolor{gray87}{rgb}{0.87,0.87,0.87}
\definecolor{gray88}{rgb}{0.88,0.88,0.88}
\definecolor{gray89}{rgb}{0.89,0.89,0.89}
\definecolor{gray8}{rgb}{0.08,0.08,0.08}
\definecolor{gray90}{rgb}{0.90,0.90,0.90}
\definecolor{gray91}{rgb}{0.91,0.91,0.91}
\definecolor{gray92}{rgb}{0.92,0.92,0.92}
\definecolor{gray93}{rgb}{0.93,0.93,0.93}
\definecolor{gray94}{rgb}{0.94,0.94,0.94}
\definecolor{gray95}{rgb}{0.95,0.95,0.95}
\definecolor{gray96}{rgb}{0.96,0.96,0.96}
\definecolor{gray97}{rgb}{0.97,0.97,0.97}
\definecolor{gray98}{rgb}{0.98,0.98,0.98}
\definecolor{gray99}{rgb}{0.99,0.99,0.99}
\definecolor{gray9}{rgb}{0.09,0.09,0.09}
\definecolor{gray}{rgb}{0.75,0.75,0.75}
\definecolor{green1}{rgb}{0.00,1.00,0.00}
\definecolor{green2}{rgb}{0.00,0.93,0.00}
\definecolor{green3}{rgb}{0.00,0.80,0.00}
\definecolor{green4}{rgb}{0.00,0.55,0.00}
\definecolor{greenyellow}{rgb}{0.68,1.00,0.18}
\definecolor{green}{rgb}{0.00,1.00,0.00}
\definecolor{grey0}{rgb}{0.00,0.00,0.00}
\definecolor{grey100}{rgb}{1.00,1.00,1.00}
\definecolor{grey10}{rgb}{0.10,0.10,0.10}
\definecolor{grey11}{rgb}{0.11,0.11,0.11}
\definecolor{grey12}{rgb}{0.12,0.12,0.12}
\definecolor{grey13}{rgb}{0.13,0.13,0.13}
\definecolor{grey14}{rgb}{0.14,0.14,0.14}
\definecolor{grey15}{rgb}{0.15,0.15,0.15}
\definecolor{grey16}{rgb}{0.16,0.16,0.16}
\definecolor{grey17}{rgb}{0.17,0.17,0.17}
\definecolor{grey18}{rgb}{0.18,0.18,0.18}
\definecolor{grey19}{rgb}{0.19,0.19,0.19}
\definecolor{grey1}{rgb}{0.01,0.01,0.01}
\definecolor{grey20}{rgb}{0.20,0.20,0.20}
\definecolor{grey21}{rgb}{0.21,0.21,0.21}
\definecolor{grey22}{rgb}{0.22,0.22,0.22}
\definecolor{grey23}{rgb}{0.23,0.23,0.23}
\definecolor{grey24}{rgb}{0.24,0.24,0.24}
\definecolor{grey25}{rgb}{0.25,0.25,0.25}
\definecolor{grey26}{rgb}{0.26,0.26,0.26}
\definecolor{grey27}{rgb}{0.27,0.27,0.27}
\definecolor{grey28}{rgb}{0.28,0.28,0.28}
\definecolor{grey29}{rgb}{0.29,0.29,0.29}
\definecolor{grey2}{rgb}{0.02,0.02,0.02}
\definecolor{grey30}{rgb}{0.30,0.30,0.30}
\definecolor{grey31}{rgb}{0.31,0.31,0.31}
\definecolor{grey32}{rgb}{0.32,0.32,0.32}
\definecolor{grey33}{rgb}{0.33,0.33,0.33}
\definecolor{grey34}{rgb}{0.34,0.34,0.34}
\definecolor{grey35}{rgb}{0.35,0.35,0.35}
\definecolor{grey36}{rgb}{0.36,0.36,0.36}
\definecolor{grey37}{rgb}{0.37,0.37,0.37}
\definecolor{grey38}{rgb}{0.38,0.38,0.38}
\definecolor{grey39}{rgb}{0.39,0.39,0.39}
\definecolor{grey3}{rgb}{0.03,0.03,0.03}
\definecolor{grey40}{rgb}{0.40,0.40,0.40}
\definecolor{grey41}{rgb}{0.41,0.41,0.41}
\definecolor{grey42}{rgb}{0.42,0.42,0.42}
\definecolor{grey43}{rgb}{0.43,0.43,0.43}
\definecolor{grey44}{rgb}{0.44,0.44,0.44}
\definecolor{grey45}{rgb}{0.45,0.45,0.45}
\definecolor{grey46}{rgb}{0.46,0.46,0.46}
\definecolor{grey47}{rgb}{0.47,0.47,0.47}
\definecolor{grey48}{rgb}{0.48,0.48,0.48}
\definecolor{grey49}{rgb}{0.49,0.49,0.49}
\definecolor{grey4}{rgb}{0.04,0.04,0.04}
\definecolor{grey50}{rgb}{0.50,0.50,0.50}
\definecolor{grey51}{rgb}{0.51,0.51,0.51}
\definecolor{grey52}{rgb}{0.52,0.52,0.52}
\definecolor{grey53}{rgb}{0.53,0.53,0.53}
\definecolor{grey54}{rgb}{0.54,0.54,0.54}
\definecolor{grey55}{rgb}{0.55,0.55,0.55}
\definecolor{grey56}{rgb}{0.56,0.56,0.56}
\definecolor{grey57}{rgb}{0.57,0.57,0.57}
\definecolor{grey58}{rgb}{0.58,0.58,0.58}
\definecolor{grey59}{rgb}{0.59,0.59,0.59}
\definecolor{grey5}{rgb}{0.05,0.05,0.05}
\definecolor{grey60}{rgb}{0.60,0.60,0.60}
\definecolor{grey61}{rgb}{0.61,0.61,0.61}
\definecolor{grey62}{rgb}{0.62,0.62,0.62}
\definecolor{grey63}{rgb}{0.63,0.63,0.63}
\definecolor{grey64}{rgb}{0.64,0.64,0.64}
\definecolor{grey65}{rgb}{0.65,0.65,0.65}
\definecolor{grey66}{rgb}{0.66,0.66,0.66}
\definecolor{grey67}{rgb}{0.67,0.67,0.67}
\definecolor{grey68}{rgb}{0.68,0.68,0.68}
\definecolor{grey69}{rgb}{0.69,0.69,0.69}
\definecolor{grey6}{rgb}{0.06,0.06,0.06}
\definecolor{grey70}{rgb}{0.70,0.70,0.70}
\definecolor{grey71}{rgb}{0.71,0.71,0.71}
\definecolor{grey72}{rgb}{0.72,0.72,0.72}
\definecolor{grey73}{rgb}{0.73,0.73,0.73}
\definecolor{grey74}{rgb}{0.74,0.74,0.74}
\definecolor{grey75}{rgb}{0.75,0.75,0.75}
\definecolor{grey76}{rgb}{0.76,0.76,0.76}
\definecolor{grey77}{rgb}{0.77,0.77,0.77}
\definecolor{grey78}{rgb}{0.78,0.78,0.78}
\definecolor{grey79}{rgb}{0.79,0.79,0.79}
\definecolor{grey7}{rgb}{0.07,0.07,0.07}
\definecolor{grey80}{rgb}{0.80,0.80,0.80}
\definecolor{grey81}{rgb}{0.81,0.81,0.81}
\definecolor{grey82}{rgb}{0.82,0.82,0.82}
\definecolor{grey83}{rgb}{0.83,0.83,0.83}
\definecolor{grey84}{rgb}{0.84,0.84,0.84}
\definecolor{grey85}{rgb}{0.85,0.85,0.85}
\definecolor{grey86}{rgb}{0.86,0.86,0.86}
\definecolor{grey87}{rgb}{0.87,0.87,0.87}
\definecolor{grey88}{rgb}{0.88,0.88,0.88}
\definecolor{grey89}{rgb}{0.89,0.89,0.89}
\definecolor{grey8}{rgb}{0.08,0.08,0.08}
\definecolor{grey90}{rgb}{0.90,0.90,0.90}
\definecolor{grey91}{rgb}{0.91,0.91,0.91}
\definecolor{grey92}{rgb}{0.92,0.92,0.92}
\definecolor{grey93}{rgb}{0.93,0.93,0.93}
\definecolor{grey94}{rgb}{0.94,0.94,0.94}
\definecolor{grey95}{rgb}{0.95,0.95,0.95}
\definecolor{grey96}{rgb}{0.96,0.96,0.96}
\definecolor{grey97}{rgb}{0.97,0.97,0.97}
\definecolor{grey98}{rgb}{0.98,0.98,0.98}
\definecolor{grey99}{rgb}{0.99,0.99,0.99}
\definecolor{grey9}{rgb}{0.09,0.09,0.09}
\definecolor{grey}{rgb}{0.75,0.75,0.75}
\definecolor{honeydew1}{rgb}{0.94,1.00,0.94}
\definecolor{honeydew2}{rgb}{0.88,0.93,0.88}
\definecolor{honeydew3}{rgb}{0.76,0.80,0.76}
\definecolor{honeydew4}{rgb}{0.51,0.55,0.51}
\definecolor{honeydew}{rgb}{0.94,1.00,0.94}
\definecolor{hotpink}{rgb}{1.00,0.41,0.71}
\definecolor{indianred}{rgb}{0.80,0.36,0.36}
\definecolor{ivory1}{rgb}{1.00,1.00,0.94}
\definecolor{ivory2}{rgb}{0.93,0.93,0.88}
\definecolor{ivory3}{rgb}{0.80,0.80,0.76}
\definecolor{ivory4}{rgb}{0.55,0.55,0.51}
\definecolor{ivory}{rgb}{1.00,1.00,0.94}
\definecolor{khaki1}{rgb}{1.00,0.96,0.56}
\definecolor{khaki2}{rgb}{0.93,0.90,0.52}
\definecolor{khaki3}{rgb}{0.80,0.78,0.45}
\definecolor{khaki4}{rgb}{0.55,0.53,0.31}
\definecolor{khaki}{rgb}{0.94,0.90,0.55}
\definecolor{lavenderblush}{rgb}{1.00,0.94,0.96}
\definecolor{lavender}{rgb}{0.90,0.90,0.98}
\definecolor{lawngreen}{rgb}{0.49,0.99,0.00}
\definecolor{lemonchiffon}{rgb}{1.00,0.98,0.80}
\definecolor{lightblue}{rgb}{0.68,0.85,0.90}
\definecolor{lightcoral}{rgb}{0.94,0.50,0.50}
\definecolor{lightcyan}{rgb}{0.88,1.00,1.00}
\definecolor{lightgoldenrod}{rgb}{0.93,0.87,0.51}
\definecolor{lightgoldenrod}{rgb}{0.98,0.98,0.82}
\definecolor{lightgray}{rgb}{0.83,0.83,0.83}
\definecolor{lightgreen}{rgb}{0.56,0.93,0.56}
\definecolor{lightgrey}{rgb}{0.83,0.83,0.83}
\definecolor{lightpink}{rgb}{1.00,0.71,0.76}
\definecolor{lightsalmon}{rgb}{1.00,0.63,0.48}
\definecolor{lightsea}{rgb}{0.13,0.70,0.67}
\definecolor{lightsky}{rgb}{0.53,0.81,0.98}
\definecolor{lightslate}{rgb}{0.47,0.53,0.60}
\definecolor{lightslate}{rgb}{0.47,0.53,0.60}
\definecolor{lightslate}{rgb}{0.52,0.44,1.00}
\definecolor{lightsteel}{rgb}{0.69,0.77,0.87}
\definecolor{lightyellow}{rgb}{1.00,1.00,0.88}
\definecolor{limegreen}{rgb}{0.20,0.80,0.20}
\definecolor{linen}{rgb}{0.98,0.94,0.90}
\definecolor{magenta1}{rgb}{1.00,0.00,1.00}
\definecolor{magenta2}{rgb}{0.93,0.00,0.93}
\definecolor{magenta3}{rgb}{0.80,0.00,0.80}
\definecolor{magenta4}{rgb}{0.55,0.00,0.55}
\definecolor{magenta}{rgb}{1.00,0.00,1.00}
\definecolor{maroon1}{rgb}{1.00,0.20,0.70}
\definecolor{maroon2}{rgb}{0.93,0.19,0.65}
\definecolor{maroon3}{rgb}{0.80,0.16,0.56}
\definecolor{maroon4}{rgb}{0.55,0.11,0.38}
\definecolor{maroon}{rgb}{0.69,0.19,0.38}
\definecolor{mediumaquamarine}{rgb}{0.40,0.80,0.67}
\definecolor{mediumblue}{rgb}{0.00,0.00,0.80}
\definecolor{mediumorchid}{rgb}{0.73,0.33,0.83}
\definecolor{mediumpurple}{rgb}{0.58,0.44,0.86}
\definecolor{mediumsea}{rgb}{0.24,0.70,0.44}
\definecolor{mediumslate}{rgb}{0.48,0.41,0.93}
\definecolor{mediumspring}{rgb}{0.00,0.98,0.60}
\definecolor{mediumturquoise}{rgb}{0.28,0.82,0.80}
\definecolor{mediumviolet}{rgb}{0.78,0.08,0.52}
\definecolor{midnightblue}{rgb}{0.10,0.10,0.44}
\definecolor{mintcream}{rgb}{0.96,1.00,0.98}
\definecolor{mistyrose}{rgb}{1.00,0.89,0.88}
\definecolor{moccasin}{rgb}{1.00,0.89,0.71}
\definecolor{navajowhite}{rgb}{1.00,0.87,0.68}
\definecolor{navyblue}{rgb}{0.00,0.00,0.50}
\definecolor{navy}{rgb}{0.00,0.00,0.50}
\definecolor{oldlace}{rgb}{0.99,0.96,0.90}
\definecolor{olivedrab}{rgb}{0.42,0.56,0.14}
\definecolor{orange1}{rgb}{1.00,0.65,0.00}
\definecolor{orange2}{rgb}{0.93,0.60,0.00}
\definecolor{orange3}{rgb}{0.80,0.52,0.00}
\definecolor{orange4}{rgb}{0.55,0.35,0.00}
\definecolor{orangered}{rgb}{1.00,0.27,0.00}
\definecolor{orange}{rgb}{1.00,0.65,0.00}
\definecolor{orchid1}{rgb}{1.00,0.51,0.98}
\definecolor{orchid2}{rgb}{0.93,0.48,0.91}
\definecolor{orchid3}{rgb}{0.80,0.41,0.79}
\definecolor{orchid4}{rgb}{0.55,0.28,0.54}
\definecolor{orchid}{rgb}{0.85,0.44,0.84}
\definecolor{palegoldenrod}{rgb}{0.93,0.91,0.67}
\definecolor{palegreen}{rgb}{0.60,0.98,0.60}
\definecolor{paleturquoise}{rgb}{0.69,0.93,0.93}
\definecolor{paleviolet}{rgb}{0.86,0.44,0.58}
\definecolor{papayawhip}{rgb}{1.00,0.94,0.84}
\definecolor{peachpuff}{rgb}{1.00,0.85,0.73}
\definecolor{peru}{rgb}{0.80,0.52,0.25}
\definecolor{pink1}{rgb}{1.00,0.71,0.77}
\definecolor{pink2}{rgb}{0.93,0.66,0.72}
\definecolor{pink3}{rgb}{0.80,0.57,0.62}
\definecolor{pink4}{rgb}{0.55,0.39,0.42}
\definecolor{pink}{rgb}{1.00,0.75,0.80}
\definecolor{plum1}{rgb}{1.00,0.73,1.00}
\definecolor{plum2}{rgb}{0.93,0.68,0.93}
\definecolor{plum3}{rgb}{0.80,0.59,0.80}
\definecolor{plum4}{rgb}{0.55,0.40,0.55}
\definecolor{plum}{rgb}{0.87,0.63,0.87}
\definecolor{powderblue}{rgb}{0.69,0.88,0.90}
\definecolor{purple1}{rgb}{0.61,0.19,1.00}
\definecolor{purple2}{rgb}{0.57,0.17,0.93}
\definecolor{purple3}{rgb}{0.49,0.15,0.80}
\definecolor{purple4}{rgb}{0.33,0.10,0.55}
\definecolor{purple}{rgb}{0.63,0.13,0.94}
\definecolor{red1}{rgb}{1.00,0.00,0.00}
\definecolor{red2}{rgb}{0.93,0.00,0.00}
\definecolor{red3}{rgb}{0.80,0.00,0.00}
\definecolor{red4}{rgb}{0.55,0.00,0.00}
\definecolor{red}{rgb}{1.00,0.00,0.00}
\definecolor{rosybrown}{rgb}{0.74,0.56,0.56}
\definecolor{royalblue}{rgb}{0.25,0.41,0.88}
\definecolor{saddlebrown}{rgb}{0.55,0.27,0.07}
\definecolor{salmon1}{rgb}{1.00,0.55,0.41}
\definecolor{salmon2}{rgb}{0.93,0.51,0.38}
\definecolor{salmon3}{rgb}{0.80,0.44,0.33}
\definecolor{salmon4}{rgb}{0.55,0.30,0.22}
\definecolor{salmon}{rgb}{0.98,0.50,0.45}
\definecolor{sandybrown}{rgb}{0.96,0.64,0.38}
\definecolor{seagreen}{rgb}{0.18,0.55,0.34}
\definecolor{seashell1}{rgb}{1.00,0.96,0.93}
\definecolor{seashell2}{rgb}{0.93,0.90,0.87}
\definecolor{seashell3}{rgb}{0.80,0.77,0.75}
\definecolor{seashell4}{rgb}{0.55,0.53,0.51}
\definecolor{seashell}{rgb}{1.00,0.96,0.93}
\definecolor{sienna1}{rgb}{1.00,0.51,0.28}
\definecolor{sienna2}{rgb}{0.93,0.47,0.26}
\definecolor{sienna3}{rgb}{0.80,0.41,0.22}
\definecolor{sienna4}{rgb}{0.55,0.28,0.15}
\definecolor{sienna}{rgb}{0.63,0.32,0.18}
\definecolor{skyblue}{rgb}{0.53,0.81,0.92}
\definecolor{slateblue}{rgb}{0.42,0.35,0.80}
\definecolor{slategray}{rgb}{0.44,0.50,0.56}
\definecolor{slategrey}{rgb}{0.44,0.50,0.56}
\definecolor{snow1}{rgb}{1.00,0.98,0.98}
\definecolor{snow2}{rgb}{0.93,0.91,0.91}
\definecolor{snow3}{rgb}{0.80,0.79,0.79}
\definecolor{snow4}{rgb}{0.55,0.54,0.54}
\definecolor{snow}{rgb}{1.00,0.98,0.98}
\definecolor{springgreen}{rgb}{0.00,1.00,0.50}
\definecolor{steelblue}{rgb}{0.27,0.51,0.71}
\definecolor{tan1}{rgb}{1.00,0.65,0.31}
\definecolor{tan2}{rgb}{0.93,0.60,0.29}
\definecolor{tan3}{rgb}{0.80,0.52,0.25}
\definecolor{tan4}{rgb}{0.55,0.35,0.17}
\definecolor{tan}{rgb}{0.82,0.71,0.55}
\definecolor{thistle1}{rgb}{1.00,0.88,1.00}
\definecolor{thistle2}{rgb}{0.93,0.82,0.93}
\definecolor{thistle3}{rgb}{0.80,0.71,0.80}
\definecolor{thistle4}{rgb}{0.55,0.48,0.55}
\definecolor{thistle}{rgb}{0.85,0.75,0.85}
\definecolor{tomato1}{rgb}{1.00,0.39,0.28}
\definecolor{tomato2}{rgb}{0.93,0.36,0.26}
\definecolor{tomato3}{rgb}{0.80,0.31,0.22}
\definecolor{tomato4}{rgb}{0.55,0.21,0.15}
\definecolor{tomato}{rgb}{1.00,0.39,0.28}
\definecolor{turquoise1}{rgb}{0.00,0.96,1.00}
\definecolor{turquoise2}{rgb}{0.00,0.90,0.93}
\definecolor{turquoise3}{rgb}{0.00,0.77,0.80}
\definecolor{turquoise4}{rgb}{0.00,0.53,0.55}
\definecolor{turquoise}{rgb}{0.25,0.88,0.82}
\definecolor{violetred}{rgb}{0.82,0.13,0.56}
\definecolor{violet}{rgb}{0.93,0.51,0.93}
\definecolor{wheat1}{rgb}{1.00,0.91,0.73}
\definecolor{wheat2}{rgb}{0.93,0.85,0.68}
\definecolor{wheat3}{rgb}{0.80,0.73,0.59}
\definecolor{wheat4}{rgb}{0.55,0.49,0.40}
\definecolor{wheat}{rgb}{0.96,0.87,0.70}
\definecolor{whitesmoke}{rgb}{0.96,0.96,0.96}
\definecolor{white}{rgb}{1.00,1.00,1.00}
\definecolor{yellow1}{rgb}{1.00,1.00,0.00}
\definecolor{yellow2}{rgb}{0.93,0.93,0.00}
\definecolor{yellow3}{rgb}{0.80,0.80,0.00}
\definecolor{yellow4}{rgb}{0.55,0.55,0.00}
\definecolor{yellowgreen}{rgb}{0.60,0.80,0.20}
\definecolor{yellow}{rgb}{1.00,1.00,0.00}

\title[Defining a Weak Lensing Experiment]{Defining a Weak Lensing Experiment in Space}
\author[M.\ Cropper et al.]
{\noindent
Mark Cropper$^{1}$\thanks{E-mail: m.cropper@ucl.ac.uk},
Henk Hoekstra$^{2}$\thanks{E-mail: hoekstra@strw.leidenuniv.nl},
Thomas Kitching$^{1,3}$\thanks{E-mail: t.kitching@ucl.ac.uk},
Richard Massey$^{3,4}$,\and
J\'er\^ome Amiaux$^{5}$,
Lance Miller$^{6}$,
Yannick Mellier$^{7,5}$,
Jason Rhodes$^{8,9}$,\and
Barnaby Rowe$^{10,9}$,
Sandrine Pires$^{5}$, 
Curtis Saxton$^{1}$ and
Roberto Scaramella$^{11}$ \vspace*{3mm}\\ 
$^{1}$ Mullard Space Science Laboratory, University College London, Holmbury St Mary, Dorking, Surrey RH5 6NT, UK\\
$^{2}$ Leiden Observatory, Leiden University, P.O.\ Box 9513, 2300 RA, Leiden, The Netherlands\\
$^{3}$ SUPA, Institute for Astronomy, University of Edinburgh, Royal Observatory, Blackford Hill, Edinburgh EH9 3HJ, UK \\
$^{4}$ Department of Physics, Durham University, South Road, Durham, DH1 3LE, UK\\
$^{5}$ Service dÕAstrophysique, CEA Saclay, Gif sur Yvette, 91191, France\\
$^{6}$ Department of Physics, University of Oxford, The Denys Wilkinson Building, Keble Road, Oxford, OX1 3RH, UK\\
$^{7}$ Institut dÕAstrophysique de Paris, UMR7095 CNRS, Universit\'e Pierre et Marie Curie, 98 bis Boulevard Arago, 75014 Paris, France\\
$^{8}$ Jet Propulsion Laboratory, California Institute of Technology, 4800 Oak Grove Drive, Pasadena, CA 91109, USA\\
$^{9}$ California Institute of Technology, 1200 E California Blvd., Pasadena, CA 91125, USA\\
$^{10}$ Department of Physics \& Astronomy, University College London, Gower Street, London WC1E 6BT, UK\\
$^{11}$ INAF, Osservatorio Astronomico di Roma, via Frascati 33, 00040 Monteporzio Catone, Italy\\
}

\begin{document}

\newcommand{\eu}{{\it Euclid }}
\newcommand{\etal}{{\it et al., }}

\newcommand{\dg} {^{\circ}}
\outer\def\gtae {$\buildrel {\lower3pt\hbox{$>$}} \over{\lower2pt\hbox{$\sim$}} $}
\outer\def\ltae {$\buildrel {\lower3pt\hbox{$<$}} \over{\lower2pt\hbox{$\sim$}} $}
\def\gaeq{\stackrel{>}{\scriptstyle \sim}}
\def\laeq{\stackrel{<}{\scriptstyle \sim}}
\newcommand{\ergscm} {ergs s$^{-1}$ cm$^{-2}$} 
\newcommand{\ergss} {ergs s$^{-1}$}
\newcommand{\ergsd} {ergs s$^{-1}$ $d^{2}_{100}$}
\newcommand{\pcmsq} {cm$^{-2}$}
\def\rchi{{${\chi}_{\nu}^{2}$}}
\newcommand{\Msun} {$M_{\odot}$}
\newcommand{\Mwd} {$M_{wd}$}
\newcommand{\Mbh}{$M_{\bullet}$}
\def\Mdot{\hbox{$\dot M$}}
\def\mdot{\hbox{$\dot m$}}

\newcommand{\ellipticity}{\varepsilon}
\newcommand{\mcm}{\mathcal{M}}
\newcommand{\mca}{\mathcal{A}}
\newcommand{\mcmp}{\mathcal{M}'}
\newcommand{\mcap}{\mathcal{A}'}
\newcommand{\eg}{\bmath{\ellipticity}_{\rm gal}}
\newcommand{\ep}{\bmath{\ellipticity}_{_{\rm C}}}
\newcommand{\ec}{\bmath{\ellipticity}_{_{\rm NC}}}
\newcommand{\eo}{\bmath{\ellipticity}_{\rm obs}}
\newcommand{\es}{\bmath{\ellipticity}_{\rm sys}}
\newcommand{\egi}{\ellipticity_{{\rm gal},i}}
\newcommand{\epi}{\ellipticity_{_{\rm C},i}}
\newcommand{\eci}{\ellipticity_{_{\rm NC},i}}
\newcommand{\eoi}{\ellipticity_{{\rm obs},i}}
\newcommand{\esi}{\ellipticity_{{\rm sys},i}}
\newcommand{\rg}{R_{\rm gal}}
\newcommand{\rp}{R_{_{\rm C}}}
\newcommand{\rc}{R_{_{\rm NC}}}
\newcommand{\ro}{R_{\rm obs}}
\newcommand{\bd}{\bmath{\delta}}
\newcommand{\pec}{P_{\ellipticity_{C}}}
\newcommand{\penc}{P_{\ellipticity_{NC}}}

\newcommand{\be}{\begin{equation}}
\newcommand{\ee}{\end{equation}}
\newcommand{\nn}{\nonumber \\}
\newcommand{\com}[1]{{#1}}
\newcommand{\red}[1]{\textcolor{black}{#1}}
\newcommand{\reda}[1]{\textcolor{black}{#1}}
\newcommand{\bluea}[1]{\textcolor{red}{#1}}
\newcommand{\blue}[1]{\textcolor{black}{#1}}
\newcommand{\green}[1]{\textcolor{green}{#1}}
\newcommand{\orange}[1]{\textcolor{orange}{#1}}
\newcommand{\dsg}[1]{\textcolor{black}{#1}}
\newcommand{\litg}[1]{\textcolor{LightGrey}{#1}}

\def\variance#1{\left\langle\left|#1\right|^2\right\rangle}

\newcommand{\xb}{\mbox{\bmath$x$}}\def\simlt{\lower.5ex\hbox{$\; \buildrel < \over \sim \;$}}
\def\simgt{\lower.5ex\hbox{$\; \buildrel > \over \sim \;$}}
\newcommand\solidrule[1][1cm]{\rule[0.5ex]{#1}{.4pt}}
\newcommand\dashedrule{\mbox{
\solidrule[1mm]\hspace{2mm}\solidrule[1mm]\hspace{2mm}\solidrule[1mm]\hspace{2mm}\solidrule[1mm]\hspace{2mm}\solidrule[1mm]\hspace{2mm}\solidrule[1mm]\hspace{2mm}\solidrule[1mm]\hspace{2mm}\solidrule[1mm]\hspace{2mm}\solidrule[1mm]\hspace{2mm}\solidrule[1mm]\hspace{2mm}\solidrule[1mm]\hspace{2mm}\solidrule[1mm]\hspace{2mm}
 }}

\date{}


\maketitle

\label{firstpage}

\begin{abstract}
This paper describes the definition of a typical next-generation space-based weak gravitational lensing experiment. We first adopt a set of top-level science requirements from the literature, based on the scale and depth of the galaxy sample, and the avoidance of systematic effects in the measurements which would bias the derived shear values. We then identify and categorise the contributing factors to the systematic effects, combining them with the correct weighting, in such a way as to fit within the top-level requirements. We present techniques which permit the performance to be evaluated and explore the limits at which the contributing factors can be managed. Besides the modelling biases resulting from the use of weighted moments, the main contributing factors are the reconstruction of the instrument point spread function (PSF), which is derived from the stellar images on the image, and the correction of the charge transfer inefficiency (CTI) in the CCD detectors caused by radiation damage. 
\end{abstract}

\begin{keywords}
gravitational lensing: weak -- methods: statistical -- space vehicles: instruments -- cosmological parameters -- cosmology: observations. 
\end{keywords}

\section{Introduction}

In the current ``Concordance Model'' of cosmology, approximately three quarters of the energy density of the Universe consists of Dark Energy, and one fifth of Dark Matter. If this model is correct, the implications are significant, because the nature of both these dark components is unknown. If some other explanation for the  appearance of the Universe is to be sought, then the implications are also momentous for our current understanding of physics and cosmology.

Dark Energy is a  relatively new entity in our understanding of Cosmology. It has been known since the 1920s that the typical separation between galaxies is growing with time -- the Universe is expanding. But it might be supposed that in a Universe made up of only matter, this expansion is decelerating: the galaxies will move apart at a decreasing rate owing to their mutual gravitational interaction. However, a little more than a decade ago,
observations comparing different distance measures for supernovae (Riess \etal 1998; Perlmutter \etal 1999) revealed that this is not the case, and the expansion is in fact speeding up. The cause of this acceleration is unseen, but has the characteristics of an extra energy density in the Universe; hence we label the entity as ``Dark Energy''. The importance of Dark Energy can scarcely be exaggerated. Most immediately, it represents the largest source of energy density in the Universe, $\sim75$\%. It is expected to dominate the future dynamics of the Universe, so the origins and nature of the Universe cannot be understood without some assessment of what Dark Energy is and what its physical characteristics are.

The next most significant constituent of the Universe, Dark Matter, exceeds the normal baryonic matter in a ratio of four or five to one. As Dark Matter structures form under gravitational collapse, baryonic matter follows. Hence the Dark Matter drives the formation and evolution of the structures we observe directly, because the behaviour of stars, galaxies and gas depends  on the underlying gravitational potential created by it. While Dark Matter apparently interacts gravitationally in  the same way that normal baryonic matter does, it seems not  to interact through the electromagnetic force. Observations must therefore rely on inferring its presence through the gravitational effect it has on light or baryonic matter, by which means it has been inferred on a range of scales, galactic and larger.


That it is critical to achieve an understanding of the nature of Dark Energy and of Dark Matter, and of the examining and testing the alternative conceptual structures for the observed characteristics of the Universe, has been recognised for some time. Summaries are available in the Dark Energy Task Force and ESA/ESO reports (Albrecht \etal 2006, Peacock \etal 2006), the ASTRONET Infrastructure Roadmap (Bode, Cruz \& Molster 2009) and most recently by the US Decadal Survey report (Blandford \etal 2010). 

Initiatives are under way to further the accuracy and precision of the observations in order to address these questions, using facilities on ground (for example BOSS -- Schlegel \etal 2009; BigBOSS -- Schlegel \etal 2011; KiDS -- de Jong \etal 2012; DES -- DES Collaboration 2005, HSC\footnote{http://anela.mtk.nao.ac.jp/hypersuprime/proposal/hs050626.pdf}, LSST -- Tyson \etal 2003  and SKA -- Blake \etal 2007) and in space (\eu -- Laureijs \etal 2011; and WFIRST -- Green \etal 2012). These employ a combination of techniques, including weak gravitational lensing, galaxy clustering (which incorporates baryonic acoustic oscillations) and supernovae, among others, both to distinguish between possible cosmologies, and also to ensure that systematic effects in the measurements are identified and quantified at the required level of accuracy. Control of systematic effects is critical. Because of the inherently stable conditions that can be achieved, space missions provide the best opportunities for controlling systematics, and payloads can be designed also to include the capability to make observations using several techniques.

Weak gravitational lensing uses statistical measurements of the distortions of galaxy shapes to study the clustering of matter in the Universe. Early studies were made by Wittman \etal (2000), van Waerbeke \etal (2000), Mellier \etal (2000), Bacon, R\'{e}fr\'{e}gier \& Ellis (2000) and Kaiser, Wilson \& Luppino (2000). Reviews can be found in Hoekstra \& Jain (2008) and Munshi \etal (2008), while more recent work includes that by Schrabback \etal (2010) and Heymans \etal (2012).  The rate at which the large-scale structure has grown depends on the expansion rate of the Universe, so the nature of the acceleration can be characterised by making these shear measurements at different redshifts, looking back in time (Hu, 1999). 

In this paper we develop a framework by which weak lensing measurements in particular can be realised in a space mission, and its likely performance anticipated (many of these considerations apply also to ground-based weak-lensing surveys). The formalism for the critical systematic effects has been developed in a series of papers (Vale \etal 2004, Mandelbaum \etal 2005, Huterer \etal 2006, Stabenau \etal 2007, Amara \& R\'{e}fr\'{e}gier, 2007, 2008, Paulin-Henriksson \etal 2008, Kitching, Taylor \& Heavens 2008a, Amara, R\'{e}fr\'{e}gier \& Paulin-Henriksson 2010) with the most contemporary development given in Massey \etal (2013) (hereafter MHK13), and we use these here as a basis.

The work was carried out in the framework of the \eu mission\footnote{http://www.euclid-ec.org}, under the auspices of the European Space Agency Cosmic Vision programme. An overview of its capabilities can be found in the \eu Red Book (Laureijs \etal 2011; this a consolidated summary of the mission at the end of the Definition Phase) and Amendola \etal (2012). With respect to its weak lensing capabilities, \eu can be considered an example of a next-generation cosmic shear survey mission.  \eu is designed to carry out both weak lensing and galaxy clustering cosmological measurements, using a payload comprising a visible imager, with which the weak lensing measurements are made, and a near-infrared spectrograph-imager. For visible measurements, CCD detectors are the currently leading technology for large focal planes, and they are assumed for this paper. However, the methodology by which we address the realisation of a successful experiment is  general, and although we will sometimes use \eu as an example, the purpose of this paper is to set out the applicable principles.

Section 2 of this paper sets the requirements and describes how an allocation can be made to the main factors contributing to the performance degradations. Section 3 briefly describes simulations and data processing. How the weak lensing performances may be evaluated is set out in section 4.

\section{Setting the Requirements}

\subsection{Mission Driving Parameters}
\label{sec:top-level}

The power of a weak lensing survey depends on five main factors:
\begin{enumerate} 
\item the size of the survey; 
\item the limiting magnitude of the survey; 
\item the size and shape of the instrument point spread function (PSF); 
\item how well this PSF is known and
\item how well we can correct for the sources of systematics. 
\end{enumerate}

Parameters (i) and (ii) set the total number of galaxies that may be available for the weak lensing shear measurements, and their range in redshift. Hence they set the maximum achievable statistical precision. This drives the area of the sky that should be observed, and consequently the field of view of the instrument, and the mission duration. A wide survey is  required to ensure the measurements are representative of the observable Universe. A deeper limiting magnitude, by providing increased signal-to-noise ratios on each individual galaxy measurement,  also determines the size of the sample, and enables higher redshifts to be accessed. Given that most of the cosmic acceleration has taken place in more recent epochs, the emphasis of most lensing surveys is on galaxies with redshifts $z\laeq2$ which makes them sensitive to structure at $z\sim0.5$ to 1, halfway between source and observer. Mitigating the confusing effects of intrinsic alignments between galaxies (resulting from the flows of material during the formation of structure in the Universe) also requires a sufficient survey depth (Joachimi \& Bridle, 2010). Because the lensing signal is cumulative along a line of sight, the more distant sources contain information about Dark Energy at low redshift as well as the information about the growth of structure at high redshift. More distant galaxies generally are, however, fainter and smaller, which makes the measurement of their shear from the weak gravitational lensing more difficult for a given instrumental PSF. Measuring their redshift is, in addition, more difficult. 

The depth of the survey drives the collecting aperture of the telescope, its throughput, the width of the observational bandpass, and the sensitivity of the detectors. Achieving a desirable size and shape of the PSF drives all of these contributors: a larger telescope reduces the size of the PSF, the optical design drives its shape, the stability of the satellite pointing modifies the PSF, and the need for adequate sampling of the PSF drives the detector pixellisation to be small. This, in combination with the need for a large field of view, requires a large detector matrix. 

The overall ``system'' PSF is a combination of PSFs produced by the optical system, satellite pointing stability, detector pixellisation and detector effects. The detector effects arise as a result of the physical realisation of the detectors (for example charge spreading in the pixel grid) and because of damage effects in space, particularly radiation damage (Holland \etal 1990). The first three contributions to the total PSF can be modelled by convolutions, while the fourth, generated by the detector, generally has a combination of characteristics, only some of which can be modelled by a convolution. 

Limitations on the number of detector pixels may drive the observing strategy, for example requiring multiple exposures to recover spatial resolution from undersampled images (Section~\ref{sec:sampling}). This also has the benefits of allowing cosmic rays to be detected and removed in the data processing. If there are small displacements between the different exposures, then additionally, the impact of cosmetic defects in the detectors can be minimised, and the gaps between the individual detectors can be filled in. Because of the detector gaps, some galaxies will have more exposures than others, and the effect of this on the best-fitting cosmological parameters requires evaluation. With somewhat larger displacements, perhaps up to half of a detector, a further benefit is obtained in that the radiation damage effects (Section~\ref{sec:CTI_correc}) which increase with distance from the CCD readout node, are formally separable from the cosmic shear. Typically, therefore, more than one exposure is taken of each field, and these will be combined to reach the depth required for the survey. The multiple exposures however impact operational considerations negatively, require more fuel for the spacecraft repointing and require more telemetry bandwidth. 

For a typical advanced weak lensing survey (such as that discussed in Laureijs \etal 2011), 
$\sim2\pi$ sr will be covered to a depth AB~$\sim25$ at $10\sigma$, yielding $\sim30$ galaxies arcmin$^{-2}$ with suitable characteristics for the survey -- a total exceeding $10^9$ galaxies. The survey will generally be limited to Galactic latitudes $|l| \gaeq30^{\circ}$, and will concentrate at least initially on regions furthest from the ecliptic plane, in order to minimise the Zodiacal background light. The pattern with which the fields are exposed will generally be constrained in order to maintain stable conditions within the payload.

The survey and instrumentation must also be planned to minimise the systematic biases in the weak lensing measurements. Factors (iii) and (iv)  strongly impact these systematic biases, and therefore on the accuracy (as opposed to the precision) of the measurements. The size and shape of the PSF influence which fraction of the observed galaxies may be useful for shear measurements. The PSF blurs images: the shape of galaxies with smaller sizes relative to the PSF will be measurable with reduced accuracy and hence smaller PSFs are desirable. For a given encircled energy width, a PSF with broad wings and narrow core will have a different effect on the shape measurement from one with narrower wings and a broader core. In addition, the detection limit of the survey will depend on the PSF, which influences the sample to some extent. 

Within these general constraints, the typical PSFs normally achieved with standard astronomical instrumentation in space-borne observatories will be acceptable (and certainly smaller and more stable than through a turbulent atmosphere). The particular and stringent aspect of the weak lensing measurements is contained in the fourth and fifth main factors: the {\it knowledge} of this PSF, and how the biases can be corrected. The ultimate power of the weak lensing measurements will depend on the level with which the PSF is known, and future generation weak lensing surveys such as those considered in MHK13 require this to be known to an unprecedented level of accuracy. Included in this knowledge is the way the PSF will change with time, with position on the focal plane and with source galaxy characteristics. 

MHK13 provides the top-level context for this investigation, while this paper provides a more detailed examination of the multiple contributing effects for each bias, and how they might be combined in a practical experiment.

\subsection{Quantifying the Biases}
\label{sec:biasquantification}

In a typical future generation weak lensing survey from space (for example, Laureijs \etal 2011), with observations of $>10^9$ galaxies, the errors on the linearly varying Dark Energy equation of state (Chevallier \& Polarski, 2001; Linder 2003) $w(z)=w_p+w_a(z-z_p)/[(1+z_p)(1+z)]$ are $w_p\laeq 0.05$ and $w_a\laeq 0.2$, to give a Figure of Merit (FoM) $1/[\Delta w_p \Delta w_a] \gaeq100$ from lensing only ($z_p$ is the redshift at which the error on $w(z)$ minimises).
\footnote{It should be recalled that the FoM is only one of the measures used for the effectiveness of Dark Energy investigations and the linear parameterisation in $w$ also is limiting. The form of the structure growth factor (Laureijs \etal 2011) and other measures to be tested for the cosmology are also relevant. However, the FoM is a standard generally used for the comparison of surveys.} 
As the precision increases as a result of combining such large samples of galaxies, the control of the systematic effects becomes more and more important.  

Effective control of the systematic effects requires first an understanding of what effects may be present, and how they combine with each other to introduce biases. Then it is important to understand how significant each effect may be in the overall performance. Some effects can be minimised by better design of the instrument and survey, and by better calibrations; others by alternative approaches in the data processing and analysis. Each of these carries implications for the viability of the experiment and for the cost and duration of the mission. For example, improved control of some biases may be achieved through newer technologies which carry more risk. Alternatively more conventional technologies could be used and the gains sought in the data analysis algorithms.

In this section we first summarise how the biases affect the derived cosmological parameters. Then we identify the factors which contribute to these biases and quantify their relative importance. Each factor generally has contributions from other  sources. We organise these into a structure which allow the effect of each to be assessed; this attempts also to clarify the relationships of the contributing factors. In respect of each lowest-level factor, an initial analysis may suggest that a certain level of knowledge can be reached, but these may require revision in order to remain below the permitted total bias, which will lead to further more detailed analyses. The purpose of this section is not to identify the values of the factors for any particular experiment, but to rather illustrate a structure by which the performance of an experiment in terms of the control of systematic effects can be assessed, and the effects of changes in any aspect can be propagated to the top level. This allows the optimisation of the experiment to be achieved.

The procedure for quantifying the biases is as follows. MHK13 and references therein consider that the {\it true} shear $\gamma$ of a galaxy will differ from that actually measured, $\widehat\gamma$, by additive and multiplicative biases $c$ and $m$ (in the survey, instrument and measurement process) as
\begin{equation}\label{eq:mandc}
\widehat\gamma = (1+m)\gamma + \boldmath{c} .
\end{equation}
The two-point ellipticity correlation function is 
\begin{equation}\label{eq:xi}
\reda{\xi_{ij}(\theta) \equiv \sum_{A,B} \gamma^A_i \gamma^B_j  (\theta)} = \langle \gamma^A_i \gamma^B_j \rangle_{(\theta)} ,
\end{equation}
where $\theta$ is the angular scale and  $i,j$ refer to  redshift bin pairs averaged over all pairs of galaxies $A, B$. This can be used (Hu, 1999) to constrain a set of cosmological parameters usually through the corresponding Fourier transform  power spectrum $C_{ij}(\ell)$ \reda{in the spherical harmonic $\ell\equiv 2\pi/\theta$}. As a consequence of the biases $c$ and $m$ in Equation~\ref{eq:mandc}, $C_{ij}(\ell)$ will  be modified (Kitching \etal 2012) by additive $\mca$ and multiplicative $\mcm$ biases into an {\it observed} 
\begin{equation}\label{eq:Chat}
\widehat{C_{ij}(\ell)} = (1+\mathcal{M}(\ell))C_{ij}(\ell) + \mathcal{A}(\ell).
\end{equation}
Kitching \etal (2012) \reda{and MHK13} find that 
\begin{align}\label{eq:switch}
\reda{\mathcal{A}(\ell)} & = \reda{\left<c\right>_{(\ell)}^2}{\rm \,\,\, \,\,usually\,\,written\,\,}\sigma^2[|\boldmath{c}|] \nn
\reda{\mathcal{M}(\ell)} & =  2\left<m\right>_{(\ell)} + \left< m \right>_{(\ell)}^2 ,
\end{align}
%
\reda{where the angle brackets $\langle \,\, \rangle_{(\ell)}$ are the Fourier transform of the ensemble average in real space over galaxy pairs separated by angle $\theta$ in Equation~\ref{eq:xi}:
\begin{equation}
\left \langle f \right \rangle_{(\ell)} = \int \left \langle f \right \rangle_{(\theta)} {\rm e}^{i\ell\theta} {\rm d}\theta .
\end{equation}
The subscripts are usually suppressed for brevity. We will include them here explicitly to ensure clarity. 
}

Non-zero $\mca$ and $\mcm$ lead to a bias in the maximum likelihood values of measured cosmological parameters (see MHK13) and an decrease in the FoM (through an increase in the covariance). As noted above, the contributors to $\sigma[|c|]$ and $m$ must be derived through a careful process of identifying all of the biases, including the imperfections in the galaxy modelling and other effects.

We adopt the formulation in section 3.3 of MHK13, based on that in Paulin-Henriksson \etal (2008), Paulin-Henriksson, R\'{e}fr\'{e}gier, \& Amara, (2009):  
\begin{align} 
\mathcal{A}\reda{(\ell)} &= \frac{1}{P_\gamma^{2}P^2_R\pec^2} \left\langle\frac{\rp^4}{\rg^4}\right\rangle 
\sigma^2[|\ep|] ~~~~~~~~~~~~~~~~~~~~~~~~~~~ \nn
&+ \frac{1}{P_\gamma^{2}\penc^2} \left\langle 1 + \frac{2}{P_R}\frac{\rp^2}{\rg^2}+\frac{1}{P^2_R}\frac{\rp^4}{\rg^4}\right\rangle 
\sigma^2[|\ec|] ~~~~~~ \nn
&+ \frac{\langle|\ep|^2\rangle}{P_\gamma ^{2}P^2_R\pec^2} \left\langle\frac{\rp^4}{\rg^4}\right\rangle \left( \frac{\left\langle\delta(\rp^2)\right\rangle^2}{\left\langle\rp^4\right\rangle} + \frac{\sigma^2[\rp^2]}{\rp^{4}} \right) \, \nn
&+ \, 4 \frac{\langle|\ep|^2\rangle}{P_\gamma^{2}P^2_R\pec^2} \left\langle\frac{\rp^4}{\rg^4}\right\rangle \left( \frac{\left\langle\delta(\rc)\right\rangle^2}{\left\langle\rc^2\right\rangle} + \frac{\sigma^2[\rc]}{\rc^{2}} \right) \nn
& +  \frac{\langle|\ep|^2\rangle}{P_\gamma ^{2}P^2_R\pec^2}\left\langle\frac{\rp^4}{\rg^4}\right\rangle\alpha^2
\label{eq:A}
\end{align}
where
\begin{align}\label{eq:alpha}
\alpha^2 ~&=    \frac{\left\langle\delta(\ro^2)\right\rangle^2}{\left\langle\ro^4\right\rangle}   \nn
&+ \, \left\langle\frac{\rg^4}{\rp^4}\right\rangle \left\langle\left(\frac{P_R\rg^2}{P_R\rg^2+\rp^2}\right)^2\right\rangle \frac{\left\langle \delta P_R \right\rangle^2}{\left\langle P_R^2\right\rangle}
\end{align}
and then also 
\begin{align}
\mathcal{M}\reda{(\ell)}& = \frac{2}{P_R}\left\langle\frac{\rp^2}{\rg^2}\right\rangle   \Bigg\{ \frac{\left\langle\delta(\rp^2)\right\rangle}{\left\langle\rp^2\right\rangle} +   2\frac{\left\langle \delta \rc \right\rangle}{\left\langle \ro \right\rangle}   \Bigg\} \nn
                       &  + \,  \frac{1}{P^2_R}\left\langle\frac{\rp^4}{\rg^4}\right\rangle \Bigg\{{\frac{\sigma^2[\rp^2]}{\langle\rp^4 \rangle}} 
                                         ~+ 4{\frac{\sigma^2[\rc]}{\langle\ro^2\rangle}} \Bigg\} \nn
                       & + ~ \frac{2}{P_R} \left\langle\frac{\rp^2}{\rg^2}\right\rangle \mu
\label{eq:M}
\end{align}
where 
\begin{align}\label{eq:mu}
\mu & = - 
 \left\langle \vphantom{\left\langle\frac{\rp^2}{\rg^2}\right\rangle} \frac{\delta(\ro^2)}{\ro(\ro - \rc)} \right\rangle \nn
       & ~~~ -           P_R \left\langle\frac{\rg^2}{\rp^2}\right\rangle \left\{ \left\langle \vphantom{\frac{\rg^2}{\rp^2}}\frac{\delta P_{\gamma}}{P_{\gamma}} \right\rangle + \left\langle\frac{\rg^2}{P_R\rg^2+\rp^2}\frac{\delta P_R}{P_R} \right\rangle\right\}.
\end{align}


$R$ refers to the size of the PSF or galaxy image and the $\bmath{\ellipticity}$ to the polarisation, generally referred to as the `ellipticity', defined in terms of the unweighted second order  moments in the image of the galaxy (Seitz \& Schneider 1995, Bonnet \& Mellier 1995). Explicitly, for a PSF $\Phi(x_i,x_j)$ and a weight function $w(x_i,x_j)$ these moments are
\begin{equation}
\label{eq:quad}
Q_{ij}=\frac{\int\int\Phi(x_i,x_j) w(x_i,x_j) (x_i-\bar{x_i})(x_j-\bar{x_j}) dx_idx_j}{\int\int\Phi(x_i,x_j)w(x_i,x_j)dx_idx_j}.
\end{equation}
Then, size
\begin{equation}
\label{eq:R}
R^2=Q_{11}+Q_{22}
\end{equation}
and, ellipticity
\begin{align}
\label{eq:ellip}
\bmath{\ellipticity} = [\ellipticity_1,\ellipticity_2]&=\left[\frac{Q_{11}-Q_{22}}{R^2}, \frac{Q_{12}+Q_{21}}{R^2}\right]; \nn
|\bmath{\ellipticity}|&=\sqrt{ \ellipticity_1^2+\ellipticity_2^2}.
\end{align}
Returning to Equations~\ref{eq:A} and \ref{eq:M}, the subscript $_{\rm C}$ refers to those components of the PSF which can be combined by convolution, and $_{{\rm NC}}$ to those that cannot. MHK13 use the subscript $_{\rm PSF}$ rather than $_{\rm C}$, but we use $_{\rm C}$ to distinguish the convolutive part of the system PSF clearly, and as a reminder that we consider the term PSF here to refer to the end-to-end system PSF.  The system PSF properties change with wavelength, the spectral energy distribution (SED) $f(\lambda)$ and the transmission as a function of wavelength $T(\lambda)$. Taking into consideration the integrated flux in the band of measurement, the size and ellipticity of the convolutive components of the PSF are given by:
\begin{equation}\label{eq:Rlambda}
R^2_{_{\rm C}}=\frac{1}{f_{\rm tot}}\int d\lambda T(\lambda) \lambda f(\lambda)R_{_{\rm C}}^2(\lambda),
\end{equation}
\noindent and
\begin{equation}\label{eq:elambda}
\bmath{\ellipticity}_{_{\rm C}}=\frac{1}{f_{\rm tot}}\int d\lambda T(\lambda)\lambda f(\lambda) \bmath{\ellipticity}_{_{\rm C}}(\lambda),
\end{equation}
\noindent where $f_{\rm tot}=\int d\lambda T(\lambda)\lambda f(\lambda)$ is the total number of photons. Note the extra factor $\lambda$ converts from energy to photons. 

$R_{{\rm gal}}$ and $R_{{\rm obs}}$ refer to the original and observed size of the galaxy, relating as (MHK13) 
\begin{equation}\label{eq:robs}
R_{\rm obs}\equiv\sqrt{(R^2_{\rm gal}+R^2_{_{\rm C}})}+R_{_{\rm NC}}. 
\end{equation}

The `shear polarizability' 
\begin{equation}\label{eq:Pgamma}
P_\gamma=2-\left<|\bmath{\ellipticity}|^2\right> 
\end{equation}
relates the ellipticity to the shear in the galaxy image caused by the weak lensing, and, although to some extent dependent on the galaxy sample and wavelength range used for the survey, when aggregated over galaxy samples out to $z>2$ is found to be approximately a constant factor 1.86 (Leauthaud et al.~2007). 

The quantities $P_R$, $\pec$ and $\penc$ are compensations for the necessity of using weighted quadrupole measurements rather than the unweighted moments that would be ideal in theory (see MHK13 for details).  
The weighting function $w(x_i,x_j)$ is introduced inside the integrals in Equation~\ref{eq:quad} to control the increasing noise fraction as the integration moves outwards from the centre of the galaxy or star image.
Indeed, the use of weighted moments is the origin of the additional $\alpha^2$ and $\mu$ terms in Equations~\ref{eq:A} and \ref{eq:M} by comparison with Equation~\ref{eq:switch}. We define 
\begin{align}
&R^2_{\rm unweighted} = P_R R^2_{\rm weighted} \nn
&\bmath{\ellipticity}_{\rm unweighted_{C}} = \pec \bmath{\ellipticity}_{\rm weighted_{C}} \nn
&\bmath{\ellipticity}_{\rm unweighted_{NC}} = \penc \bmath{\ellipticity}_{\rm weighted_{NC}}.
\end{align}
In general $\pec$ and $\penc \simeq 1$ while $P_R$ is larger (in the \eu case, $P_R\simeq 2$).

In Equations~\ref{eq:A} and \ref{eq:M}, $\left<|\delta|^2\right>$ terms  have been decomposed into a systematic bias of a model value away from the truth $\left<\delta\right>^2$, and uncertainties $\sigma^2$ in this. To elaborate this important point, many biases in the measurements will be corrected by detailed modelling: an example may be the change in the size of the PSF resulting from out-of-band leakage, which may be calculated from the measured characteristics of the bandpass and the spectral energy distribution of the particular source. This model calculation will not, of course, produce exactly the true value. From uncertainties in the inputs to the modelling (in the example above this might include the uncertainties in the transmission at each wavelength), and inadequacies in the physical model (again in this example, this might arise from codes used to predict a stellar spectrum), the modelling will produce a slightly incorrect prediction $\left<\delta\right>$, and an associated uncertainty on this prediction $\sigma$.

We can write Equations~\ref{eq:A} and \ref{eq:M} as follows:
\begin{align}\label{eq:As}
\mca\reda{(\ell)}&= a_{1}\left(\left<\bd\boldmath{\ellipticity}_{_{\rm C}}\right>^2 + \sigma^2\left[|\bmath{\ellipticity}_{_{\rm C}}|\right]\right)\nn
&+ a_{2}\left(\left<\bd\bmath{\ellipticity}_{_{\rm NC}}\right>^2 + \sigma^2\left[|\bmath{\ellipticity}_{_{\rm NC}}|\right]\right)\nn
&+ a_{3}\left(\frac{\left<\delta(\rp^2)\right>^2}{\left<\rp^4\right>} + \frac{\sigma^2[\rp^2]}{\rp^4}\right)\nn
&+ a_{4}\left(\frac{\left<\delta(\rc)\right>^2}{\left<\rc^2\right>}+ \frac{\sigma^2[\rc]}{\rc^2}\right)\nn
&+ \,a_{5}\,\left(\left\langle \delta(\alpha)\right\rangle^2 + \sigma^2[\alpha]\right)
\end{align}
and
\begin{align}\label{eq:Ms}
\mcm\reda{(\ell)}& =  m_{1} \frac{\left<\delta(\rp^2)\right>}{\left<\rp^2\right>}  \nn
&+ m_{2}\frac{\left<\delta(\rc)\right>}{\left<\rc\right>} \nn
&+ m_{3}~\frac{\sigma^2[\rp^2]}{\rp^4}\nn
&+ m_{4}~\frac{\sigma^2[\rc]}{\rc^2}  \nn
&+ \,m_{5}\,\left(\left\langle \delta(\mu) \right\rangle + \sigma[\mu]\right).
\end{align}

To reiterate, the last terms of Equation~\ref{eq:A}  and \ref{eq:M} relate to the modelling error for the galaxies resulting from the fact that weighted, rather than unweighted quadrupole moments are used in practice. We have swept up all of the additive galaxy modelling errors in Equation~\ref{eq:A} into a model error $\alpha$, split into bias and knowledge errors $\left\langle\delta(\alpha)\right\rangle^2$ and $\sigma^2[\alpha]$. We have done the same for the multiplicative errors in Equation~\ref{eq:M}, with the model error $\mu$ split into $\left\langle \delta(\mu) \right\rangle$ and $\sigma[\mu]$, so that
\begin{align}\label{eq:alpha-mu}
\blue{\left\langle\alpha^2\right\rangle} & \rightarrow \left\langle \delta(\alpha) \right\rangle^2 + \sigma^2[\alpha]   
\nn
\rm{and} & \nn
\left\langle\mu\right\rangle & \rightarrow \left\langle \delta(\mu) \right\rangle + \sigma[\mu].
\end{align}
In practice, the $\sigma$ terms in Equation~\ref{eq:alpha-mu} can be reduced to insignificant levels by ever-larger simulations, but these are ineffective in correcting for the $\delta$ terms. Hence we will ignore the $\sigma^2[\alpha]$ and $\sigma[\mu]$ terms in carrying forward any allocations for imperfections in the modelling.

Also, we should note at this point that the first term in Equation~\ref{eq:mu} 
\begin{equation} 
\left\langle \frac{\delta P_{\gamma}}{P_{\gamma}} \right\rangle \rightarrow 0 \nn
\end{equation}
for a sufficiently large survey. The third term captures the error which will arise from the use of the weighting function in the quadrupole moment integral (Equation~\ref{eq:quad}) as a result of the \blue{absence of perfect knowledge of} higher-order multipoles, while the second term is a knowledge error arising from imperfect measurements, and hence is dependent on signal-to-noise ratios.

Note also that Equation~\ref{eq:As} differs slightly from Equation~\ref{eq:A}  to the extent that we introduce $\delta$ terms in the first two lines. This accommodates the biases that will, in practice, occur in a weak lensing experiment which employs the use of bright stars to define the PSF used for faint galaxies, with the associated non-linearity and wavelength mismatching. 

So, Equations~\ref{eq:As} and \ref{eq:Ms} contain terms for knowledge bias $\left<\delta\right>^2$ and knowledge uncertainties $\sigma^2$ in the following categories: convolutive, and non-convolutive errors in the PSF sizes; convolutive, and non-convolutive errors in the ellipticities; and bias errors $\alpha$ and $\mu$ on the transformation from ellipticity to shear resulting from the fact that we use weighted moments of the PSF. The coefficients $a_i$ and $m_i$ are now seen to be weighting factors, whose values depend on the characteristics of the instrument and the galaxies being measured. Equations~\ref{eq:As} and \ref{eq:Ms} provide the prescription by which these contributing effects can be combined. 

We are now in a position to quantify the impact on the cosmology from systematic effects in the weak lensing measurements given the knowledge biases $\left<\delta\right>$, knowledge uncertainties $\sigma$ and weighting functions $a_i$ and $m_i$.

\subsection{\reda{Quantifying Requirements}}

\reda{
To simplify the requirements for a practical experiment, and to make them less dependent on the assumption of a cosmological model, we will integrate over the range of spatial scales of interest taking into account the density of the sampling of the modes available from the survey to recover the discrete nature of the summation over the modes sampled by it. Under the assumption of isotropy, an integral over two dimensional spatial scales reduces to a single dimension, $\ell$, as
\be
 \int \ell(\ell+1)\rho_{\rm survey} 2\pi \ell {\rm d}\ell
\ee
The sampling is set by the largest angular scale $\Delta\theta$ in the survey, and this gives the density of modes $\rho_{\rm survey} =(\Delta\theta/2\pi)^2$. Factors of $2\pi$ partially cancel. The factor  $\Delta\theta^2$ is effectively the solid angle for the survey and is by convention subsumed into the definition of the $C_{ij}(\ell)$ (see for example Kitching, Heavens \& Miller 2011 equation A6) and therefore all of the following is conditional on the assumed area of the fiducial survey. We then
construct integrated quantities over $\ell^2{\rm d}\ln\ell$ because 
\[ \frac{1}{2\pi}\int\ell^2(\ell+1){\rm d}\ell  \simeq  \frac{1}{2\pi}\int\ell^3{\rm d}\ell   \rightarrow \frac{1}{2\pi}\int \ell^2 {\rm d}\ln\ell.\]
}

\reda{Using this formalism, we  redefine (again) the angle brackets $\langle \,\, \rangle$ as expectation values over such integrals, for example
\begin{equation}
\left \langle f \right \rangle_{(\int\!\!\ell)} = \frac{1}{2\pi} \int \left \langle f \right \rangle_{(\ell)} \ell^2{\rm d}\ln\ell  .
\end{equation}
Then, the integrals of each of the terms on the right hand sides of the requirements equations ({\it i.e.} Equations~\ref{eq:A} to \ref{eq:mu}) will contribute to an integral over the systematic contribution of the power spectrum $\widehat{C_{ij}(\ell)} - {C_{ij}(\ell)}$ (Equation~\ref{eq:Chat}). 
And now we also need to integrate the left hand side of the $\ell$-dependent equations consistently. For the additive term (setting $\mcm(\ell)=0$) this gives
\begin{align}\label{eq:Apri}
\frac{1}{N_{ij}}\sum_{ij}  \frac{1}{2\pi} & \int [\widehat{C_{ij}(\ell)} - C_{ij}(\ell)] \ell^2{\rm d}\ln\ell     \nn
& = \frac{1}{N_{ij}}\sum_{ij} \frac{1}{2\pi} \int {\mathcal A}(\ell) \ell^2{\rm d}\ln\ell \nn 
& \equiv \frac{1}{N_{ij}}\sum_{ij}{\mathcal A}'
\end{align}
where $N_{ij}$ is the number of redshift bin pairs $ij$. Similarly for the multiplicative term (setting $\mca(\ell)=0$):
\begin{align}\label{eq:Mpri}
\frac{1}{N_{ij}} \sum_{ij}\frac{1}{2\pi}&  \int  [\widehat{C_{ij}(\ell)} - C_{ij}(\ell)] \ell^2{\rm d}\ln\ell  \nn
&= \frac{1}{N_{ij}}\sum_{ij}\frac{1}{2\pi}\int {\mathcal M}(\ell)C_{ij}(\ell) \ell^2{\rm d}\ln\ell   \nn 
& \equiv \frac{1}{N_{ij}}\sum_{ij}{\mathcal M}'.
\end{align}
}

\reda{The next step is to assign values for $\mca$ and $\mcm$ by which the impact on $\widehat{C_{i,j}(\ell)}$ (Equation~\ref{eq:Chat}) remains within acceptable values. From the definition of $\bar{\mca}$ in MHK13, but considering it per redshift bin, we see that 
\be 
{\mathcal A}' \,\,\, = \,\,\, \bar{\mathcal A}\frac{1}{2\pi}\int \ell^2{\rm d}\ln\ell \,\,\, \leq 2.6 \times 10^{-7}
\ee
where the requirement on $\bar{\mathcal A}$ from MHK13 of $1.8\times 10^{-12}$ has been multiplied by its denominator $\int \ell^2{\rm d}\ln\ell /2\pi$ which has a value of $1.43\times 10^5$ for the range $10 \leq \ell \leq 5000$, a fiducial survey of $15000$ square degrees and $30$ galaxies per square arc minute. This is per redshift bin pair, and there are $55$ power spectra used in setting the requirements in MHK13; an integrated requirement over all power spectra would further multiply the above value by $55$. Next, from MHK13:
\[\bar{\mathcal M} \,\,\, = \,\,\, \frac{\frac{1}{2\pi}\int {\mathcal M}(\ell) \ell^2{\rm d}\ln\ell}{ \frac{1}{2\pi}\int \ell^2{\rm d}\ln\ell} \]
where again we consider it per redshift bin pair. Consequently (MHK13 appendix B):
\begin{equation}
\mcmp = \frac{1}{2\pi}\int {\mathcal M}(\ell) C(\ell) \ell^2{\rm d}\ln\ell \simeq \bar{\mathcal M}\frac{1}{2\pi}\int C(\ell) \ell^2{\rm d}\ln\ell.  
\end{equation}
Substituting the mean weighted power spectrum $\bar{C}$ 
\be
\bar{C} \,\,\, = \,\,\, \frac{\int C(\ell) \ell^2{\rm d}\ln\ell }{\int \ell^2{\rm d}\ln\ell}
\ee
then
\be \label{eq:Aprime}
{\mathcal M}' = \,\,\,  \bar{\mathcal M}\bar{C}\frac{1}{2\pi}\int \ell^2{\rm d}\ln\ell \,\,\,\, \laeq 1.4\times10^{-2}
\ee
where the requirement on $\bar{\mcm}$ of $4.0\times10^{-3}$ from MHK13 has been multiplied by $\bar{C}\int \ell^2{\rm d}\ln\ell /2\pi \simeq 3.58$ (this value is somewhat cosmology-dependent) for the range $10 \leq \ell \leq 5000$.\hspace*{-1mm}
\footnote{\reda{Alternatively, adapting the definition of $\bar{\mcm}$ from MHK13 so that it  includes $C_{ij}(\ell)$ :
\[\bar{\mathcal M} \,\,\, = \,\,\, \frac{\int {\mathcal M}(\ell) C(\ell) \ell^2{\rm d}\ln\ell}{ \int C(\ell) \ell^2{\rm d}\ln\ell} \]
where again we consider $\bar{\mcm}$ per redshift bin pair. Then
\be \label{eq:Mprime}
{\mathcal M}' = \,\,\,  \bar{\mathcal M}\frac{1}{2\pi}\int C(\ell) \ell^2{\rm d}\ln\ell \,\,\,\, \simeq 1.4\times10^{-1},
\ee
 where the requirement on $\bar{\mcm}$ of $3.9\times10^{-2}$ (recalculated for this definition of $\bar{\mcm}$ as in MHK13) has been multiplied by $\int C(\ell)\ell^2{\rm d}\ln\ell/2\pi\simeq 3.58$  for the range $10 \leq \ell \leq 5000$. Note that this is a factor 10 larger than the  calculation in Equation~\ref{eq:Mprime}. Comparing to the GREAT10 results, as is done in MHK13 section 5.1,
these updates result in previous statements on the
performance of methods relative to the requirement to be improved by a factor of 10 in the case
of $\mcm$. In GREAT10 the best methods achieved $\bar{\mca}=7.4\times10^{-11}$ and $\bar{\mcm}=5.6\times10^{-3}$ at S/N=10.}}}

%

\reda{MHK13 calculated these requirements over
scales $10\leq\ell\leq 5000$ in order to avoid non-linear
scales that may be potentially difficult to model as a result of
non-linear effects (e.g. Smith et al., 2003) or baryon feedback (e.g.
Semboloni et al., 2013). This is the main factor in the relaxation of the requirement values in MHK13 compared to those in the previous analyses by Amara \& Refregier (2008) and Kitching et al., (2008a), which extended to $\ell\leq 20$,$000$. In order to allow for a future potential
increase of scope in the use of the non-linear modes, in this paper we
will retain the more stringent requirements of
${\mathcal A'}\leq 1\times 10^{-7}$ and ${\mathcal M'}\leq 4\times
10^{-3}$ from Amara \& Refregier (2008) and Kitching et al., (2008a)
respectively. 
}

\reda{
In a practical experiment, we will be working with angular measures $\theta$ rather than the Fourier transform variable $\ell$. We therefore take the Fourier transform of both sides of Equations~\ref{eq:A}--\ref{eq:mu} integrated over  $\ell$  (Equations~\ref{eq:Apri} and \ref{eq:Mpri}), replacing $\mca'$ and $\mcm'$ with $A$ and $M$ to indicate this. In the interests of clarity, we retain the notation on the right hand side of these equations, noting that now these values are effectively integrated over $\theta$ and that they still relate redshift pairs $(ij)$ (Equation~\ref{eq:xi}). Angle brackets now refer to the Fourier transforms in Equations~\ref{eq:Apri} and \ref{eq:Mpri} which are constants:
\begin{align}
\left \langle f \right \rangle_{\mathcal{F}(\int\!\!\ell)} & = \frac{1}{2\pi}\int \left \langle f \right \rangle_{(\int\!\!\ell)} {\rm e}^{-i\ell\theta} {\rm d}\ell 
 =  \frac{\left\langle f \right \rangle_{(\int\!\!\ell)} }{2\pi}\int  {\rm e}^{-i\ell\theta} {\rm d}\ell \nn
& =\left \langle f \right \rangle_{(\int\!\!\ell)}.
\end{align}
Hence, the values used for the requirements in Equation~\ref{eqn:AMlim} are unaffected, and now refer equally also to $A$ and $M$.
We will therefore use
\begin{align}\label{eqn:AMlim}
\mathcal{A}' & = A & &  \leq 10^{-7} & \Rightarrow & & \sigma^2[|\boldmath{c}|] & \leq 10^{-7}  \nn
\mathcal{M}' & = M & & \leq 4.0\times10^{-3} & \Rightarrow & & 2\left<m\right> & \simeq 4.0\times10^{-3} .
\end{align}
}

\subsection{A Hierarchical Structure by which Systematic Effects may be Identified, Evaluated and Controlled}

We start by identifying lower-level contributing factors which might contribute to the knowledge biases  and  uncertainties. These may be grouped into categories, such as the imperfect knowledge of the source characteristics, calibration errors, residual effects in correcting for detector effects, and the imperfect modelling of the PSF itself. Some categories will pertain to the instrument design and others to the data processing. In order to minimise the interconnectedness of the different factors, and hence to maximise the visibility and control of the bias effects introduced by each, some thought is required as to this categorisation of factors, and to the organisational hierarchy relating the categories.

Within each of these categories there are several factors to be considered: for example, within the PSF modelling there are fundamental imperfections of the model, and then inaccuracies in the parameters derived for the model resulting from photon statistics and pixelisation effects (which need to take account of multiple exposures, if these are used). In the category of calibrations, an example may be imperfect subtraction of the electronic reference level, an allowance for the effect of imperfect identification of cosmic rays, and so on. In the category of detector effects, an example may be the imperfect correction for radiation damage effects or detection chain non-linearity. This process must be continued to lower levels: in the last example the further contributing factors may include the output node linearity of the CCD, the linearity of the analog electronics associated with that node, the characteristics of the analog-to-digital conversion, and so on. In addition, it may be necessary to consider the stability of these different parts, and the accuracy with which any factor can be established -- this is also connected to the calibrations. 

As the level of accuracy required from the experiment increases, more and more factors must be considered, each of which will contribute to a degradation in performance. At some level, however, the factors become negligible, or can be made so by design or through operational strategies. For example, the flat fielding of the detector can be made unimportant by combining a large number of flat field calibration exposures. In practice, these less significant effects may not be fully evaluated, at least in the early stages of a programme.

We must now quantify the weighting functions $a_i$ and $m_i$. The first thing to notice is that we will want to calculate these in real space, because directly measured values will be used for $\rg$, $\rp$ {\it etc.} 
To relate these to the limits derived in Fourier space, the simple scaling from \reda{${\mathcal A}'\rightarrow A = \sigma^2[|c|]$} and \reda{ ${\mathcal M}' \rightarrow M = 2\left< m \right>$} using Equation~\ref{eqn:AMlim} is used. The $a_i$ weighting functions will remain unchanged, while the $m_i$ can simply be halved. We will therefore make this adjustment, and from this point work in real space, using primes to designate the real space $a'_i=a_i$ and $m'_i=m_i/2$.

If we assume a limiting magnitude $m_{\rm AB} = 24.5$, then we can adopt $R_{gal}=0.20$ arcsec (MHK13). The value for $R_{_{\rm C}}$ and $R_{_{NC}}$ will depend on the experiment: here we will use $R_{_{\rm C}}= 0.22$ and $R_{_{NC}}=0.05$ arcsec as in the \eu mission. Then $R_{\rm gal}/R_{_{\rm C}}=0.91$, and $R_{\rm obs}/R_{_{\rm C}}=1.6$. If we further set $|\ep| <0.1$, which is generally achievable in practice, then $\langle|\ep|^2\rangle^{1/2} < 0.1$. We will use $\langle|\ep|^2\rangle^{1/2} = 0.1$ and fix $P_\gamma=1.86, P_R = 2.0$ and  $\pec = \penc = 1.0$. We can now calculate the $a_i$ and $m_i$:
\begin{align}
\label{eqn:a_values}
a'_1& =  \frac{1}{P^2_ \gamma P^2_R\pec^2} \left\langle\frac{\rp^4}{\rg^4}\right\rangle & = ~ & 0.10  \nn
a'_2 & =\frac{1}{P^2_\gamma\penc^2} \left\langle 1 + \frac{2}{P_R}\frac{\rp^2}{\rg^2}+\frac{1}{P^2_R}\frac{\rp^4}{\rg^4}\right\rangle & = ~ & 0.74 \nn
a'_3& =\frac{\langle|\ep|^2\rangle}{P^2_\gamma P^2_R\pec^2} \left\langle\frac{\rp^4}{\rg^4}\right\rangle  & = ~ & 1.0\times10^{-3}  \nn
a'_4& = 4a'_3 & = ~ & 4.2\times10^{-3}  \nn
a'_5 & = a'_3 & =~ & 1.0\times10^{-3} 
\end{align}
and
\begin{align}
\label{eqn:m_values}
m'_1 &= \frac{1}{P_R}\left\langle\frac{\rp^2}{\rg^2}\right\rangle & = ~ & 0.60 \nn
m'_2 & = \frac{2}{P_R}\left\langle\frac{\rp^2}{\rg^2}\right\rangle \left\langle\frac{\rc}{\ro}\right\rangle& = ~ &  0.17 ~~~~~~~~~~~~~~~~~~~~~~~~~~ \nn
m'_3 & =  m'^2_1/4 &  = ~ & 9.0\times10^{-2} \nn
m'_4 & =  m'^{2}_2/4 & = ~ & 7.5\times10^{-3} \nn
m'_5 & \,\, = m'_1  &  = ~ & 0.60
\end{align}
It is evident that the dominant weighting factors to $\sigma[|c|]$ are $a'_1$ and $a'_2$, and these act through the PSF ellipticity. The lower values of $a'_3$ and $a'_4$ allow the knowledge error of the PSF size to be more relaxed than the knowledge error of the ellipticity. For $\left\langle m\right\rangle$, $m'_1$, acting through the PSF size, and $m'_5$ are dominant, with a contribution from $m'_2$. Hence for the additive bias, the ellipticity error will need the closest attention, while for the multiplicative bias, the size is more important.

Having identified contributing factors and a way of organising them, together with their combinatorial rules and weightings, values must now be assigned to these terms in Equation~\ref{eq:As} and \ref{eq:Ms} in order to quantify their effects and to identify which are the more significant. Initially any values can be assigned in order to verify the combinatorial rules in the hierarchical structure. The next step is then to include reasonable values for the factors. Many different considerations will bear on the values adopted for each factor, including the mission and instrument design, cost, calibration strategies, data analysis techniques, risk, organisational resources and many others. Once an approach is identified, the factor values will typically be established though calculations and simulations. In some cases, the tools may not be available to realise a value directly and a judgement must be made on the basis of experience as to what reasonable allocations should be made for each factor, until at some later stage the value can be established more quantitatively. These calculations may indicate that some factors have disproportionate effects, while others are easily realised. This generally leads to a rebalancing with measures introduced to address the disproportionate effects (for example by a change in the technology used) and to simplify the approach in respect of those factors that are easily realised, until a viable functioning point is achieved. 

In order to illustrate this process with an example, we have provided a hierarchical structure in Table~\ref{tab:aln}. This identifies how the terms in Equations~\ref{eq:As} and \ref{eq:Ms} could be related to a model of instrumental or data-processing biases and uncertainties. Table~\ref{tab:aln}  contains numerical factors for most of the categories (we provide example values from the \eu case).
 
\begin{table*}
\begin{tabular}{l}
\begin{minipage}{0.9\columnwidth}
\vspace*{-4mm}
\begin{tabbing}
\hspace*{7mm} \= \hspace*{40mm} \= \hspace*{20mm} \= \hspace*{30mm} \=    \\
$ \sigma[|c|]$ \> $1.3\times10^{-4}$ Additive bias	 \>  \blue{(=$\sqrt{\mathcal{A}'}~\,~\leq3.2\times 10^{-4}$ -- Equation~\ref{eqn:AMlim})} \\		
$\left<m\right>$ \> $1.5\times10^{-3}$ Multiplicative bias \> \blue{(= $\mathcal{M}'/2\leq2.0\times 10^{-3}$ -- Equation~\ref{eqn:AMlim})} \\		
\end{tabbing}	
\end{minipage}
\\[-4mm] \dashedrule \\ [-1mm]
\begin{minipage}{0.9\columnwidth}
\vspace*{-3mm}
\begin{tabbing}
\hspace*{10mm} \= \hspace*{5mm} \= \hspace*{30mm} \= \hspace*{25mm} \= \hspace*{20mm} \= \hspace*{10mm} \=   \\
\> \> $|\epsilon_{_{\rm C}}|$\' 	 \blue{$a'_1$}  $7.6\times10^{-5}$ Weighted convolutive contribution to ellipticity	\\			
 \> \> : \>	$\sigma$ \' :	  \red{$2.3\times10^{-4}$} PSF model ellipticity knowledge {\it [Figure~\ref{fig:Lance_sigma}]}	\\		
\> \>  :	\> :	\> 	\blue{Q} \red{$2.1\times10^{-4}$} PSF model truncation error	{\it [Figure~\ref{fig:n_PCA}]}	\\
\> \>  :	\> :	\> 	\blue{Q} \red{$9.7\times10^{-5}$} PSF model coefficient  error		\\
\> \>  :	\> :	\> :	\> 	\blue{Q} \red{$7.0\times10^{-5}$} Photon noise	\\
\> \>  :	\> :	\> :	\> 	\blue{Q} \red{$6.7\times10^{-5}$} PSF model coefficient calibration error	\\
\>  \> :	\> :	\> :	\> :	\> 	\blue{Q} \red{$3.0\times10^{-5}$} Bias residual \\
\>  \> :	\> :	\> :	\> :	\> 	\blue{Q} \red{$3.0\times10^{-5}$} Flat field residual \\
\>  \> :	\> :	\> : 	\> :	\> 	\blue{Q} \red{$3.0\times10^{-5}$} Sky subtraction residual \\
\>  \> :	\> :	\> :	\> :	\> 	\blue{Q} \red{$3.0\times10^{-5}$} Linearity residual \\
\>  \> :	\> :	\> :	\> :	\> 	\blue{Q} \red{$3.0\times10^{-5}$} Cosmic Ray subtraction residual \\
\>  \> :	\> :	\> :	\> 	\blue{Q} \red{\rule{14mm}{0.2mm}} Pixelisation effects \it{[Figure~\ref{fig:4-3exp}]}	\\
\>  \> :	\> :	\> 	\blue{Q} \red{\rule{14mm}{0.2mm}} Spacecraft pointing		\\
\> \> :	\> :	\> :	\> :	\blue{Q} \red{\rule{14mm}{0.2mm}} Displacement between guider and instrument	\\
\> \> :	\> :	\> :	\> :	\blue{Q} \red{\rule{14mm}{0.2mm}} Guider measurement precision	 \\
\> \> : \>	$\delta$ \' :	  \red{$5.2\times10^{-5}$} Model transfer to object			\\
\> \> :	\> :	\> 	\blue{Q} \red{$3.5\times10^{-5}$} Wavelength variation of PSF contribution		\\
\> \> :	\> :	\> 	\blue{Q} \red{$3.5\times10^{-5}$} Linearity residual		\\
\> \> :	\> :	\> 	\blue{Q} \red{$1.0\times10^{-5}$} Out-of-band transmission {\it [Figure~\ref{fig:outofband}]} \\
\> \> $R^2_{_{\rm C}}$\' 	 \blue{$a'_3~$} \red{$2.2\times10^{-5}$} Weighted convolutive contribution to PSF size 				\\
\>  \'  \>  	 \blue{$m'_1$}  \red{$3.0\times10^{-4}$} Weighted convolutive contribution to PSF size 				\\
\> \'	\>  	\blue{$m'_3$}  \red{$2.1\times10^{-8}$} Weighted convolutive contribution to PSF size 				\\
\> \>:  \> 	$\sigma$ \'  : \red{$4.8\times10^{-4}$} PSF model size knowledge	{\it [Figure~\ref{fig:Lance_sigma}]}		\\
\> \> :	\> :	\> 	\blue{Q} \red{$3.0\times10^{-4}$} PSF model truncation error		\\
\> \> :	\> :	\> 	\blue{Q} \red{$3.7\times10^{-4}$} PSF model coefficient  error		\\
\> \> :	\> :	\> :	\> 	\blue{Q} \red{$3.0\times10^{-4}$} Photon noise	\\
\> \> :	\> :	\> :	\> 	\blue{Q} \red{$2.2\times10^{-4}$} PSF model coefficient calibration error	\\
\> \> :	\> :	\> :	\> :	\> 	\blue{Q} \red{$1.0\times10^{-4}$} Bias residual \\
\> \> :	\> : 	\> :	\> :	\> 	\blue{Q} \red{$1.0\times10^{-4}$} Flat field residual \\
\> \> :	\> :	\> :	\> :	\> 	\blue{Q} \red{$1.0\times10^{-4}$} Sky subtraction residual \\
\> \> :	\> :	\> :	\> :	\> 	\blue{Q} \red{$1.0\times10^{-4}$} Linearity residual \\
\> \> : 	\> :	\> :	\> :	\> 	\blue{Q} \red{$1.0\times10^{-4}$} Cosmic Ray subtraction residual \\
\> \> :	\> :	\> :	\> 	\blue{Q} \red{\rule{14mm}{0.2mm}} Pixelisation effects	\it{[Figure~\ref{fig:4-3exp}]}\\
\> \> :	\> :	\> 	\blue{Q} \red{\rule{14mm}{0.2mm}} Spacecraft pointing		\\
\> \> :	\> :	\> :	\> 	\blue{Q} \red{\rule{14mm}{0.2mm}} Displacement between guider and instrument	\\
\> \> :	\> :	\> :	\> 	\blue{Q} \red{\rule{14mm}{0.2mm}} Guider measurement precision	 \\
\> \> :\> $\delta$ \'  : \red{$5.0\times10^{-4}$} Model transfer to object			\\
\> \> :	\> :	\> 	\blue{Q} \red{$3.5\times10^{-4}$} Wavelength variation of PSF contribution		\\
\> \> :	\> :	\> 	\blue{Q} \red{$3.5\times10^{-4}$} Linearity residual		\\
\> \> :	\> :	\> 	\blue{Q} \red{$2.0\times10^{-5}$} Out-of-band transmission  {\it [Figure~\ref{fig:outofband}]}\\ 
\end{tabbing}	
\end{minipage}
\\[-5mm] \dashedrule \\ [-1mm]
\begin{minipage}{0.9\columnwidth}
\vspace*{-3mm}
\begin{tabbing}
\hspace*{10mm} \= \hspace*{5mm} \= \hspace*{30mm} \= \hspace*{25mm} \= \hspace*{20mm} \= \hspace*{10mm} \=   \\
\> \> $|\epsilon_{_{\rm NC}}|$ \' 	 \blue{$a'_2$}  \red{$9.5\times10^{-5}$} Weighted non-convolutive contribution to ellipticity	(CTI) \\		
\> \> :	\> :	\red{$1.1\times10^{-4}$} Non-convolutive contribution to ellipticity	(CTI) {\it [Figures~\ref{fig:CTI_correc} -  \ref{fig:CTI_fom}]} \\		
\>  \> : 	\> : 	\> 	$\sigma$ \' :  \hspace*{0.5mm} \red{\rule{14mm}{0.2mm}} Model  calibration error	\\
\>  \> :	\> :	\> 	$\delta$ \'  :  \hspace*{0.5mm} \red{\rule{14mm}{0.2mm}} Fidelity of physical model		\\
\> \> $R_{_{\rm NC}}$ \' 	 \blue{$a'_4~$} \red{\rule{13mm}{0.2mm}} Weighted non-convolutive contribution to size	 (CTI)		\\
\> \' \> 	 \blue{$m'_2$} \red{\rule{13mm}{0.2mm}} Weighted non-convolutive contribution to size	 (CTI)		\\
\> \' \> 	 \blue{$m'_4$} \red{\rule{13mm}{0.2mm}} Weighted non-convolutive contribution to size	 (CTI)		\\
\> \> :	\> : \red{\rule{13mm}{0.2mm}} Non-convolutive contribution to size	 (CTI)		\\
\> \> :	\>:	\> 	$\sigma$ \'  : \hspace*{0.5mm} \red{\rule{14mm}{0.2mm}} Model  calibration error		\\
\> \> :	\>:	\> 	$\delta$ \'  : \hspace*{0.5mm} \red{\rule{14mm}{0.2mm}} Fidelity of physical model		\\
\end{tabbing}	
\end{minipage}
\\[-5mm] \dashedrule \\ [-1mm]
\begin{minipage}{0.9\columnwidth}
\vspace*{-3mm}
\begin{tabbing}
\hspace*{10mm} \= \hspace*{5mm} \= \hspace*{30mm} \= \hspace*{25mm} \= \hspace*{20mm} \= \hspace*{10mm} \=  \\
\> \> $\alpha$ \' \blue{$a'_5$} \red{$5.2\times10^{-7}$} Weighted additive model bias knowledge	{\it [MHK13]}\\
\> \> :	\>	 $\delta$ \'  :   \red{$5.0\times10^{-4}$} Additive model bias knowledge	\\
\> 	\>: 	\>: \> 	\blue{L} \red{\rule{14mm}{0.2mm}} Model error on simulated data with monochromatic galaxies \\
\> 	\>:	\>: \> 	\blue{L} \red{\rule{14mm}{0.2mm}} Residual CTI correction error \\
\> 	\>:	\>: \> 	\blue{L} \red{$4.2\times10^{-5}$} Colour gradients {\it [Semboloni \etal (2013)]} \\
\> \> $\mu$ \' 	\blue{$m'_5$} \red{$1.2\times10^{-3}$} Weighted multiplicative model bias knowledge {\it [MHK13]}	\\
\> 	\>: \> 	$\delta$\' :  \red{$2.0\times10^{-3}$} Multiplicative model bias knowledge	\\
\> 	\>:	\>: \> 	\blue{L} \red{$1.0\times10^{-3}$} Model error on simulated data with monochromatic galaxies \\
\> 	\>:	\>: \> 	\blue{L} \red{$5.0\times10^{-4}$} Residual CTI correction error \\
\> 	\>:	\>: \> 	\blue{L} \red{$5.0\times10^{-4}$} Colour gradients {\it [Semboloni \etal (2013)]} \\
\end{tabbing}	
\end{minipage}
\\[-4mm] \dashedrule \\ [-1mm]
\begin{minipage}{0.9\columnwidth}
\vspace*{-3mm}
\begin{tabbing}
\hspace*{10mm} \= \hspace*{5mm} \= \hspace*{30mm} \= \hspace*{30mm} \= \hspace*{20mm} \= \hspace*{10mm} \=  \\
\> \> :	~~\red{$3.0\times10^{-5}$} Residual distortion				
\end{tabbing}	
\end{minipage}
\end{tabular}
\caption{Potential contributors to the knowledge error terms in Equations~\ref{eq:As} and \ref{eq:Ms}. Where the allocations are preceded by Q they terms are added quadratically to generate the next level up and where they are preceded by L they are added linearly. In the higher levels (if there are no Q or L designators) they are weighted as in Equations~\ref{eq:As}, \ref{eq:Ms}, \ref{eqn:a_values} and \ref{eqn:m_values} to provide final values for $\sqrt{\mathcal{A}'}$ and $\mathcal{M}'/2$. See text for details. Certain entries contain references to figures in later sections, where there is an evaluation of their feasibility.} 
\label{tab:aln}
\end{table*}

There are three broad categories in Table~\ref{tab:aln} reflecting those in Equations~\ref{eq:As} and \ref{eq:Ms}: convolutive contributions; non-convolutive contributions; and model bias knowledge, with a final direct contribution of the residual spatial distortion. Horizontal lines indicate that there is no allocation in this category in this example, though in general an allocation should be made. Their  indentation in the table indicates the organisation of the contributing terms.  The penultimate of the broad categories contains the allocations within $\alpha$  and $\mu$ of method errors, reflecting the final lines in Equations~\ref{eq:As} and \ref{eq:Ms}. However, we have not included bias error allocations $\sigma^2[\alpha]$ and $\sigma[\mu]$ as these can be minimised by simulations. The simulations will not reflect reality at some level, so there are residual knowledge biases, the remaining terms $\left<\delta(\alpha)\right>^2$ and  $\left<\delta(\mu)\right>$, for which an allocation is assigned.

It is easiest to follow the combination rules in Table~\ref{tab:aln} by starting from the right-most values. In the absence of any other information, the factors can be combined to calculate the next value in the hierarchy: quadratically if they are independent, or linearly if they are not. They have therefore been prefixed with a Q or an L. Moving leftwards, eventually the combinations will lead to a row labelled as a $\delta$ or a $\sigma$, depending on whether categories of knowledge bias or knowledge uncertainty have been combined. Now these knowledge errors must be combined with the weighting factors $a'_i$ and $m'_i$ in Equations~\ref{eq:As} and \ref{eq:Ms}, so the $\sigma$ and $\delta$ values are multiplied by their corresponding $a'_i$ and $m'_i$ to produce the values in the second column. These are then added, as prescribed in Equations~\ref{eq:As} and \ref{eq:Ms}, together with the residual distortion after correction factor, to provide the final values for $ \sigma[|c|]$ and $\left<m\right>$. Note that because the values are in real space,  the total multiplicative bias is halved in converting from $\langle m \rangle$ to $\mathcal{M}'$, as prescribed by Equation~\ref{eqn:AMlim}, and also that because the table propagates $\sigma$ rather than $\sigma^2$, due regard is required in calculating $\mathcal{A}'$.

We will evaluate the values in Table~\ref{tab:aln} in the subsequent sections of this paper to examine whether they are reasonable to use as a basis for the knowledge bias and knowledge uncertainties that may be achievable in a practical experiment. The values assigned to the different factors will be different for different experiments, and we emphasise that the purpose of Table~\ref{tab:aln} is not directly to prescribe any values in particular, nor primarily to justify the values used, but rather to provide a conceptual structure for  a weak lensing experiment in space.

\subsection{Absolute Characteristics of the PSF and Further Breakdown within Subsystems}

Recall that the PSF characteristics (factor (iii) in the discussion in Section~\ref{sec:top-level}) must be suitable. In particular this applies to the PSF  size in $R^2$ terms, and its ellipticity $\bmath{\ellipticity}$. In addition to $R$, $F$, the full width half maximum (FWHM) is also often used to provide an additional constraint on the PSF, specifying the width of its core. This is primarily to guard against PSFs which might be problematic in some characteristics while nevertheless conforming with the requirement on $R$ (for example annular PSFs, resulting from an out-of-focus condition).  The PSFs can be evaluated using standard procedures to examine whether they meet the  requirements. The only non-standard component in the breakdown is the contribution of the detector radiation damage effects to size and ellipticity, as these are not generally rendered in size and ellipticity terms. However, they can be calculated using Equations~\ref{eq:quad} -- \ref{eq:ellip}. 

When {\it further} breaking down contributions within the system, for example to constrain the individual contributors of different subsystems within the experiment to the overall system PSF, care is required in their combination. In particular, the ellipticities must be weighted taking into account the values of the FWHM $F_i$ for that particular contribution to the PSF: this is because a contributing factor may be intrinsically strongly elliptical, but with small associated $F_i$ it will be relatively unimportant. A reasonable approximation by which to combine the ellipticities is as
\begin{equation}\label{eqn:eps_c_combine}
\bmath{\ellipticity}_{\rm tot} = \sum_i (F_i^2/F^2_{\rm tot}) \bmath{\ellipticity}_i 
\end{equation}
for the terms that can be represented by a convolution and 
\begin{equation}\label{eqn:eps_nc_combine}
\bmath{\ellipticity}_{\rm tot}=\sum_j \bmath{\ellipticity}_j
\end{equation}
for those that cannot. The knowledge residuals in Table~\ref{tab:aln} should be similarly combined:
\begin{equation}\label{eqn:eps_kn_c_combine}
\sigma^2[ \bmath{\ellipticity}_{\rm tot}]=\sum_i (F_i^2/F^2_{\rm total})^2 \sigma^2[\bmath{\ellipticity}_{i}]
\end{equation}
for the terms that can be represented by a convolution and
\begin{equation}\label{eqn:eps_kn_nc_combine}
\sigma^2 [\bmath{\ellipticity}_{\rm tot}]=\sum_j \sigma^2[\bmath{\ellipticity}_{j}]
\end{equation}
for those that cannot. While all $\bmath{\ellipticity}_i$ in Equations~\ref{eqn:eps_c_combine} and \ref{eqn:eps_nc_combine} have equivalent $F_i$ categories, this may not be the case for the $\sigma[\bmath{\ellipticity}_{i}]$ ellipticity knowledge residuals in Equations~\ref{eqn:eps_kn_c_combine} and \ref{eqn:eps_kn_nc_combine}. In this case an appropriate mapping must be assigned, possibly through experience.\\[2mm]

\subsection{Recapitulation}

To recap, the area covered by the survey, and its limiting magnitude, together with the overall characteristics of the PSF, will set the number of galaxies that can be used for the weak lensing measurements. This will provide a level of random error. In order to achieve a systematic error which is some moderate fraction of this random error set by the Poisson noise, a requirement then arises for the control of systematic effects. This can be apportioned between additive and multiplicative effects, the impact of each of which depends on lower level uncertainties -- the weighted individual knowledge bias $\left<\delta\right>$ and knowledge uncertainties $\sigma$ of the convolutive, non-convolutive and model error size and ellipticity. 

For the  practical experiment, an evaluation must be carried out for these individual component factors, to establish whether they are achievable using techniques at hand, or from reasonable projections of what techniques may become available in the timeframe of the mission. The most significant factors must be quantified through an appropriately detailed assessment, for which Table~\ref{tab:aln} may be used as a starting point. The evaluation is likely to entail large scale simulations and evaluations. These will establish whether the value assigned to the different factors can be reached in any concrete design or procedure. If not, the values can be adjusted to relax the constraint on any one contributing factor, but then others must be tightened accordingly, in order to remain within the required levels of $\mcap$ and $\mcmp$. In this way, a balance may or may not be achieved, depending on the characteristics of the mission, with implications for its feasibility.  

So far we have provided the rationale by which the systematic biases can be identified, organised and their combined effects evaluated, and provided an example hierarchical structure in Table~\ref{tab:aln}. Table~\ref{tab:aln} contains example numerical values for the different factors, derived from the \eu programme. We will examine in Section~\ref{sec:PSFmodelling} and beyond some of the methods by which these values can be calculated, with the aim of illustrating the process, rather than providing a justification for any particular case. To do this, we first need to create simulated data, and process these in a representative manner, so we discuss briefly how these may be done, concentrating on aspects of particular importance to a weak lensing experiment.
\vspace*{3mm}

\section{Simulations and Data Processing}
\label{sec:sims}

\subsection{Procedure}
\label{sec:simproc}

To quantify the impact of biases we must create simulated data with the appropriate level of fidelity, incorporating information from laboratory and prototype tests.  The simulations typically proceed as follows:
\begin{enumerate}
\item The telescope \reda{optical model is used to produce many different instances of the optical PSF at different locations on the field-of-view and at different wavelengths (to explore the wavelength dependence). The PSFs should be super-sampled by a sufficient factor compared to the detector grid. The imperfections in the alignments of the optical elements and also the manufacturing errors for the optical elements are generally included via perturbations to the optical model: these can be generated by  Monte-Carlo sampling over the likely range of  the misalignments and manufacturing errors. For part of the field of view, a finer sampling of field-of-view points may be used to explore variations on smaller spatial scales until no further variation is found.} The aim is to sample the instrument states and the spatial and spectral variation of the PSF at many points on the field of view in a representative fashion.  
\item \reda{Photons incident on a pixel will generally be recorded in that pixel, but there is a finite possibility that they will be recorded in an adjacent pixel owing to charge spreading within the CCD. This effect is characterised from laboratory measurements as a function of wavelength, using an optical spot which is as small a fraction of the pixel size as possible. The charge spreading is then added as a convolution to the PSF for each wavelength.}
\item \reda{The pointing of the satellite will not be completely stable, blurring the PSF slightly. A sequence of optical PSFs  are displaced slightly in the focal plane according to a time series of simulated pointings  from the satellite attitude control system, and then integrated over the exposure duration. This completes the generation of the system PSF at any  point in the field of view, for any particular pointing displacement time series and for each wavelength.}
\item \reda{These system PSFs represent stellar images, and they are then pixelised onto the detector grid with the appropriate intensity, \reda{spectral and positional} distribution according to real sky data (or an appropriate Galactic model) and the expected instrumental throughput. If real-sky data are used for the simulations, then the optical distortion map should be applied, and the star positions displaced to take account of the individual CCD positions and rotations in the focal plane (as derived from simulations or engineering measurements).}
\item Galaxy images are \reda{produced using} galaxy models, or from real data from deep field observations, and then scaled and rotated individually and with a number distribution consistent with observed number counts. Weak lensing shears are also added at this stage if required. Each image is convolved with the system PSF derived from the steps above, again taking account of the instrument throughput. \reda{They should be distorted using the telescope optical model.} These images are also pixelised onto the detector grid.
\item  \reda{The internal and external (cosmic) background, and the CCD thermal  (dark) noise are modelled and added.  Poisson noise is added to the accumulated signals and backgrounds for each pixel. The pixel-pixel non-uniformity is then applied through multiplication by the flat-field map. Detector cosmetic effects (hot and dead pixels) are then included.} Saturation ceilings \reda{appropriate to the full-well capacity of the CCD} are applied and the associated pixel bleeding calculated.
\item CCD radiation damage models are then applied to reproduce the radiation damage effects, again using representative laboratory data from radiation testing to ensure their fidelity. Readout noise for \reda{each} CCD readout node is finally added, together with \reda{its} electronic bias level. 
\end{enumerate}
\reda{Some aspects of the performance evaluation may not require all of these stages, and, in particular, subfields of view are often sufficient to examine many effects.}

The survey strategy may also require simulation \reda{to impose the correct displacements between successive exposures and} to ensure that the anticipated mission samples the sky adequately to allow the information about the galaxy shears to be recovered to the required level of accuracy.

\subsection{Radiation Damage Effects}
\label{sec:radsims}

While most of the procedures above are relatively standard, the modelling of the radiation damage effects in the CCDs requires particular attention, as this is where most of the non-convolutive effects in the system arise. 

The radiation environment above the Earth's atmosphere will gradually degrade the performance of all electronics. The principal impact for a CCD-based weak lensing experiment will be changes to image shapes as a result of radiation-induced lattice damage in the CCDs. This will introduce inefficiencies in the charge transfer during readout (Charge Transfer Inefficiency; CTI). As electrons are transferred to an amplifier at the edge of the device, they can be temporarily captured by lattice defects (ÔtrapsÕ) and released only after a time delay (e.g. Holland \etal 1990). These electrons then appear in pixels subsequently read out, as a spurious trail behind the image in both the column (parallel) and row (serial) directions. The degradation is negligible in pixels adjacent to the readout node, because electrons undergo few transfers before being read out, and worst at the positions furthest from a node. The effect will modify the size and introduce a spurious elongation of galaxies (Massey \etal 2010), dependent on flux and its position on the CCD, directly modifying the cosmological weak lensing signal if not accounted for. CTI trailing is particularly troublesome because the trailed flux is a nonlinear function of the total flux (signal plus sky background), and of the size and shape of a source, and therefore contributes as a non-convolutive effect.  

There are two main types of CTI models used in the simulations. In the first, the charge transfer process is modelled statistically in detail, including the interactions between the charge cloud and the electric field structure within the pixels, by Monte Carlo techniques (e.g. Seabroke, Holland \& Cropper, 2008; Prod'homme \etal 2011). These potentially offer the highest fidelity description of the radiation damage effects. However because the parameters in the model (mainly the capture and release times, and the capture cross sections) are determined by iterative fitting to laboratory data, these parameters are in practice not well constrained, as  the models are computationally intensive. The other approach is to capture as well as possible in a simplified model the essential physical interactions while modelling the statistical effects: this enables rapid iterative parameter fitting at the cost of a reduced fidelity. Examples include the model developed for the {\it Hubble Space Telescope} Advanced Camera for Surveys (Massey \etal 2010) and the Charge Distortion Model ({\sc cdm03}) (Short \etal 2010) used in the {\it Gaia} programme, both of which have variants explicitly tailored for weak lensing surveys. Unfortunately, the approaches are currently sufficiently different that some of  the parameters determined in the second approach are not yet directly useful for the Monte Carlo approach. 

The radiation environment to be experienced by the detectors is mainly parameterised by the mission duration, as well as the fluence and energy spectrum at the orbital location. All laboratory test data should be as representative of the flight condition as possible. Because lattice damage effects are different at ambient and cold temperatures, irradiations of test devices should be made at operational temperature, and the devices should be maintained at that temperature for the subsequent characterisation: this is logistically difficult. The main operational dependencies are the temperature and the CCD parallel and serial transfer rates, and these will be the main parameter space to be explored in the testing. Both are affected by the shape of and voltage levels used in the actual waveforms to read out the device, and generally this will be explored beforehand and should be agreed and standardised for all further tests, to permit the inter-comparison of results. The results will also be different for different device types, because of the different physical pixel structure, manufacturing procedures and raw Si characteristics. In particular, different doping regimes will lead to different trap populations. 

Once the parameters are determined by fitting to laboratory test data, the models are used to include the radiation effects in the simulated images in step (vii) above. Generally, the test data improve as the programme proceeds, partly because actual flight-design CCDs may not initially be available. Consequently these simulations will evolve.

\subsection{Cosmic Rays}

The effect of high-energy ionizing particle fluxes (electrons, protons, and ions)  on the detectors in terms of induced transient tracks on the images must be included in the simulations because they effectively reduce survey area when they are excised, or the local exposure duration if there are multiple exposures of each field. There is also an allocation made to any systematic effects this may induce in Table~\ref{tab:aln}. The fluxes can be simulated using Monte-Carlo codes such as the {\sc stardust} code (Rolland \etal 2007). These are able to compute realistic samples of images and the statistical properties of the induced particle tracks. They incorporate solutions of the diffusion equation and take into account the nuclear reactions, the shield anisotropy description, the propagation of energetic electrons and the generation of delta electrons.

The main input data are the information about the detector structure, the environment particle spectra and a thickness table describing the shield around the detector.
Generally a particle travelling inside the detector is assumed to lose energy along its trajectory according to the continuous slowing down approximation or via the production of delta electrons or by nuclear reactions (for the protons and the ions). The number of deposited electron-hole pairs can be obtained by dividing the deposited energy by the energy necessary to create a thermalized electron-hole pair (3.6 eV in Si). Charges, primary or secondary deposited in depleted zones must be directly collected; otherwise, they are subject to diffusion.

\subsection{Normal processing steps}

The raw data from a weak lensing experiment is not used directly. A data processing sequence is carried out to reach the required data quality from which the galaxy shears can be measured. This incorporates external information (such as parameters from laboratory tests, or astrometric source parameters) and internal calibration data. The information can be used either directly, or through a model. This adds to the information content of the data, but very great care must be taken in the quality of the external information, and in the acquisition of the calibration data to ensure that these are taken in a representative manner. At the extreme level of accuracy required for the weak lensing measurements, this incorporation procedure will always, at some level, introduce bias effects feeding into $\mcap$ and $\mcmp$, and this is reflected in the need for an allocation for incorrect values in Table~\ref{tab:aln}. Other calibrations of, for example dense star fields, can potentially be used to examine PSF spatial variability on small scales, if these exposures can be considered sufficiently representative in order not to introduce spurious biases. Wherever possible, the calibration information should be from within the data frames themselves. 

The processing follows a process of electrical bias (zero light level) subtraction, correction for linearity, correction for CTI effects, flat-fielding, correction for detector cosmetic defects, cosmic ray subtraction and astrometric correction. Scattered light contributions may be removed and the background modelled. Because the knowledge of the PSF  is one of the most critical aspects in a weak lensing experiment, the steps requiring the most attention are those that impact the PSF. 

\subsection{Correction for Radiation Damage}
\label{sec:CTI_correc}

The absolute density of charge traps will gradually increase during the mission as radiation damage accumulates. Laboratory tests indicate that there are different species of charge traps. While all of these species contribute to the total CTI, not all of the species equally degrade weak lensing measurements. Following Rhodes \etal (2010), charge traps with a characteristic release time much shorter than the charge transfer speed at which rows and columns of pixels are read out, move electrons by at most one pixel, and hence affect astrometry. Charge traps with long characteristic release times remove electrons entirely from a source, and degrade photometry. As shown by Massey \etal (in preparation), charge traps with characteristic release times a few times the charge transfer speed move electrons from the core of an astrophysical source into its wings, and primarily affect its size and ellipticity, and these have the most negative effects on the weak lensing measurements. 

As noted above, one of the early steps in the data processing is to minimise as far as possible the radiation damage CTI effects in the data. This is a critical step, as the residuals from this process will be the largest non-convolutive knowledge contributor in Table~\ref{tab:aln}: the effects modify the magnitude, ellipticity and position of the object in a intensity- and size-dependent manner. In this respect, the CTI also creates non-linearity which is in addition to that caused by the detection chain (Section~\ref{sec:lin}).

Ideally, the trailed electrons should be returned back to the pixel to which they belong. Fortunately, the trailing is typically a small perturbation around the true image, so an inverse operation can be achieved via a rapidly-converging iteration of the forward algorithm (Bristow \& Alexov 2002, Massey \etal 2010). This is done by taking the real data, passing them through the best model available for CTI effects -- generally those used in the simulations in Section~\ref{sec:radsims} -- to create double, and higher multiples of the damage. Linear combinations of these images are subtracted from the original data to remove the effects to the required level. The procedure is shown in table~1 of Massey \etal (2010).

The level of correction that is possible depends upon the accuracy of the CCD CTI model and the level of readout noise. Readout noise places a fundamental, hardware limit on the correction accuracy because it is added to an image after the charge transfer, and is therefore not trailed (Anderson \& Bedin 2010, Massey \etal (in preparation). This effect leads to correlated noise in the final corrected image.
 
Note that during charge transfer, charge capture and release are stochastic events, the exact location of each trailed electron cannot be accurately predicted. Individual galaxies on individual exposures may not be perfectly corrected, but statistical measurements of an ensemble galaxy population can be corrected to an unbiased level (Rhodes \etal 2007, Massey \etal (in preparation). However, with advanced calibration techniques known as pocket-pumping in which an image is shifted forwards and backwards by a small number of pixels, the location and characteristics of each trap may be ascertained (Janesick 2001).

\section{Modelling the System PSF}
\label{sec:PSFmodelling}

\subsection{Overview}

After the data processing, including the calibrations and the CTI correction, a set of optimised images containing stars and galaxies is available for each part of the sky. We are now in a position to calculate the shear maps to determine the cosmological parameters. 

In order to reach the accurate measurement of galaxy shape needed for reconstructing the shear maps, classical deconvolution approaches do not deal adequately with the effects of noise and finite sampling by the detector. There has been continuous progress over the past decade in the accuracy with which shear can be measured (for example Heymans \etal 2006, Massey \etal 2007, Bridle \etal 2010, Kitching \etal 2010, 2012). In one approach, for each particular galaxy shape measurement, a galaxy model is constructed from a combination of intensity profiles. These profiles are convolved with the model PSF, and compared in a model fitting process to the observed galaxy, the true image of which has been convolved with the true (but not fully known) instrument PSF (for example Miller \etal 2007, 2012, Kitching \etal 2008b). In another, the ellipticity is computed more directly using the quadrupole moments as in Equation~\ref{eq:ellip} (for example Kaiser, Squires \& Broadhurst 1995, Luppino \& Kaiser 1997, Hoekstra \etal 1998).

An allocation for the uncertainty in this model-fitting process is in the model bias knowledge $\left<\delta(\alpha)\right>^2$ and  $\left<\delta(\mu)\right>$ lines of Table~\ref{tab:aln} and some of these aspects are discussed briefly further in Section~\ref{sec:galaxy}. We continue in this section with the error arising from the fact that the instrument PSF is not fully known: calculated or estimated values are in the lines of Table~\ref{tab:aln}. Stars provide measures of the PSF at different points on the field of view, which will enable the PSF for any particular galaxy to be calculated. Each exposure will also have been taken under slightly (perhaps minutely) different conditions, for example changed payload temperatures resulting in different optical alignments. The PSF model for each galaxy must be reconstructed from the stellar PSFs in the field of view, and the task is to model the PSF to reach a level of fidelity to the true PSF such that the biases in the shear measurements must be within the levels in Equation~\ref{eqn:AMlim} for the cosmological goals to be met.

As already related in respect of  Table~\ref{tab:aln}, this modelling process has several categories. Some are related to the amount of information that is available to reconstruct the PSF for any particular galaxy.  For example, the spatial sampling must be adequate (Section~\ref{sec:sampling}). Another is the precision available in the calibrating PSFs simply from their photon shot noise. Other categories include the mathematical form of the model used to characterise the PSF, the accuracy with which the coefficients of the model can be derived in order to construct any particular PSF and the accuracy with which calibrations can be transferred to the particular stellar PSF being modelled. 

We now examine the main categories of convolutive effects, non-convolutive effects, and galaxy modelling in Table~\ref{tab:aln} in sequence.

\subsection{Convolutive effects in the PSF modelling}
\label{sec:conv}

\subsubsection{Sampling Issues}
\label{sec:pix}
\label{sec:sampling}

Assuming a central wavelength of $\lambda$ for the instrument bandpass and a primary mirror of diameter $D$, all images are fundamentally bandlimited at a spatial frequency of $u_{max} = D/\lambda $ even in the presence of spatial blurring in the remainder of the instrument.  The telescope PSF can contain no modes at higher frequency than this value, and thus no higher frequency signal remains in images after convolution with this PSF. A band-limited image can fully be recovered as a continuous function, without loss of information or accuracy, using Sinc function interpolation between a discrete set of samples provided these samples are spaced at a greater spatial frequency than the critical sampling rate, or Nyquist rate, $2u_{\rm max}$.
An output image must therefore be constructed at a resolution of  $< 1/(2u_{max})$ radian per sample in order to be fully-sampled and allow full reconstruction of the sky. 

In general, for a given optical system, and number of pixels in the focal plane, there is a tradeoff between maximising the survey area (more arcseconds per pixel) and maximising the PSF sampling (fewer). If the system PSF is fully sampled by the detector pixel grid, then it is not degraded by the sampling. If on the other hand some compromise is made to enlarge the field of view, with the expectation that some spatial resolution can be regained through multiple exposures, which are often required for other reasons in any case, then an analysis of the effect of this undersampling is required.  Further, while undersampling is an important consideration in the modelling of the PSF, aliasing also affects galaxy shape measurement for methods which do not directly fit parametric models to the data.

\begin{figure*}
\includegraphics[width=0.805\columnwidth]{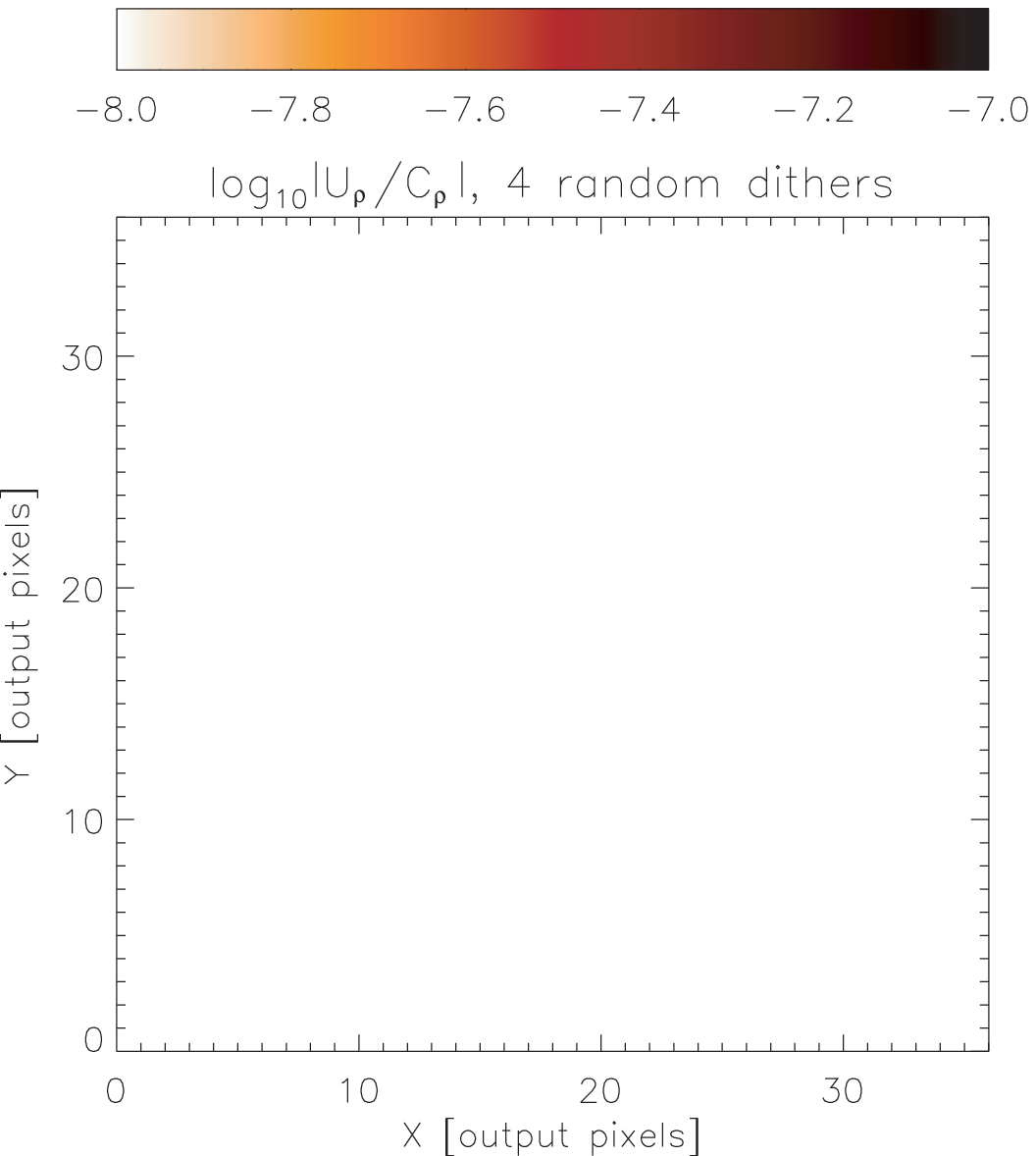}\hspace*{2mm}
\includegraphics[width=0.775\columnwidth]{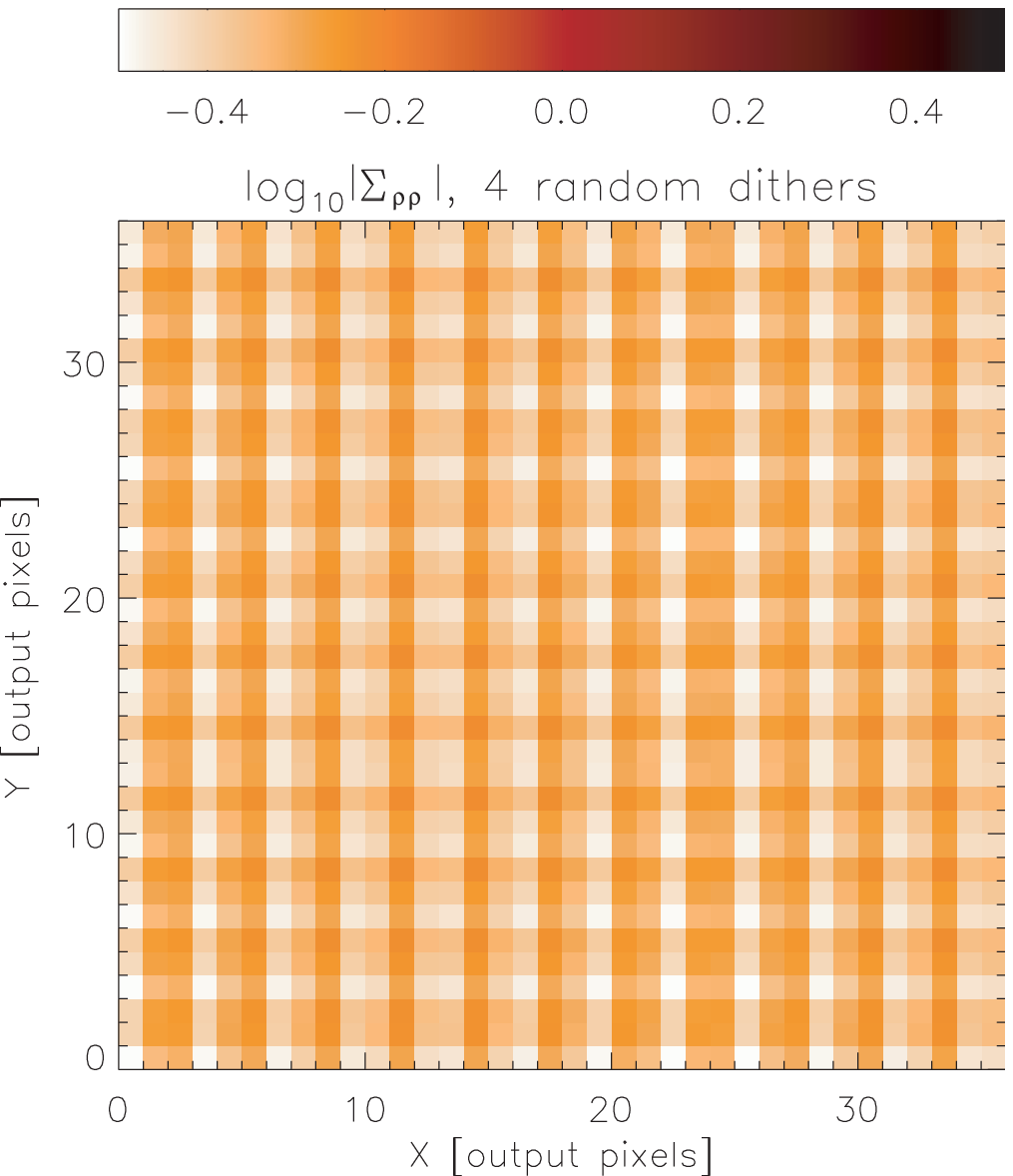} \\
\includegraphics[width=0.8\columnwidth]{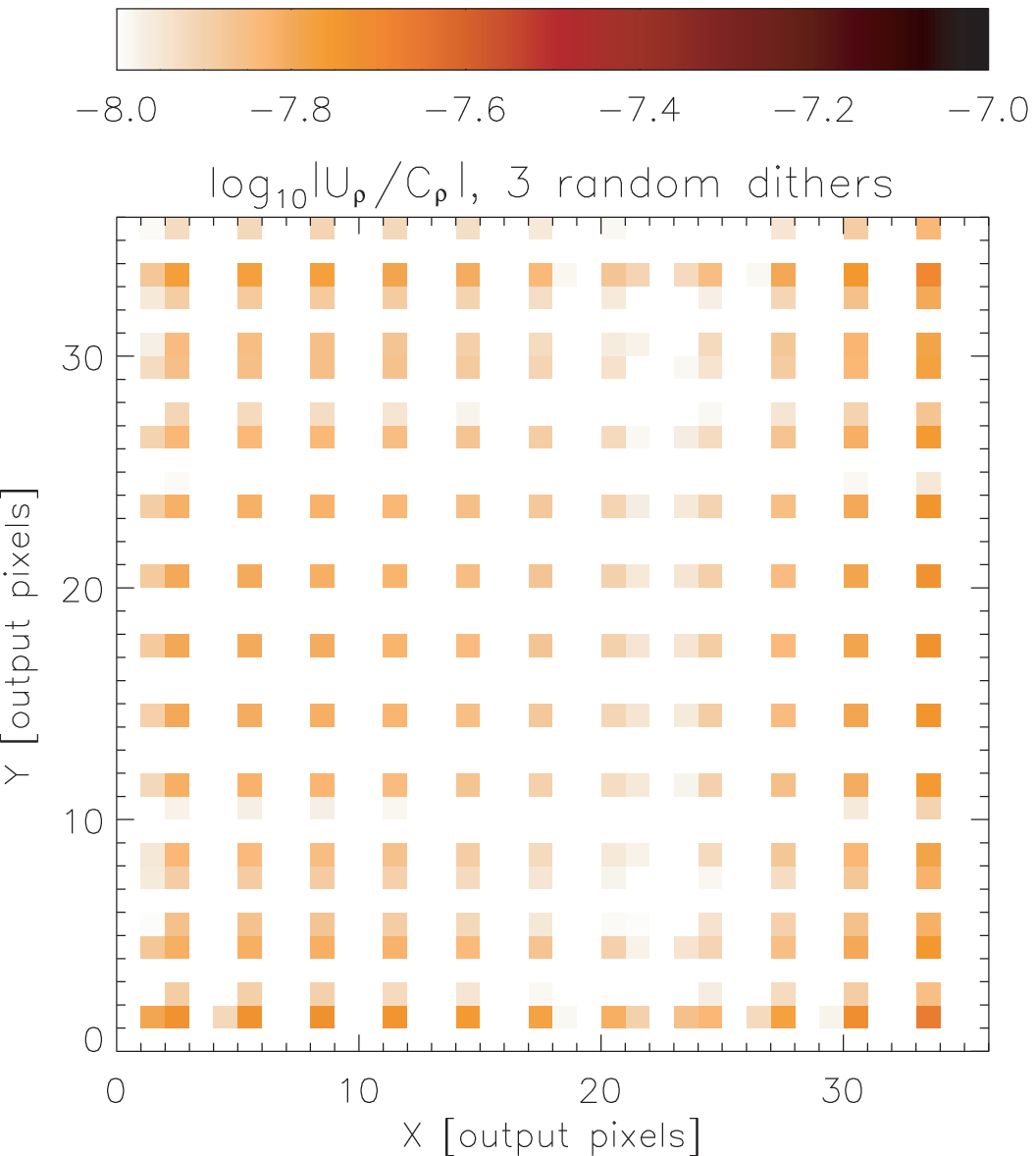}
\includegraphics[width=0.8\columnwidth]{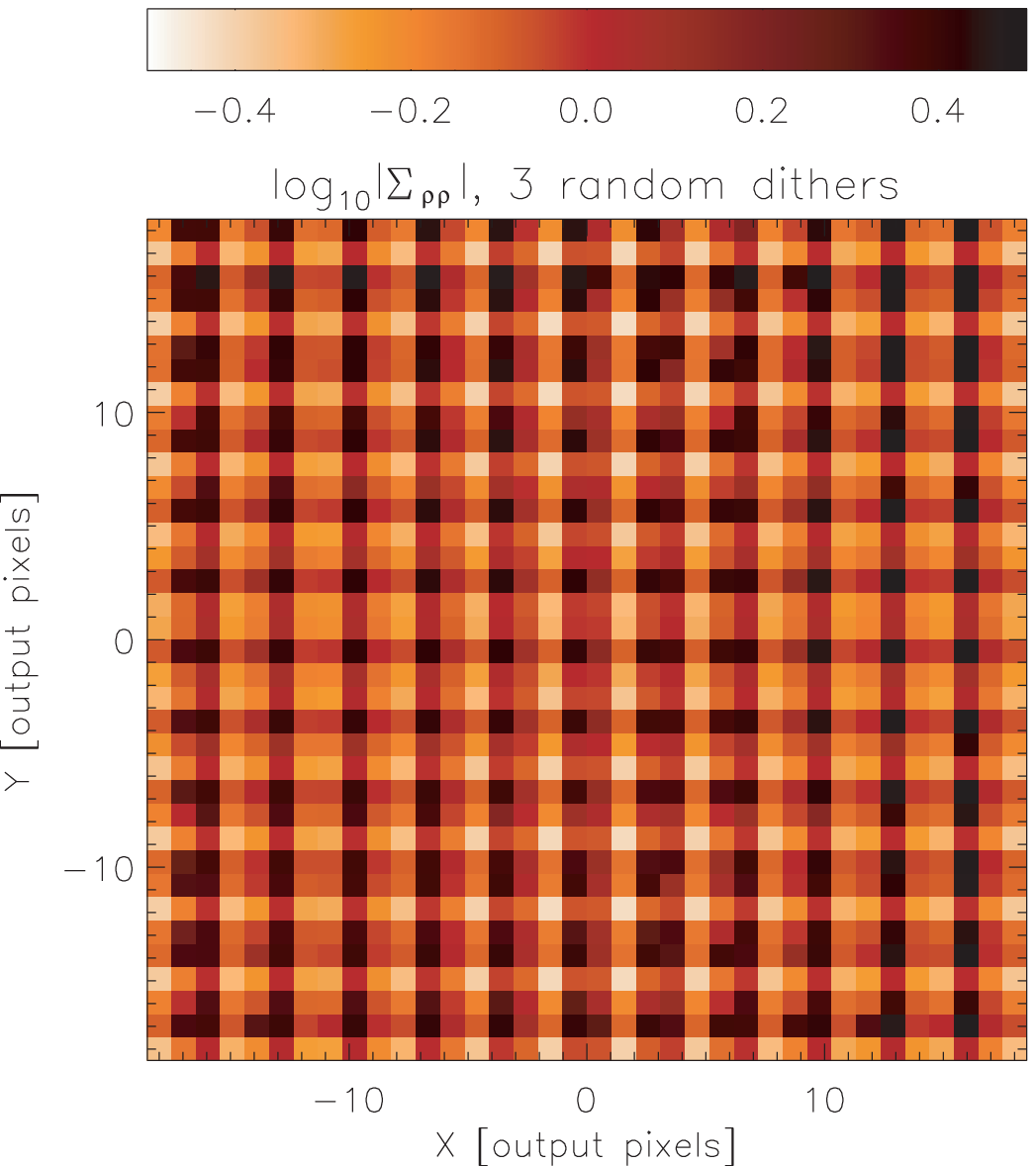}
\caption{Maps of normalised output leakage objective $U_{\rho} / C_{\rho}$ (left column) and noise variance $\Sigma_{\rho\rho}$ (relative to a unit input variance; right column) after linear image combination from single realizations of a four (upper row) and three (lower row) random dither configuration.  These output maps are taken from the central regions of large images, showing the central regions of primary interest (see Rowe \etal\ 2011). }
\label{fig:4-3exp}
\end{figure*}

We start by examining the effect of moderate undersampling with 3 or 4 exposures, as such a sequence may be typical to recover gaps in the detector matrix. The full system PSF should be used to test the sampling characteristics and these sampled on the detector pixel grid, with successive exposures offset to mimic the multi-exposure survey strategy. 
We use an optical PSF based on the \eu example with an input focal plane sampled at 0.688 of the Nyquist rate. We then convolved this PSF with an additional circular Gaussian of standard deviation $0.196 / (2 u_{\rm max})$, to approximate the dispersive 
effects of charge diffusion within the CCD pixels. Finally, we add an additional jitter component to
the combined PSF using a time series of 216  jitters of  displacement $0.145 / (2 u_{\rm max})$
(Gaussian rms).

We use the optimal linear image combination formalism of Rowe, Hirata \& Rhodes (2011), sampling the output at the Nyquist rate. This formalism aims to minimize two contributions to imperfect image reconstruction: the leakage objective $U_{\rho}$ and the output noise variance $\Sigma_{\rho\rho}$.  The former measures the fidelity of the output image to the target PSF: in this test a low value for $U_{\rho}$ indicates that unwanted changes to the PSF from the linear combination process have been small.  To set an absolute tolerance on this quantity, it is useful to consider the normalized leakage objective $U_{\rho}/C_{\rho}$, where $C_{\rho}$ is a measure of the integrated PSF autocorrelation (see Rowe \etal\ 2011). A tolerance value for a normalized leakage objective of $U_{\rho} / C_{\rho} < 10^{-8}$ approximately corresponds to controlling unwanted changes to the PSF to better than one part in $10^4$ and ensures that such changes are a minimal contribution to the PSF uncertainty budget in Table~\ref{tab:aln}.  The output noise variance $\Sigma_{\rho\rho}$ is specified in units of the variance of noise in the input images.  An output $\Sigma_{\rho\rho}< 1$ therefore demonstrates that the noise variance in the output pixels is reduced relative to the noise in the inputs, and this can be taken as an indication of stable control of noise in the reconstructed, fully-sampled output image. Following the methodology of Rowe \etal\ (2011), we can test to see whether the PSF and multiple-exposure strategies will allow linear combination of input images to generate output images that are unbiased at the $U_{\rho} / C_{\rho} < 10^{-8}$ level while simultaneously keeping output noise to an acceptable level. 

In the upper panels of Figure~\ref{fig:4-3exp} we show maps of normalized $U_{\rho} / C_{\rho}$ and $\Sigma_{\rho\rho}$ (given in units of the input noise variance) for a single realization of a four randomly-offset exposure system (dithers) in which the sampling is 0.688 of the Nyquist rate.  This realization was one of 30 realizations tested, and results were typical.  $U_{\rho} / C_{\rho}$  is found to be $< 10^{-8}$ everywhere.  The output maps shown come from the central regions of the input images, where data coverage is good and edge effects do not impair results. As discussed in Rowe \etal (2011), the effects in edge regions can be mitigated in real data by tessellating many small regions of reconstructed output such as those shown.

For the ensemble of 30 realizations tested, the average $U_{\rho} / C_{\rho}$ in the reconstructed output was  $9.95 \times 10^{-9}$, in the very centre of the specified tolerance range, demonstrating a desired level of control over unwanted distortions in the output image.  The average noise variance $\Sigma_{\rho\rho}$ in the reconstructed output was 0.663 (in units of the input variance). Because there is no background in the images, this variance will be an upper limit with respect to real sky exposures. This demonstrates that for a PSF sampled at 0.688 of Nyquist, with four input exposures, a linear combination of images can be used to generate fully-sampled output while maintaining acceptable levels of noise in the output.

The lower panels of Figure~\ref{fig:4-3exp} show the normalized $U_{\rho} / C_{\rho}$ and $\Sigma_{\rho\rho}$ for a single realization of a three exposure pattern.  Here the optimal linear combination produces  $U_{\rho} / C_{\rho} > 10^{-8}$ at some points in the output image: these can be seen as red squares.  It can also be seen that that output noise variance for this reconstruction is significantly greater than was the case for four input exposures. 

As for the four exposure case, a total of 30 realizations of the three exposure scenario were investigated. The average $U_{\rho} / C_{\rho}$ in the reconstructed output pixels across this ensemble  was  $1.0583 \times 10^{-8}$, slightly larger than the desired maximum for reconstruction, and the average $\Sigma_{\rho\rho}$ was 1.4405 (in units of the input variance).  The results from a single realization shown in Figure~\ref{fig:4-3exp} are typical of these tests, but the variation in reconstruction quality between realisations was noticeably greater than in the four exposure case.

\begin{figure}
\includegraphics[width=0.8\columnwidth]{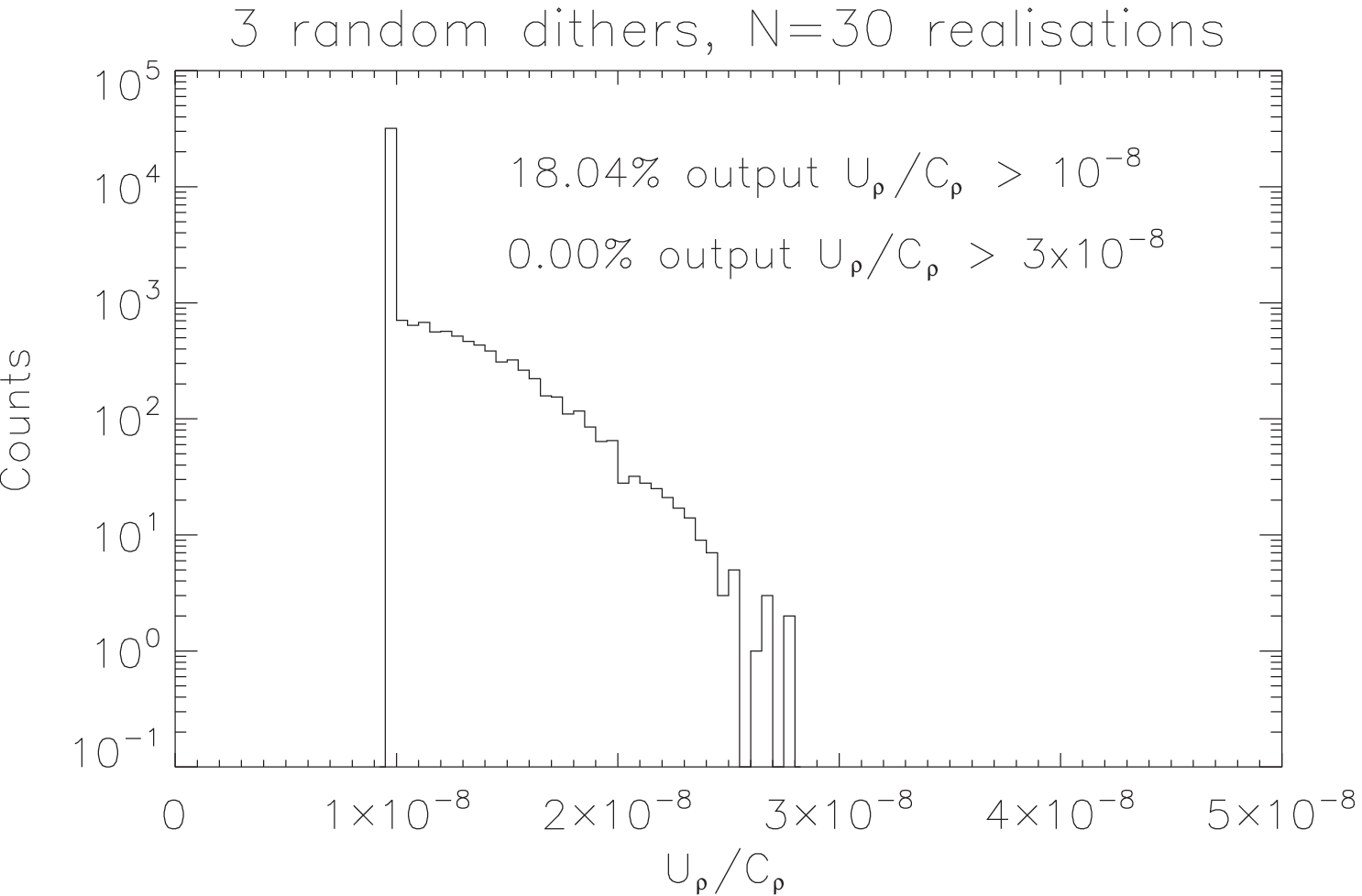}
\includegraphics[width=0.8\columnwidth]{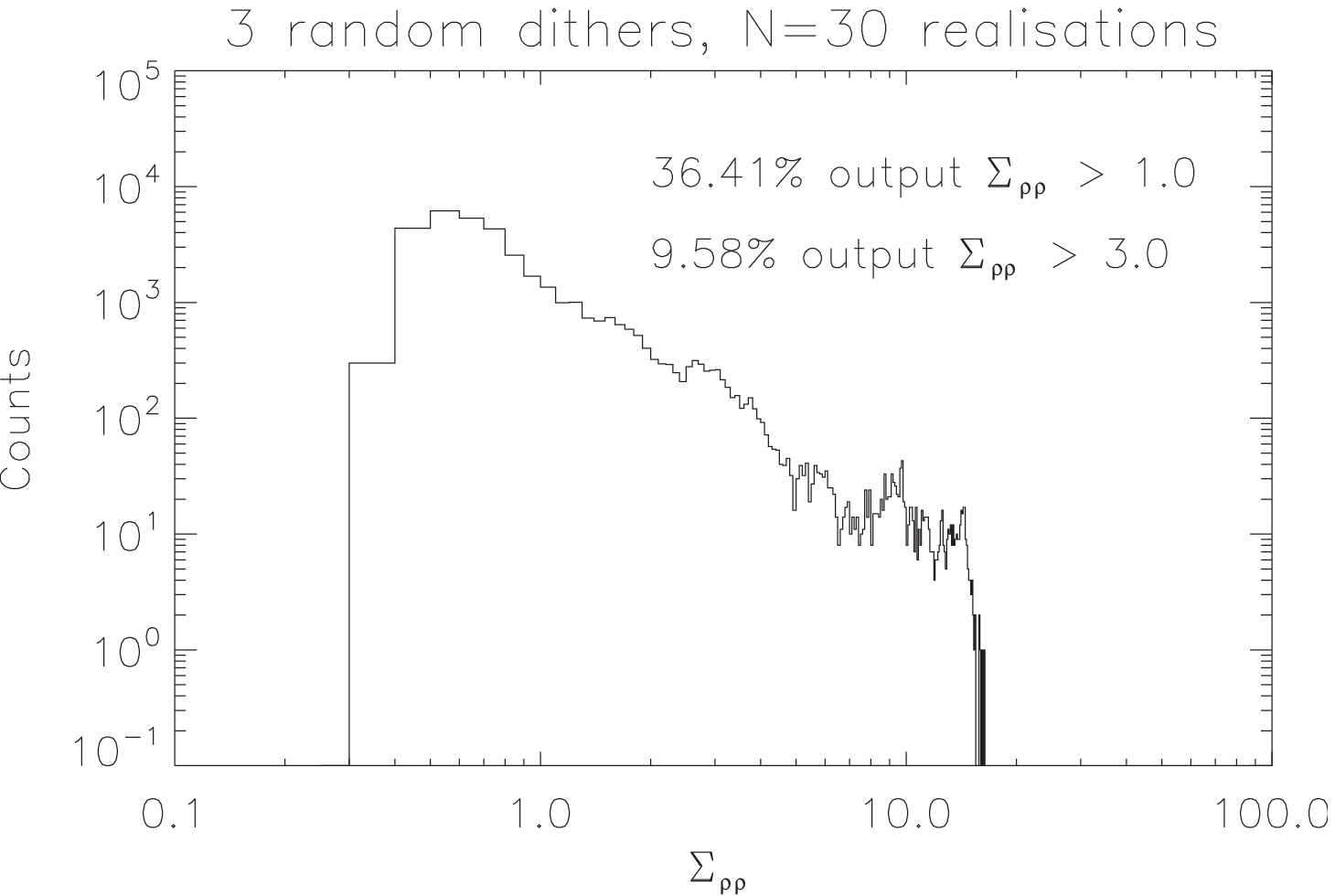} 
\caption{Histograms of normalized leakage objective $U_{\rho} / C_{\rho}$ (left panel) and noise variance $\Sigma_{\rho\rho}$ (in units of input variance; right panel) for a total of 38880 output pixels from 30 realizations of the three input exposure scenario.}
\label{fig:UShists}
\end{figure}

In Figure \ref{fig:UShists} we plot histograms showing the distribution of $U_{\rho} / C_{\rho}$ and $\Sigma_{\rho \rho}$ output pixel values for the full ensemble of 30 realizations of the three exposure random dither pattern.  As each output region of the type shown in Figure \ref{fig:4-3exp} consists of $36 \times 36$ output pixels, there are 38880 total output pixel locations making up the full sample for each of these histograms. We also provide some statistics of the distributions, showing that in the three exposure case, only $\simeq 18\%$ of the reconstructed output had $U_{\rho} / C_{\rho} > 10^{-8}$, but nowhere did this quantity exceed $3 \times 10^{-8}$. Nearly two thirds of the output pixels have a noise variance smaller than the noise variance on input pixels, and for less than $10\%$ of the output is the variance greater than a factor of three times the input. 

Because both $U_{\rho} / C_{\rho}$ and $\Sigma_{\rho \rho}$ are squared metrics of the quality of reconstruction (Rowe \etal\ 2011), these results suggest that while the three exposure case does not meet stated requirements in this 0.688 Nyquist-sampled case, it comes close.  Figure~\ref{fig:4-3exp} shows that this failure to meet the tolerance is spread regularly over the survey regions.  It is not clear to what extent this regularity, and the failure to meet $U_{\rho} / C_{\rho}$ will effect weak lensing measurements for this 800nm PSF.  We also note that for shorter wavelengths within the bandpass the Nyquist frequency is correspondingly increased, exacerbating the situation. On the other hand, the analysis in Section~\ref{sec:non-conv} (where the three exposure case including CTI is propagated into the shear power spectrum) indicates that such a variation has a limited impact on the Dark Energy FoM. 

\subsubsection{Construction of the PSF}
\label{sec:stability}

\begin{figure}
\includegraphics[width=\columnwidth]{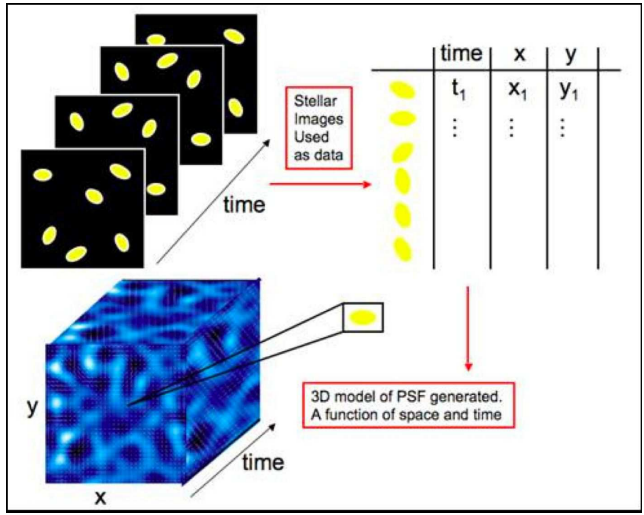}
\caption{The  PSF is expected to vary as a function of spatial position in the field of view and with time, resulting from changes in the instrument state. The instrument state may be that, for example, characterised by the primary-secondary mirror separation, or the temperature difference between certain optical elements. These individual contributions may substituted for the `time' column. An additional parameter will be the effective wavelength of the light producing the PSF.}
\label{fig:PSFcube}
\end{figure}

Because the PSF will be derived from the stars in the field of view surrounding each galaxy, a fundamental limit on the fidelity of this model is set by the photon statistical error in each pixel containing the  PSF. If there are sufficient bright stars in the field, then the form of the PSFs in each exposure can be modelled to the necessary level of fidelity,  exposure by exposure: in their analysis, Paulin-Henriksson \etal (2008, 2009) found that  $\sim50$ stellar PSFs at a signal-to-noise ratio of 500 were sufficient to determine the PSF to the accuracy required for the particular PSF that they were using. The surface density of stars which both are not nearly saturated ($i \ga 18.3$) and which would have signal-to-noise ratio greater than 500 is expected to be about $950$ deg$^{-2}$ in a survey such as \eu, for fields with Galactic latitude $b \sim 30^{\circ}$ near the North Ecliptic Pole.  If  the system is stable between slews, then measurements of stars from successive fields may  be combined appropriately to improve the PSF model. This approach is shown in Figure~\ref{fig:PSFcube}. However, this stability is not strictly necessary, if the modelling can take into account the variation of the PSF with time, capturing all of the possible states of the system. The PSF can then be reconstructed for a particular position on the field of view, and for a particular instrument state. 

The instrument state is defined through a (large) number of parameters, for example mirror separations and alignments in the optical train. Hence  the ``time'' dimension in Figure~\ref{fig:PSFcube} can be replaced with a separate dimension for each parameter. Some of these (such as the primary--secondary mirror separation in the telescope, leading to focus changes, examined by Ma \etal 2008) will, however, be dominant, and not all physical changes will induce ellipticity, so in practice the additional dimensionality should be constrained. The additional dimensionality beyond the 2 dimensions of the focal plane reduces the accuracy with which the PSF can be constructed, but with typically $>10^{9}$ suitable stars in a long survey duration, even a large number of additional dimensions can be accommodated. Then in principle {\it all} of the exposures in the survey can be used, and a multitude of PSFs will be available to construct the PSF for any galaxy.

\subsubsection{Principal Component Analysis of the PSF}

Any PSF can be modelled through a combination of functional forms. Which functional forms are optimal will depend on the criteria by which this is assessed. One simple criterion may be that each of the components making up the PSF should be orthogonal; another may be that a minimal set should be used, requiring that the series of components should converge rapidly.  

\begin{figure*}
\includegraphics[width=1.8\columnwidth]{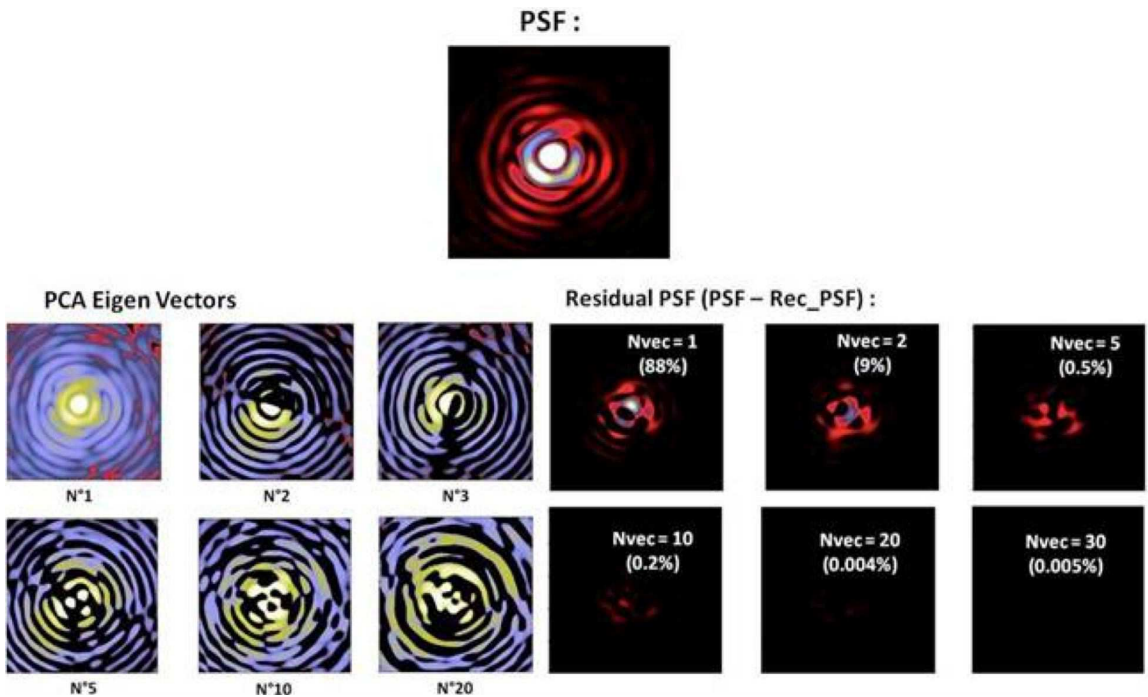}
\caption{ Components (left) and residuals (right) depending on number of components used for modelling  a  PSF (top). The intensity scales are logarithmic, with a colour table which enhances lower image levels.}
\label{fig:PCA_comp}
\end{figure*}

Principal Components Analysis (PCA) is a general statistical method that enables variation in data to be identified in a way that makes minimal assumptions about the nature of the underlying variation. More formally, the PCA methodology is a mathematical procedure that uses orthogonal transformations to convert a set of correlated variables into a set of uncorrelated variables called principal components. PCA also determines the coefficients which describe how much of each component should be used. 

PCA makes the assumption that modes of variation are additive.  This may be  restrictive when changes in PSF result from, for example, focus variation, so other, more physically-described models, such as those directly coupled to the optical modes {\it e.g.} Schechter \& Sobel Levinson, 2011) may be more efficient. 

In applying PCA we may consider the input data to be the PSFs provided by stars, and the input (correlated) variables to be the position of the PSF in the field of view, the spectral energy distribution (SED) of the photons in the bandpass and the parameters describing the instrument state (such as the focus). Each  component of the PCA basis set derived from the PSF is an image, and the components together generate an orthogonal set of 2-dimensional images. This is illustrated in Figure~\ref{fig:PCA_comp}. As there is a coefficient for each component to instruct how much of that coefficient should be used in the construction of a PSF, the coefficients are vector functions, with length corresponding to the numbers of PCA components. The dependencies in the derived component functions are the positions in the focal plane, the SED and the instrument state.

PCA PSF reconstruction has been successfully implemented on space-based weak lensing data from the {\it Hubble Space Telescope} (Jee \etal 2007) (see also Rhodes \etal 2007, Schrabback \etal 2007, 2010 who used PCA to characterise the variation of the two-component ellipticity, rather than the PSF pixel values). We could use the stellar (noisy, pixellised) images themselves  to generate the PCA components and coefficients. As described in Section~\ref{sec:pix}, the problem with undersampled data is that Fourier modes above the Nyquist sampling limit are not only lost, but are aliased to lower frequencies, resulting in corruption of all Fourier modes. 
The apparent shape of the PSF depends on the sub-pixel location with respect to the detector pixel grid, and no linear interpolation scheme can allow us to predict the PSF at one location, given its form at another. The effect of undersampling is to corrupt the PCA component calculation to make them no longer orthogonal, and the presence of noise results in spuriously high coefficients, particularly at higher-order eigenmodes.  
While this may be mitigated by the use of multiple, dithered exposures, 
in the presence of noise, it may not be possible to 
make a unique, method-independent reconstruction of a fully-sampled PSF.

\begin{figure}
\includegraphics[width=\columnwidth]{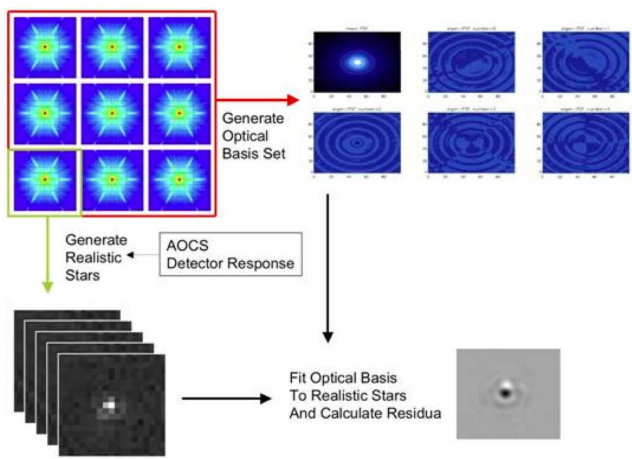}
\caption{Procedure for using simulated PSFs to generate the basis set of 2-dimensional images which are fitted to the simulated image data. In our simulations, one of the optical PSFs not used for the basis set construction is used to generate 144 realisations of realistic stars, including realistic levels of noise, AOCS and detector response. }
\label{fig:PCA_mod}
\end{figure}

One approach to estimating the high-frequency modes, beyond the
  sampling limit of an individual observation, is to create a
  super-resolution model of the PSF, fitted to the data.  We
  illustrate this approach here by creating basis-set models based on simulated
  super-sampled PSFs as a function of focal plane position and SED for
  each state of the instrument. PSF variation is created using a Monte Carlo
  approach to vary the instrument's optical characteristics. The eigenmodes of those
  PSF variations are found, and the
  coefficients for this basis set are calculated by fitting to simulated observations. 
  This process is illustrated schematically in Figure~\ref{fig:PCA_mod}.  In this
way we can test the accuracy to which a super-resolution PSF may be reconstructed
provided we have accurate PSF models: such a test investigates the information limit
of the data, but does not probe our ability to generate accurate models.

In Miller \etal 2013, the fitting procedure is treated as a Bayesian
estimation problem,
in which we use our prior information about the
statistical distribution of the eigenmodes from the simulations, 
together with measurements of the likelihood of the models
fitted to the data, to obtain the statistically most-likely PSF
reconstructions. Such a procedure
has significant advantages: it makes full use of the available information
about the system; it places the problem in a rigorous statistical
framework; being a forward-modelling process, we may include all
effects that we believe are present in the real system; and the Bayesian
approach prevents overfitting of noise. In the case that the model PSFs are too far from the actual ones in orbit, we expect that the model basis set
may be updated in-orbit as more information becomes available from star measurements.

\subsubsection{Characteristics of the Basis Set}

Before we perform the Bayesian fitting to  the noisy pixellated data, we first examine the characteristics of the 
eigenmode basis set, and investigate how many components may be required to adequately model the PSF. 
In detail, the results of this analysis will depend on the nature of the optical system, the pointing performance, and the detector characteristics, but the procedure would be similar for any realistic system.

To test this we first generate end-to-end simulations as described in Section~\ref{sec:sims} above, using the \eu case. In summary, we first determined the optical system behaviour from simulations of the optics, with variations imposed on the optical system over agreed ranges, and then with the simulated system being perturbed as expected in orbit by convolving the optics PSF with the charge spread within the detector and with a kernel arising from guiding errors. The PSFs are oversampled by a factor 12 compared with the \eu visible detector sampling.
There are no noise sources in this test.

\begin{figure*}
\includegraphics[width=0.9\columnwidth]{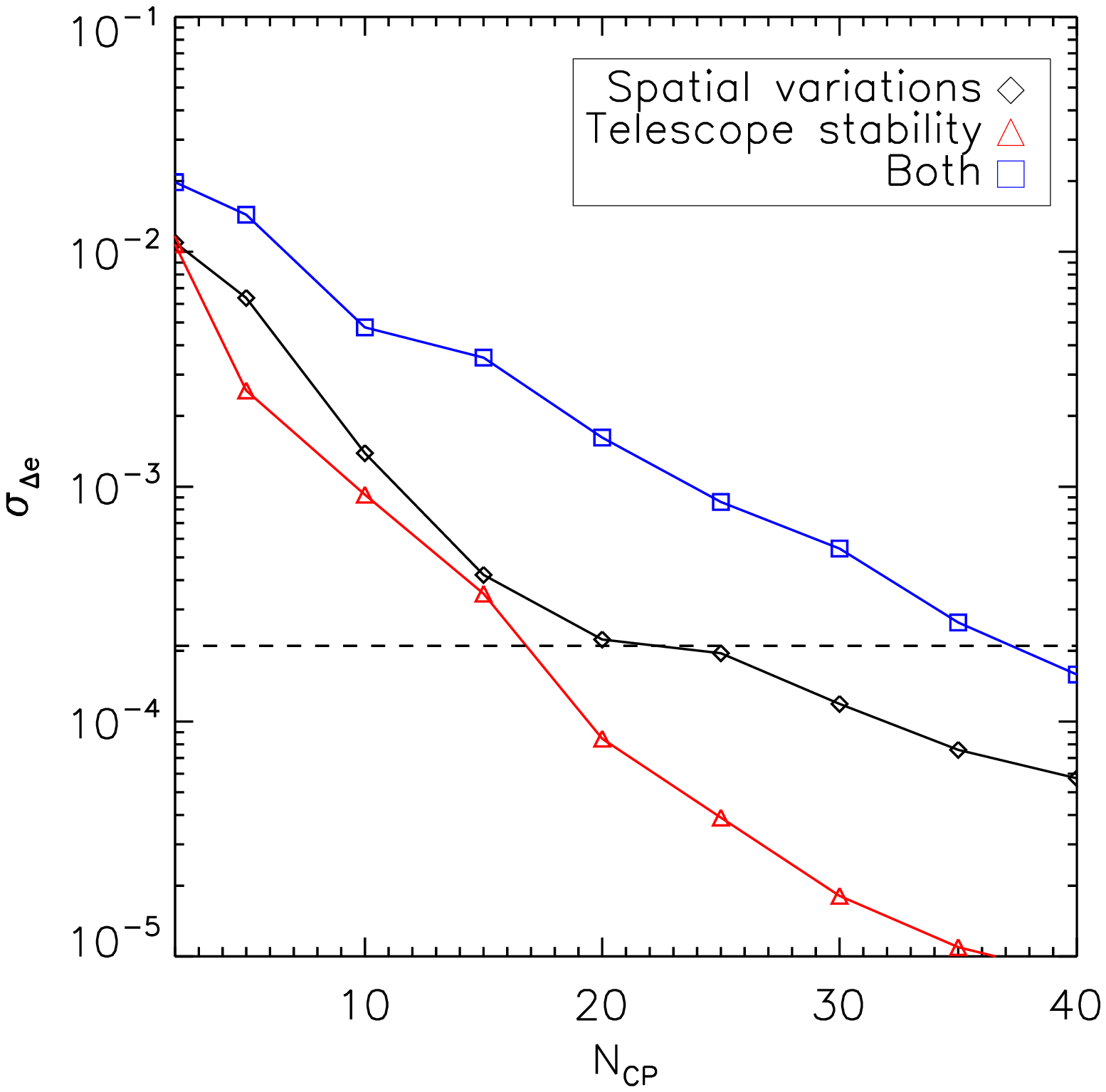}
\includegraphics[width=0.9\columnwidth]{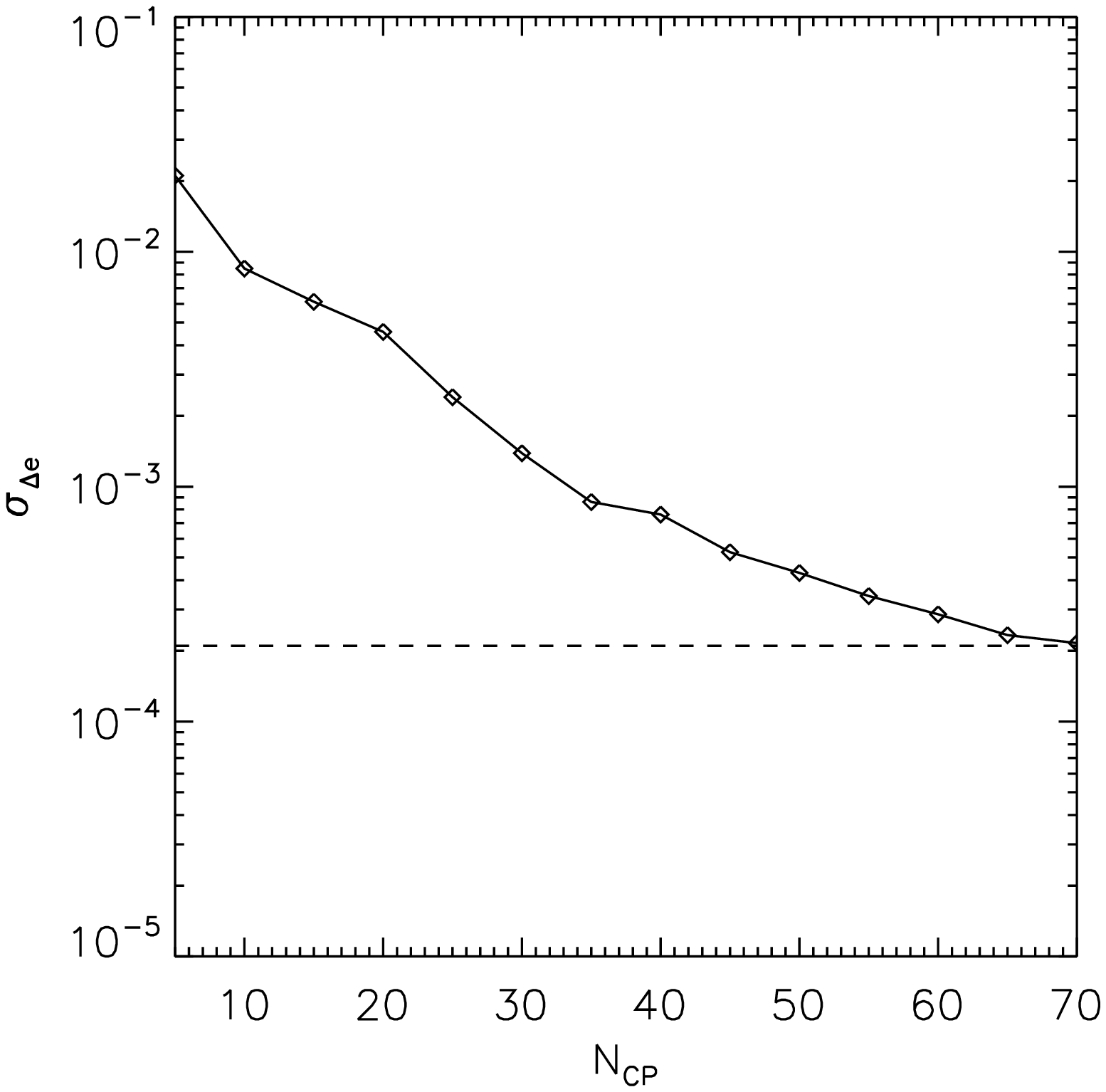}
\vspace*{-3mm}
\caption{
(Left) The difference between the known ellipticity and the modelled one as a function of the number of principal components used for the modelling. 
Diamonds represent the standard deviation of the residual error considering the spatial variations of the PSF over the field of view,
triangles represent the standard deviation of the residual error considering the variations of the PSF arising from variations in the instrumental state, and
squares represent the standard deviation of the residual error considering both together. (Right) As, for the left hand panel, but with the diamonds representing the standard deviation of the residual error considering simultaneously the wavelength dependence together with all of the variations of the PSF over the field of view and the stability of the telescope. The dashed line indicates the value assigned in Table~\ref{tab:aln}.
}
\label{fig:n_PCA}
\end{figure*}

After this we measure the ellipticity of the PSF using Equation~\ref{eq:quad} with a wide Gaussian weighting function with $\sigma(w)=4$ times the FWHM.
We first examine the required number of components needed to model the spatial variations of the PSF accurately over the full field-of-view for a monochromatic PSF at 800nm (Figure~\ref{fig:n_PCA}). For the \eu case, we  find that $\sim18$ components are enough to describe the PSF spatial variations with sufficient accuracy. This will be similar for other systems in practice. We now examine the number of components to encompass  the variations in the opto-mechanical system post-launch and in the space environment, for a single field point (derived from a Monte Carlo tolerancing analysis) and again find that $\sim20$ components are adequate. If we combine both, the number of components required to model correctly both spatial variations of the PSF and variations corresponding to the instrument state rises to $\sim38$.

\subsubsection{PSF Wavelength Dependence}

So far the PSF we have been modelling is monochromatic. In reality, the PSF will be different, depending on the spectrum of the star generating that PSF multiplied by the instrument end-to-end throughput as a function of wavelength. The largest contributor to this effect is the diffraction in the optical system, which increases linearly with wavelength. This is generally counteracted slightly by the inverse wavelength dependence of the charge spread in the CCDs because photons of redder wavelengths travel deeper into the pixel and closer to the electrode structure before they are absorbed. The other contributions (the attitude control system pointing variation and the radiation damage effects) do not have any wavelength dependence.

\blue{
We have examined the number of principal components that will be required to model a multi-wavelength PSF. 
In a new analysis, we have added the wavelength dependence effect by considering a set of monochromatic PSFs at 550 nm and 800nm.
Because the size of the core of the PSF changes approximately linearly with the wavelength, this affects significantly the ability of the eigenmodes to represent the wavelength dependence.
To reduce the number of components, a spatial rescaling by a factor 800/550 of the PSFs at 550nm has been applied. With this simple measure the number of components required to model the PSF correctly including the wavelength dependence effect, the spatial variations and the telescope stability, rises to $\sim70$. This result is shown in the right panel of  Figure~\ref{fig:n_PCA}.
}

This analysis suggests that the eigenmode approach indeed enables the full range of PSF to be modelled in a representative fashion, albeit at the price of potentially needing a large number of eigenmodes in the analysis.  However, this analysis does not take into account the relative importance of the modes at 550\,nm and 800\,nm in actual data: 
for realistic spectra, the long-wavelength parts of the spectrum dominate the PSF, and thus the actual modes needed
in practice may be fewer than would be implied by Figure~\ref{fig:n_PCA}.

\subsubsection{Bayesian Model-fitting}
\label{sec:bayes}

Having explored the approximate number of eigenmodes that may be required to construct the PSF, we now examine whether the Bayesian approach discussed above can provide sufficient information on the coefficients of this component set for reconstructing the PSF in realistic simulations, to meet the allocations in Table~\ref{tab:aln}.  For this purpose a conservative assumption is to limit the amount of temporal stability required and hence to analyse each set of exposures of a region of sky independently of any other field. The aim is to investigate the extent to which the underlying, fully-sampled PSF may be reconstructed from noisy data in a single field. We do, however, assume
that each field is observed with three dithered exposures, and that the PSF is invariant during those dithered
exposures.

The first step of this reconstruction is to define the set of basis model components that characterize the system using normal mode decomposition as described in earlier sections. We then fit these components to noisy realisations of stars. The star profiles are taken in turn from the set of model PSFs, but excluding that profile from the determination of the components of the models above. As before, the PSF used in this analysis is the \eu system PSF taking into account the opto-mechanical, detector and attitude control system pointing variation contributions. Provided the information on the pointing variation is telemetered by the spacecraft, the effect of this uncertainty on the PSF may be corrected, to a certain level of accuracy. On the other hand, we could proceed without this information, and then the guiding errors would need to be included as additional fit parameters. 
For this test, it is assumed that the CTI has been fully corrected in prior data processing 
({\it e.g.} Massey \etal 2010): the efficacy of this is described in Section~\ref{sec:non-conv}. 

The simulation uses the Besan\c{c}on  model of the Milky Way (Robin \etal 2003) to predict the number-magnitude relation of stars at the North Ecliptic Pole in the Canada-France-Hawaii Telescope system $i$ band.\footnote{\url{http://model.obs-besancon.fr/}}
There are 3.5 stars arcmin$^{-2}$ in the range $18 < i < 23$, and 6300 stars in a half square degree,
corresponding to the \eu full field of view.  The Besan\c{c}on model also allows the creation of a simulated catalogue of stars with
magnitude and spectral type, and to create the simulated \eu observations, stars were randomly selected from that
catalogue.
As the stellar PSFs used in the PSF modelling are all moderate or high signal-to-noise ratio, their colours will often be known from catalogs, such as that which ESA's {\it Gaia} mission will produce. Here, we assume that their optical and
near-infrared magnitudes can be measured from the \eu mission data alone. To model the PSF in the presence of
the varying colours of stars, a simple model was assumed for the PSF wavelength dependence, in which the optics
component alone was assumed to scale in angular scale linearly with wavelength, with respect to simulated
PSFs calculated for wavelength 800\,nm. While this model is an oversimplification of the true wavelength
dependence, it serves to capture the basic effect and allows us to test whether, in principle, the PSF could be reconstructed at the required level of accuracy. Simulated stars were created by
dividing the \eu visible instrument passband into small wavelength intervals, evaluating the expected number of detected photoelectrons
in each wavelength interval, given the SED of each simulated star, and coadding the wavelength-stretched PSFs across the bandpass with wavelength-dependent weight given by that number of photoelectrons. Star SEDs were obtain from the ``UVK'' library of Pickles (1998).\footnote{
\url{http://www.ifa.hawaii.edu/users/pickles/AJP/hilib.html}} 
In the measurement/fitting test, stars were assumed to have noisy photometric
measurements, from which an estimated SED was evaluated using the same stellar library ({\it i.e.} assuming that star SEDs 
may be obtained from broad-band photometry without systematic error), and PSF eigenmodes were adapted to the
SED of each star using the same SED-weighting procedure that was used to create the simulated observations.  The model PSFs and
simulated stars are not expected to match exactly because of the introduction of photon shot noise in the simulated
stars.
It was further assumed that 30\,percent of stars might not be measureable owing to the effects of confusion with
faint galaxies and image artefacts.

In this test some account was taken also of the variation in PSF across  the \eu field.  The field was divided
into five zones of equal area, and PSF models were calculated at five locations (the four corners and the centre of
the field).  When the PSF eigenmodes were created, pixels were included from all five PSF models, so that the normal mode
analysis generated position-dependent modes, albeit sampled only at five locations.  The PSF was assumed to be
spatially invariant within each zone: a more advanced method should allow interpolation as a function of position
in the field.

Information from three dithered simulated exposures was used by jointly fitting the PSF models to 
all three exposures.  In this test, the absolute positions of stars were assumed to be unknown, and were marginalised
over in the fitting, but the relative positions of stars on each of the three exposures was assumed to be
fixed and known.  In practice, the relative registration of multiple exposures should be determinable to
very high accuracy from joint analysis of all the stars in the field. By assuming the star positions to be unknown,
we are discarding potentially useful information on the field distortion, which also provides information on
the optical path.  In practice, accurate absolute star positions may be available from the {\it Gaia} mission data, which could
also be included in the analysis.

\begin{figure*}
\includegraphics[width=0.8\columnwidth]{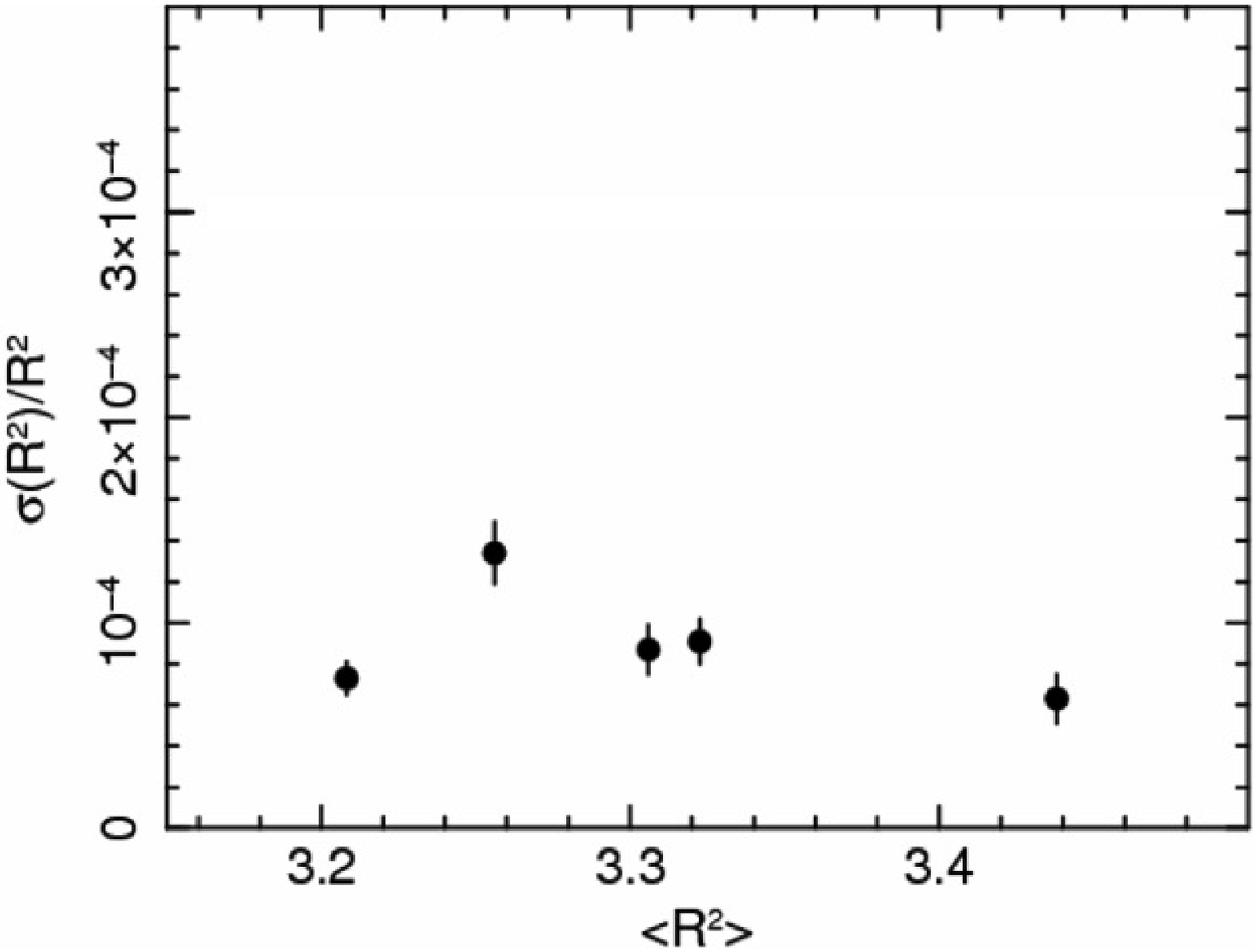}
\hspace*{7mm}
\includegraphics[width=0.82\columnwidth]{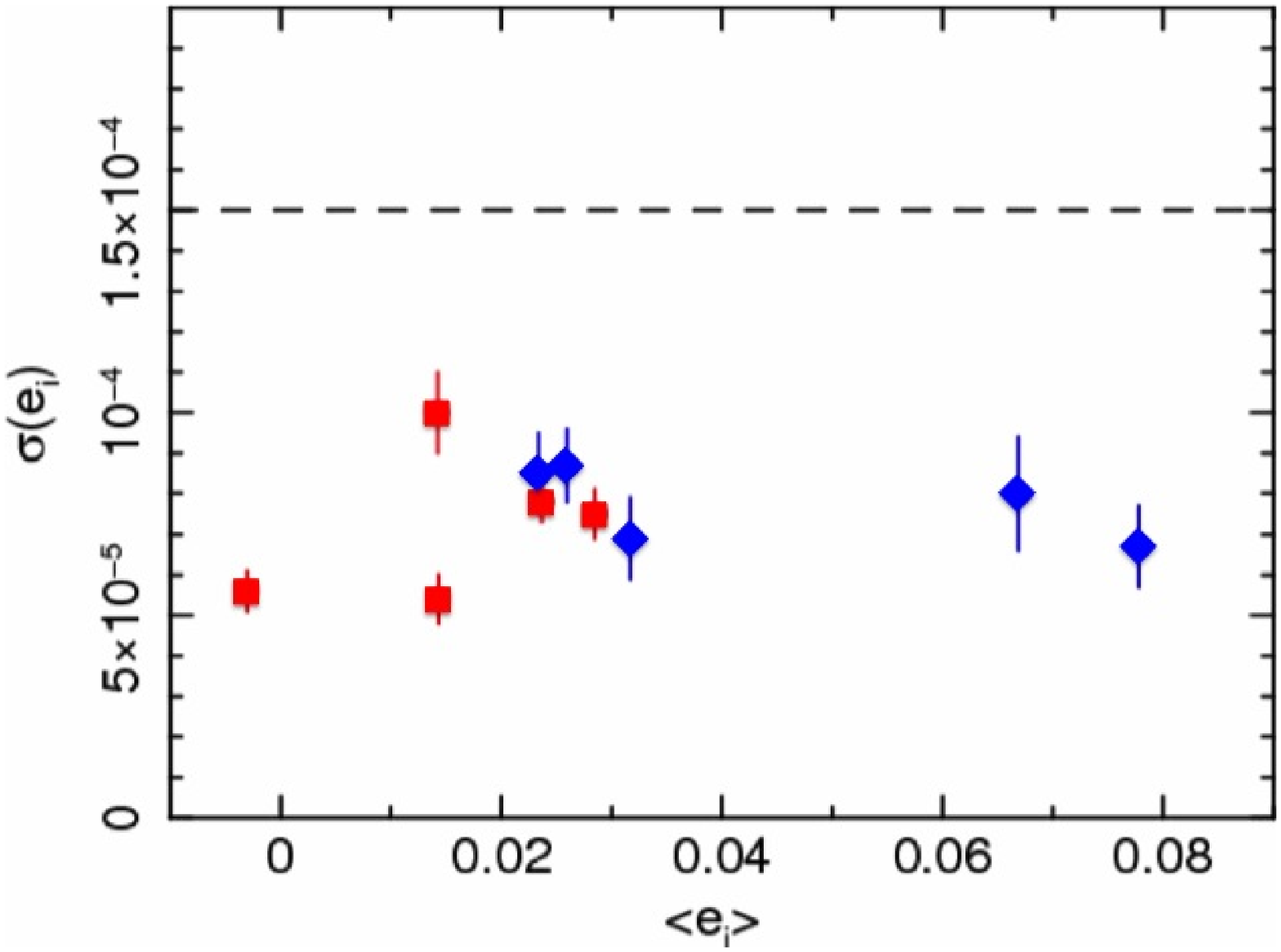}
\caption{Variation in the residuals in the knowledge of the PSF in terms of size $\sigma[\rp^2]$ (left panel) and  a single component of the convolutive ellipticity $\sigma^2[\ellipticity_{i_{\rm C}}]$ (right panel) as a function of a range in PSF size and ellipticity, derived using the normal mode modelling of 4400 stars. The square and diamond symbols correspond to the two ellipticity components. The dotted line in the right panel is the level in Table~\ref{tab:aln} below which the ellipticity requirements for a {\it single} ellipticity component allocated to this contribution is met. The equivalent level for the size in the left panel is above the top of the plot, indicating that the knowledge uncertainties for this contribution are easily met.}
\label{fig:Lance_sigma}
\end{figure*}

Further description and evaluation of the above procedure will be provided by Miller \etal (in preparation). 
In this initial evaluation,
forty modes were found to capture the PSF variation at an adequate level. This number is fewer than expected from
Figure~\ref{fig:n_PCA} because of the weighting of the contributions from differing wavelengths by the SED photon counts. In line with the formalism established above, the PSF reconstruction is evaluated by the statistics $\sigma[\rp^2]/\rp^2$ and $\sigma^2[\bmath{\ellipticity}_{_{\rm C}}]$ of the differences between the input PSF image and the reconstructed PSF, both quantities being measured from the image-weighted second moments. 
The results are shown in Figure~\ref{fig:Lance_sigma}. In this preliminary evaluation, a systematic offset was
found in the value of ellipticity in some parts of the field, which does not appear in the rms statistics
shown in Figure~\ref{fig:Lance_sigma}. While, in a full PSF modelling system, such a systematic would need to be
eliminated, the exercise presented here nonetheless shows that, {\em in principle}, sufficient information exists
in simulated observations of realistic stars fields to allow accurate PSF reconstruction to the levels assigned in Table~\ref{tab:aln}.  
For the full-field simulation of 4400 used stars, the uncertainty $\sigma[\rp^2]/\rp^2$ in size is  $<1.5\times10^{-4}$ and that in ellipticity, $\sigma^2[|\ellipticity_{i_{\rm C}}|]<1.0\times10^{-4}$ per ellipticity component. The allocations aggregated from several contributions  in Table~\ref{tab:aln} are $\sigma[\rp^2]/\rp^2<4.8\times10^{-4}$ and $\sigma^2[|\ellipticity_{i_{\rm C}}|]<1.5\times10^{-4}$ per ellipticity component.

The results presented here show that it is possible in principle to reconstruct the  PSF to sufficient accuracy to meet the science requirements set by Equation~\ref{eqn:AMlim} and organised in Table~\ref{tab:aln} by normal mode model-fitting to observations of stars in single fields with a small number (three) of dithered exposures, even if no longer-timescale temporal information is used. The requirements and values in Table~\ref{tab:aln} will be different for different experiments, so this will need to be evaluated on a case-by-case basis. Should additional margin be required, the temporal information could be exploited.

\subsubsection{Flat Fields}

Because of manufacturing tolerances, all CCDs are subject to slight variations in their pixel-to-pixel sensitivity. This is called photo-response non-uniformity (PRNU). This is at least partially caused by differences in pixel size in the photolithographic mask sets used to manufacture the CCD, but the PRNU can also show colour dependence, which indicates that other effects also contribute. 

At low signal levels the pixel-pixel variations on typical CCD exposures are dominated by the readout noise. At higher signal levels the Poisson noise, increasing as $\sqrt{N}$ (where $N$ is the number of counts in the pixel) dominates. At even higher levels the PRNU, increasing as $N$, starts to dominate, even though the intrinsic PRNU is typically only ~2\%. Generally, stars (which fall on several CCD pixels) will be used to calibrate the PSF  and many of these stars will be close to the saturation level,  where PRNU dominates. 

Typically, an on-board flat field calibration source is used to provide an even illumination over the CCDs in order to measure this PRNU. By accumulating several flat field exposures, a high signal-to-noise ratio normalised PRNU map can be accumulated. Dividing science exposures by this map -- flat-fielding -- can almost eliminate the PRNU degradation. It is essential, however, to achieve accuracies in the accumulated flat-field which are sufficient, or else the flat-fielding can instead add noise to the image. An approximate way to minimise the effect of the flat field calibration on $\mcap$ and $\mcmp$ is to ensure it is smaller by a factor $\zeta$ compared to the Poission noise, the effect of which which has been calculated by Paulin-Henriksson \etal (2008). If a number $n_{\rm PSF}$ bright stellar PSFs are used, and we assume the brightest pixels in these PSFs are filled to the same level as that provided by the flat-field illumination, then the number of flat field exposures $n_{\rm f} = \zeta \sqrt{n_{\rm PSF}}$
(here we have used the fact that the flat field under each stellar PSF is different). For $\zeta=3$, $n_{\rm PSF}=50$ (Paulin-Henriksson \etal 2008) then $n_{\rm f} \sim20$ flat field exposures are required to be combined in order to meet the levels in Table~\ref{tab:aln}. In practice the frequency of flat field exposures in order to achieve this number (with consequential operational overheads) will be driven by the timescale of the temporal changes in the PRNU, which is still unknown at this level.

\subsubsection{Pointing Accuracy Issues}

The imperfect operation of the satellite's attitude control system contributes to the PSF of an exposure because the telescope axis is not perfectly stable, and the combination of pitch, yaw and roll leads to field-dependent displacement of the images. In the analysis in Section~\ref{sec:bayes} we assumed that these pointing displacements are provided by the spacecraft. These will not be noise free, but  provided prior information for the normal mode analysis which was used in the forward modelling. 

Even in the absence of this information, Ma \etal (2008) have shown that when the pointing variation is much smaller than the width of the Airy disk of the optics PSF, its effect on the observed PSF is described by the mean displacement and the covariance matrix of the displacements. In this case the detailed pointing variation history is not important, and in principle only two stars are needed to describe this contribution anywhere in the field of view. In the presence of noise more stars are needed, but in the typical case as discussed above, all $~\sim 4000$  stars in the field of view can be used in the normal mode analysis to determine the pointing variation contribution for each exposure, requiring only a few additional components. If the pointing variation amplitudes are comparable or larger than the optics PSF, the mean and covariance of the displacements is not sufficient and the pointing history is required. 

\subsection{Residuals in the Correction for Radiation Damage}
\label{sec:non-conv}

\begin{figure}
\vspace*{-5mm}
\hspace*{-6mm}
\includegraphics[angle=90,width=1.14\columnwidth]{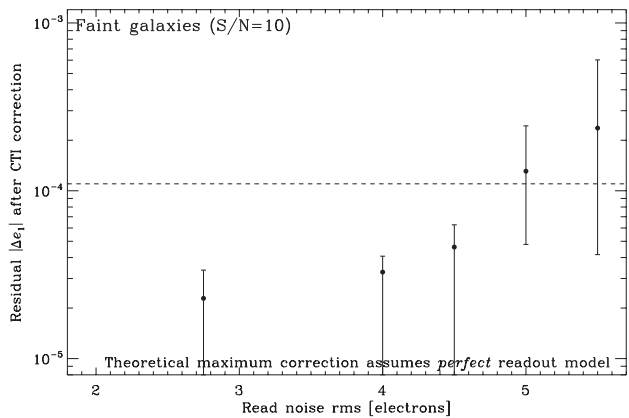}
\vspace*{-10mm}
\caption{The end-of-mission residual ellipticity $|\bmath{\ellipticity}_{NC}|$ in a single component induced by CTI
effects in the detector system after correction during image processing for the worse case configuration: faintest and smallest
galaxies located furthest from the readout node. The degree of correction is limited by readout noise, and the required level is met for a readout noise of $<4.8e^{-}$. The correction assumes a correct model of the CCD charge transfer from pixel-to-pixel during readout.  
}
\label{fig:CTI_correc}
\end{figure}

Having established  the major convolutive  effects, we now consider the extent to which the non-linear CTI effects caused by radiation damage can be corrected in the data processing described in Section~\ref{sec:CTI_correc}.

Figure~\ref{fig:CTI_correc} shows the residual ellipticity in the galaxy images after the image post-processing. This is plotted as a function of the readout noise of the detection chain on the abscissa, because  readout noise is an important limiting factor, as discussed in Section~\ref{sec:CTI_correc}. The actual value on the ordinate will depend on a number of parameters, for example the CCD characteristics, the fluence received by the CCD, background levels, signal-to-noise ratios {\it etc.} These will all be inputs to the simulations discussed in Section~\ref{sec:radsims}. Here we assume a five year mission in a deep orbit, four of which are at Solar maximum. At the end of mission, taking account of a nominal focal plane shielding and some margin, this accumulates to a fluence of $6\times10^9$ protons  cm$^{-2}$ (scaled to the effects of protons of energy 10 MeV) a typical value for such a mission. The example uses small (minimally sampled) galaxies with a signal-to-noise ratio of 10, located far from the readout node (requiring $\sim 2000$ transfers). A readout noise $<5$ electrons (which is generally reachable with careful design in a CCD detector matrix, as long as the readout speed is not too high) enables a residual ellipticity knowledge of  $\sigma[|\epsilon_{_{\rm NC}}|] < 10^{-4}$.  Because this analysis addresses the performance for the faintest galaxies in the worst position for CTI effects,  this knowledge error will be smaller if  the population ensemble of all galaxies that will be studied for the weak lensing is substituted and if they are placed randomly with respect to the readout node, and, further, if the average value of the radiation damage is used, rather than the end-of-mission level.

\begin{figure*}
\begin{center}
\includegraphics[width=.95\columnwidth]{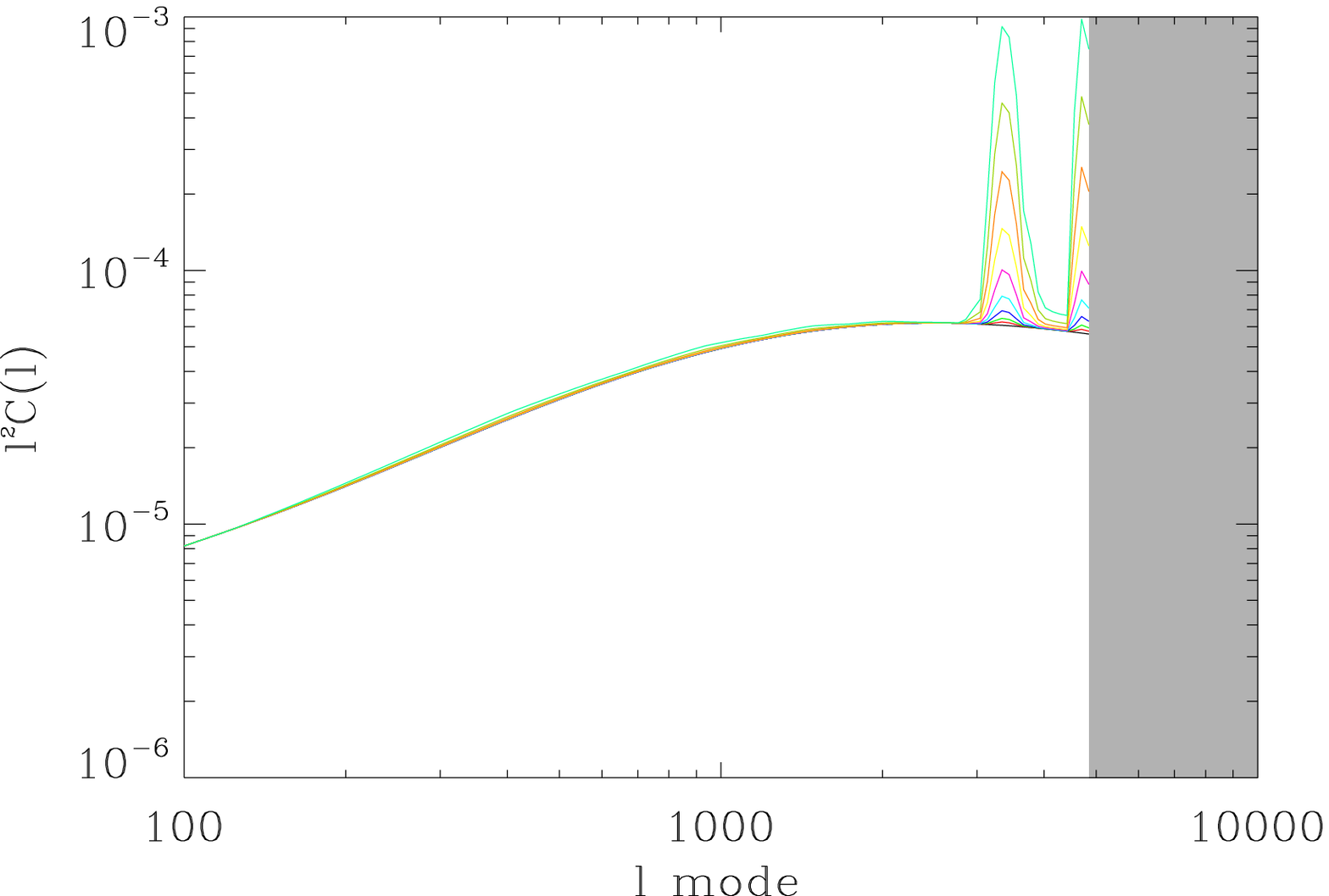}
\includegraphics[width=1.02\columnwidth]{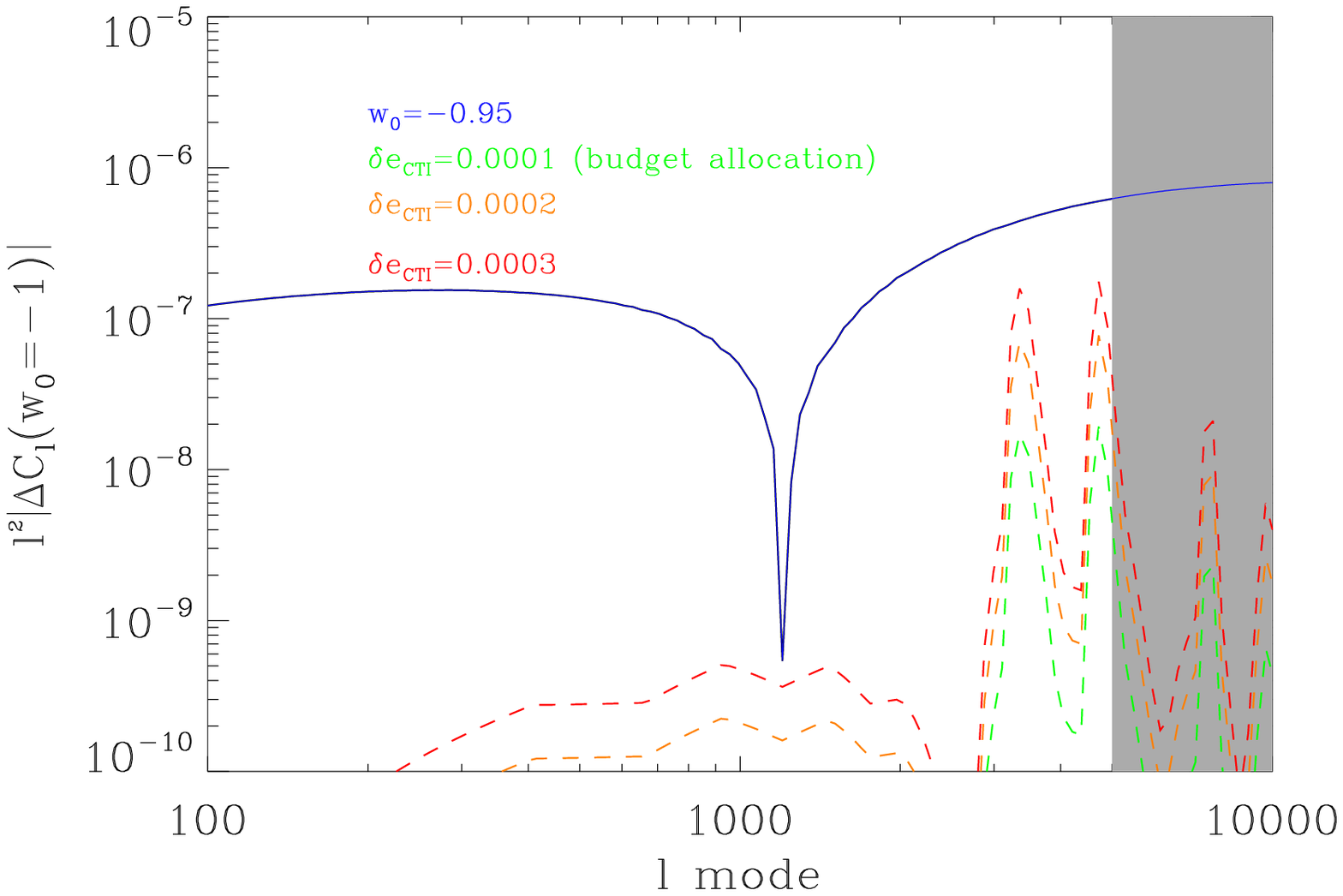}
\caption{
(left) The effect of imperfect CTI correction on the power spectrum $C_{ij}(\ell)$,  multiplied by $\ell^2$ to show equal power for different spatial scales $l$. Here we use the auto-correlation power spectrum $(i = j)$ and a mean redshift of the
tomographic bin of $z_i = 1$. We limit the power to $\ell<5000$ as described in Laureijs \etal (2011). The lines indicate progressive levels of CTI correction where $\delta e_{\rm
CTI}=x \delta_{\rm CTI,uncorrected}$ and we show 10 values of $x$
logarithmically spaced between $1$ and $100$. The impact is reduced as the correction is improved and the effect is limited to spatial scales of the CCD and smaller (corresponding to $l>2000$ in the \eu case).  (right) The absolute difference (the difference changes sign at $\ell \simeq1200$) in the $C(\ell)$ power spectrum
when the Dark Energy parameter is changed from $w_0=-1.0$ to
$w_0=-0.95$ (blue): this represents the cosmological sensitivity of
the $C(\ell)$. The dotted lines show the absolute difference beween an
unaffected $C(\ell)$, with no CTI and $w_0=-1.0$ and a power spectrum
affected by CTI with various levels of correction (dotted lines). 
}
\label{fig:CTI_spatial}
\end{center}
\end{figure*}

In this case, the CTI has been generated by a model in the simulations, and corrected using the same model: hence this conclusion is reached using a perfect CTI model. Currently, the best that has been achieved in practice is a factor 20 reduction by Massey \etal (2010). In addition to the readout noise, the efficacy of image-level CTI mitigation will therefore rely on the accuracy of the CTI model compared to the solid-state physics taking place in the real instrument, both in terms of the fidelity of the model, and the accuracy of the parameters used within it: these necessitate a substantial characterisation programme for the CCDs. The parameters include the trap density and the release times of each trap species. This is explored more fully in Massey \etal (in preparation). In orbit, the parameters could be determined from fits to injected charge lines (during calibration exposures), to cosmic ray events, by the use of pocket pumping and perhaps by direct analysis of stellar PSFs. The algorithms to do this, used at the data processing stage, require careful development.

While the requirements on the modelling and the determination of the parameters for it turn out to be challenging with respect to the allowed values of $\mathcal{A}'$ and $\mathcal{M}'$ in Equation~\ref{eqn:AMlim}, any residual shear effects in the detector coordinate system which are related to position with respect to the readout nodes (as opposed to those in the sky coordinate system) will be identified as inadequacies in the CTI model, and can potentially be iteratively nulled to negligible levels. The readout noise floor in Figure~\ref{fig:CTI_correc} is therefore an important parameter.  While any final small residual errors in the CTI correction will not be fully convolutive, in that they are magnitude and background dependent, the contributions to the error that {\it are} linear could be incorporated with the standard PSF modelling discussed above, and modelled out. 

\begin{figure}
\begin{center}
\vspace*{2mm}
\includegraphics[width=\columnwidth]{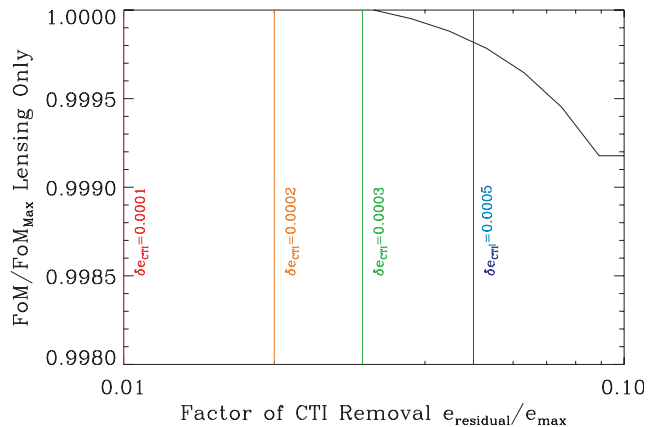}
\caption{
The reduction in figure of merit (FoM) with respect to perfect correction for the progressive levels of correction. The blue vertical line with $\delta\bmath{\ellipticity}_{\rm CTI}=0.0005$ indicates the impact on the FoM achieved with {\it HST} data. 
}
\label{fig:CTI_fom}
\end{center}
\end{figure}

Moreover, the CTI residuals will not contribute to errors in the derived power spectrum on all spatial scales, but will be limited to scales of the CCD and smaller (if the CCD is divided into sectors with separate readouts). We have quantified this by analysing the effect on the power spectrum, and on the FoM. This analysis includes the radiation damage effects assumed in the modelling above, including a fluence of $6\times10^9$ protons  cm$^{-2}$, but  with the galaxies (all again small and minimally sampled and with a signal-to-noise ratio of 10) now placed randomly with respect to the readout nodes. We use three slightly displaced exposures as in the \eu pattern of $\sim100$ arc sec (1000 pixels) in the parallel direction, one of which also has a 50 arc sec displacement (500 pixels) in the serial direction. The effect of a progressively improved correction on the power spectrum is shown in Figure~\ref{fig:CTI_spatial}.  
We also show in Figure~\ref{fig:CTI_fom} the change in the lensing-only FoM
using the same parameters as those used in MHK13 for the systematic
evaluations (and the same as those used in Laureijs \etal 2011).
At
all scales the difference between the corrected power and the
unaffected power is less that the difference in the power induced by a
change in the Dark Energy parameter $w_0=-1.0$ to $w_0=-0.95$ (start of Section~\ref{sec:biasquantification}).
This encouragingly indicates that a CTI-corrected power spectrum at this level will have limited effect on the Dark Energy measurements.

\subsection{Model Transfer-to-Object and Model Bias Knowledge Errors}
\label{sec:galaxy}

\subsubsection{Linearity}
\label{sec:lin}

The PSF model is inferred from the images of stars that are much
brighter than the faint galaxies used in the
weak lensing analysis. If the response of the detector is independent
of flux then the model can be applied directly. Real detectors, and their associated external electronics,
however, will have a non-linear response. This will lead to  systematic
errors resulting from changes in the shape of the PSF, which will either be more or less peaked than it should be. Note that this concern regarding the non-linearity is particular to the field of weak lensing measurements. The more usual concern with its effect on the overall photometric accuracy of the measurement may also be important, for example in determining the SED of the star being used for modelling the PSF.

There are two main effects which lead to the nonlinearity. The first is the classic effect of non-linearity and saturation in the detector and external electronics. This is typically limited to $\sim$ few percent. It can be addressed by calibrating the detector using multiple exposures of the same field with different exposure times. In practice this may place tight constraints on the repeatability of any shutter or readout mechanism. The main difficulty of these measurements is to obtain sufficient faint stars in order to calibrate the linearity at lower signal levels. After calibration, residual non-linearities from these effects can be constrained to extremely low levels, especially as they are expected to vary only slowly with time, depending on operating voltage levels within the electronics.

The second effect arises from the CTI in the detectors caused by radiation damage. Those traps with long release times remove photoelectrons from the PSF entirely. If the population density of traps encountered by the charge cloud as it is transferred through the CCD does not increase linearly with signal level, this effect can induce non-linearities in the relationship between the optical flux level and the charge arriving at the readout node of the CCD. The effect also depends on the level of background light, mostly Zodiacal, which has ameliorating effects particularly for these slow traps, and on the prior history of PSFs read out ahead of the PSF of interest. The effect can be corrected using the same model and technique used to correct for the faster traps (Section~\ref{sec:CTI_correc}), and again the residuals will be identified from their amplitude in the detector reference frames. This will  be a monotonically increasing effect as the mission progresses.

The residual of the linearity correction is included in different categories in Table~\ref{tab:aln}, mostly in the PSF modelling and in the transfer of model to object (because of the use of bright star PSFs to calibrate faint galaxy PSFs). These residuals are assumed to be quasi-linear. CTI-induced non-linearity can be incorporated within the non-convolutive category.

\subsubsection{Bandpass and Out-of-Band Transmission}

\begin{figure}
\includegraphics[width=\columnwidth]{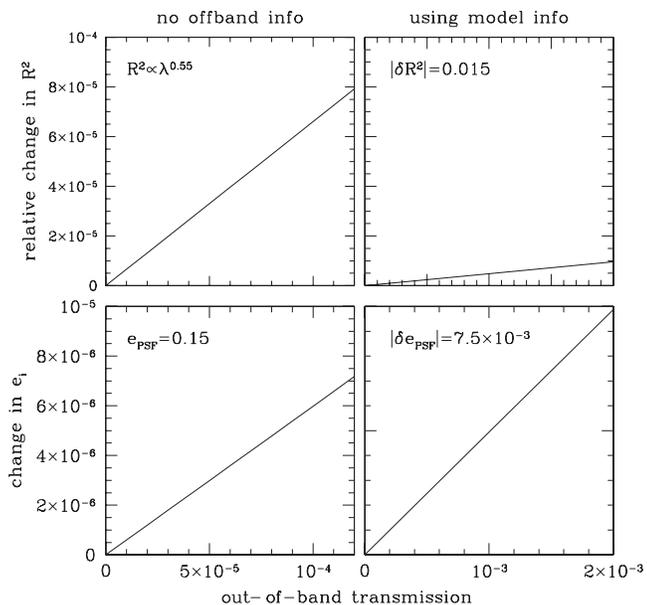}
\caption{
The relative change in $R^2_{_{\rm C}}$ (top panels) and $\epsilon_{_{\rm C}}$ (bottom panels) as a function of the level of out-of-band transmission for a flat-spectrum source for the   \eu  reference. The left panels show the impact without any off-band information, while the right panels show the impact if the off-band information is used to in the PSF modelling. }
\label{fig:outofband}
\end{figure}

The wavelength dependence of the PSF can be determined from the data
by comparing to stars that cover a range in colour for wavelengths
where the transmission is high. Outside the nominal band, where the
tranmission is low, no information can be recovered. 

The flux that is transmitted out-of-band contributes to the galaxy PSF, which is a concern. Ideally, in Equations~\ref{eq:Rlambda} and \ref{eq:elambda}, $T(\lambda)=1$ in-band and $T(\lambda)=0$ out-of-band. The transition has a finite width, and the level of out-of-band transmission  is $f_{\rm out}>0$. As a result the width of the transition region and the allowed (average) level of out-of-band transmission need to be determined. Under the assumption that the bias is small, the relative errors in the PSF size and shape can be determined by taking the ratio of the in- and out-of-band contributions to the integrals in Equations~\ref{eq:Rlambda} and~\ref{eq:elambda}, so that this is an upper limit to the impact of the out-of-band leakage.

\begin{figure*}
\center{
\includegraphics[width=0.9\columnwidth]{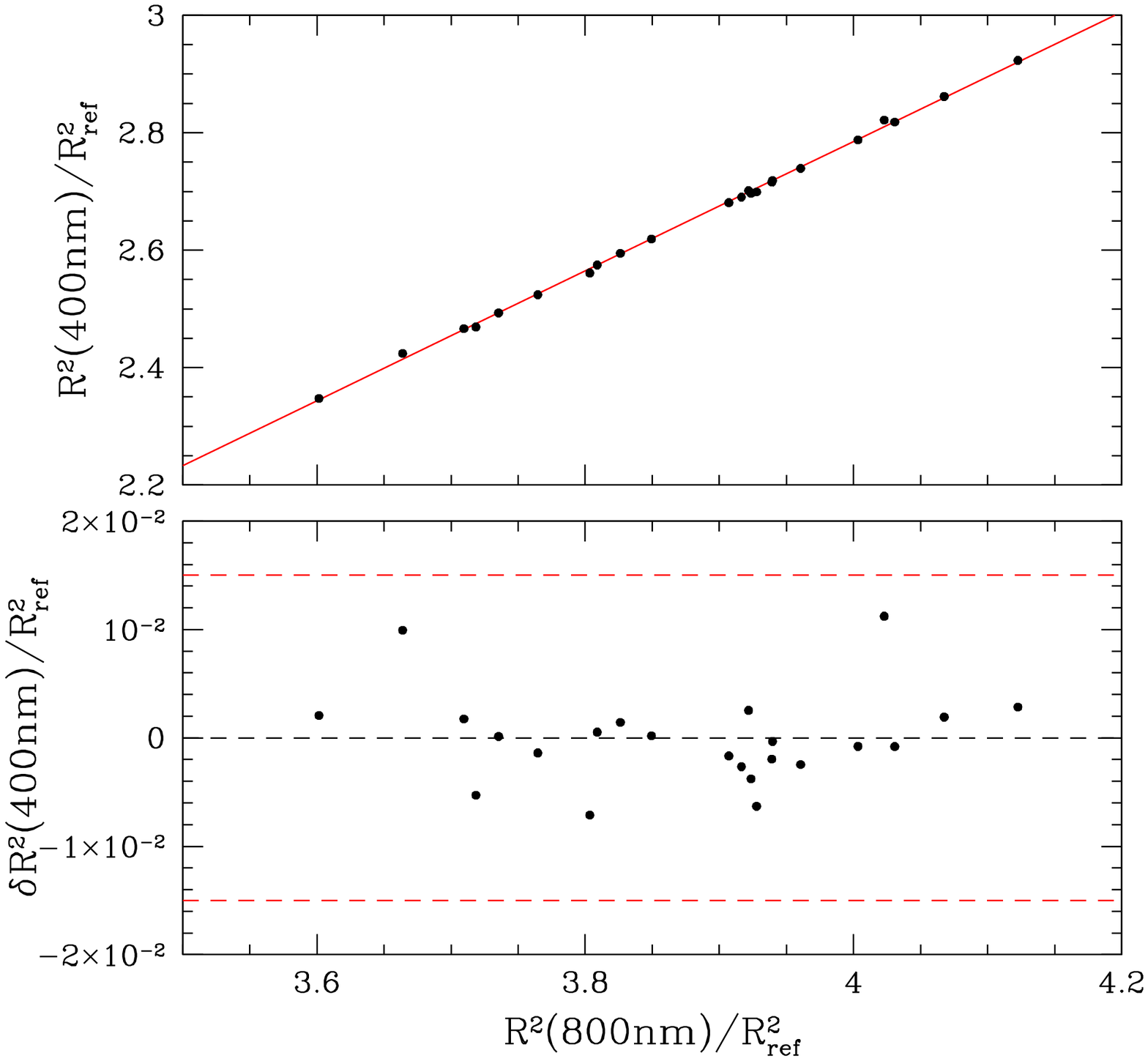}
\includegraphics[width=0.9\columnwidth]{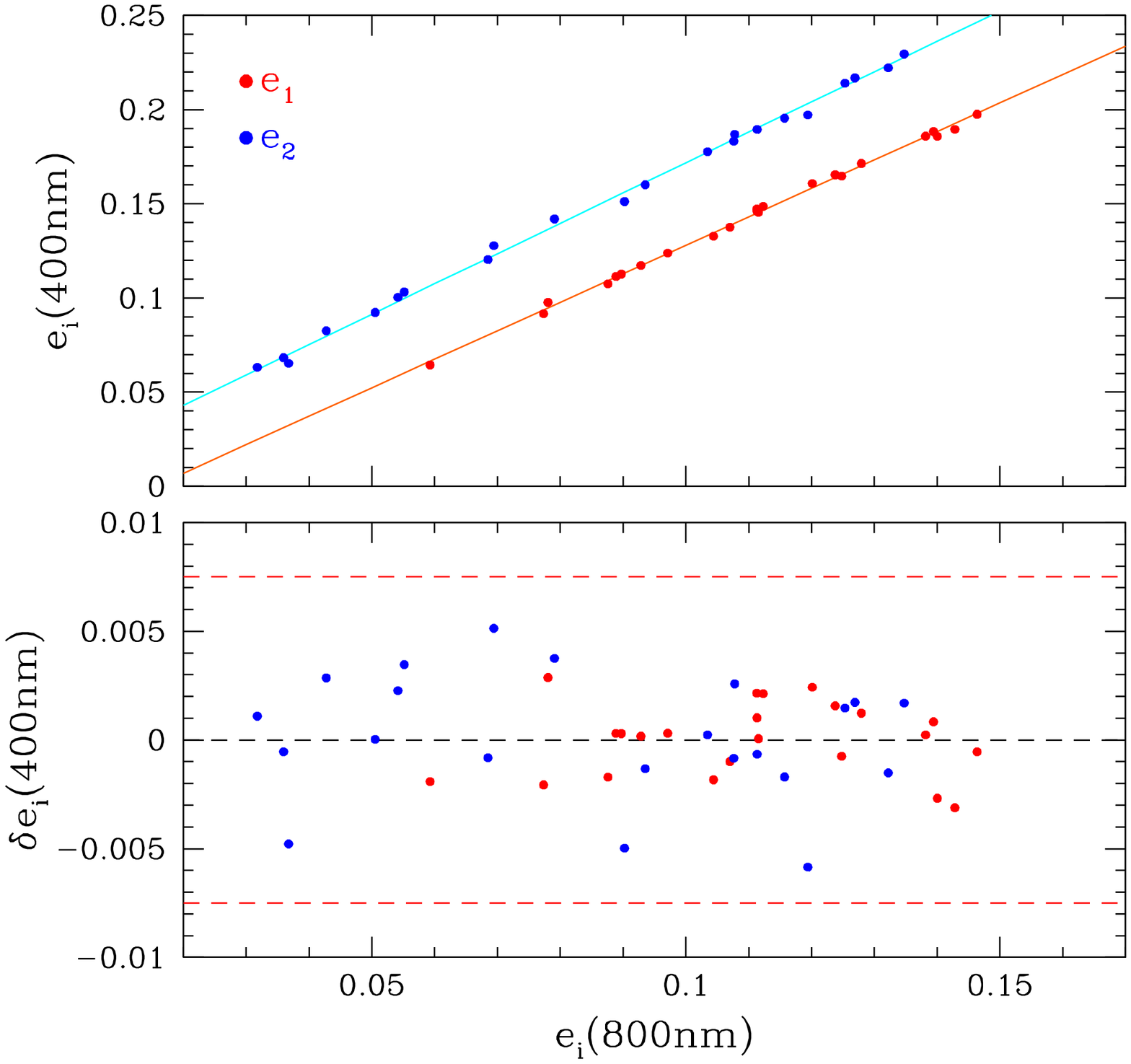}
}
\caption{The correlation between the PSF sizes at 400nm and 800nm (top left) and PSF ellipticity
components (top right) as inferred from the \eu model PSFs. The drawn lines are linear fits to the
measurements, with the results shown in the lower panels. The dashed lines indicate reasonable maximum
uncertainties. These can then be used to derive constraints on the out-of-band flux, and hence on the effect of the out-of-band transmission levels as shown in the right panel of Figure~\ref{fig:outofband}.
}
\label{fig:lambda_fit}
\end{figure*}

The most conservative approach is to assume no knowledge about the
out-of-band PSF. The left panels in Figure~\ref{fig:outofband} shows how the PSF size and
shape change for a flat spectrum source ($f(\lambda)=$constant) as a
function of $f_{\rm out}$ where we assumed $R_{_{\rm C}}^2 \propto \lambda^{0.55}$. This is the case for the \eu mission, but the results are  relatively insensitive to the assumed slope of the spectral dependance.
In practice, however, it is possible to extrapolate the observed PSF to
other wavelengths using a model to correct for the leakage more accurately. The top panels of Figure~\ref{fig:lambda_fit} show the out-of-band 400nm PSF size and ellipticity as a function of the in-band 800nm values for the\eu mission for different parts of the field of view and for different optical alignments. The residuals to linear fits are given in the lower panels, which indicates that biases in the 400nm size and ellipticity (per component) as predicted from the 800nm values are (in this case) smaller than $|\delta R^2| = 0.015$ and $|\delta e_i |= 7.5\times10^{-3}$ respectively (dashed lines in Figure~\ref{fig:lambda_fit}). With such an analysis, a more relaxed out-of-band transmission requirement can be permitted. This is  evident in the different $x$-axis scale for the right hand panels of Figure~\ref{fig:outofband} (calculated for these $|\delta R^2| = 0.015$ and $|\delta e_i |= 7.5\times10^{-3}$ values). The relaxation in the requirement is particularly true for the PSF size.

\subsubsection{Galaxy Colours}

In Sections~\ref{sec:conv} and \ref{sec:non-conv} we described how the PSF can be constructed
with high accuracy using observations of stars in the field-of-view for any position in the focal plane and for any colour of star.
The PSF
entries in Equations~\ref{eq:A} and \ref{eq:M} refer to the PSF with which each galaxy has been
convolved by the instrument. This requires an estimate of the galaxy spectral energy distribution (SED). A good estimate of the galaxy SED can be inferred from broad-band data, which
are typically available in any case because of the need to determine photometric
redshifts for the source galaxies. The actual SED, however, depends on
the star formation history of the galaxy, its metallicity, redshift,
etc.  Hence it cannot be known perfectly, which inevitably leads to an
error in the estimate of the PSF to be used for the galaxy PSF. 

The impact of uncertainties in the SED on the PSF size has been studied  by Cypriano \etal (2010). Within the assumptions of Gaussian PSFs with fully sampled data, they find that the PSF size can be recovered with a relative uncertainty less than $\sim 2\times 10^{-4}$. 
Spatially varying SEDs (spatial colour gradients within the galaxy) resulting from a broad bandpass have been shown to cause $\mcm$-like biases at the level of $\laeq 5\times10^{-4}$ (see Voigt \etal 2012, Semboloni \etal 2013). This is comparable to the allocation in Table~\ref{tab:aln}.

\section{Summary}

We have set out with this paper  to extend in a practical scheme the more general treatments of weak lensing measurements in the literature, and particularly that of MHK13. This can be used as a framework to define a next-generation space-based weak lensing experiment.

We have started with the main requirements. It is necessary to observe a large enough number of galaxies through a wide-area survey with sufficient photometric sensitivity, range of redshift and spatial resolution to ensure that the parameters in different cosmological models can be tightly constrained (these parameters may be those in the standard Concordance Model, or from alternative models). The size of the next-generations survey must be very large, some 15000 square degrees, observed to $m_{\rm{AB}}>24.5$, in order to make available more than $10^9$ galaxies. With such a survey, the large intrinsic variation within the galaxy population can be averaged to produce very precise measurements of the cosmological parameters. However, with such precision, systematic effects in the measurements potentially become the limiting factor. This leads to the other main requirement which is that the parameters used in the derivation of the shear information, principally the shape of the point spread function (PSF), are known with sufficient accuracy. These, together with the biases introduced by imperfect shear measurement methodologies, are constrained to be less than a small factor (a bias-over-error ratio of $\leq 0.31$; MHK13) of the uncertainties arising from the finite size of the survey, thus ensuring that the cosmological parameters will be derived with the required accuracy.

The main requirements are therefore survey size, depth, spatial resolution, the knowledge of the instrument characteristics and the extent to which the biases can be corrected. The first three are relatively conventional, although demanding: they drive the size of the telescope, the field of view, the detector pixel scale and noise levels, the survey duration, and so on. The remaining two constitute the different and particularly challenging aspect of a weak lensing experiment. This requires a detailed cataloging of all of the potential effects which affect our knowledge of the instrument, and particularly the PSF, the classification of these effects into different categories, and the appreciation of how and to what extent each will impact this knowledge.

We have therefore examined these effects, first following MHK13 in considering additive and multiplicative biases $\mcap$ and $\mcmp$ in  separating out the linear and non-linear contributions to each, together with the biases introduced by the weight function in the modelling ($\alpha^2$ and $\mu$ terms in Equations~\ref{eq:A} and \ref{eq:M}). Linear contributions have been represented by convolutions, while non-linear effects (which generally arise in the detectors and electronics) are non-convolutive. We then examined the different scale of the contributions in each, and with the consequent weighting, combine them with a permitted error in our knowledge of the ellipticity and size of the system PSF to calculate the impact on the total $\mcap$ and $\mcmp$. These knowledge errors can be adjusted and balanced, based on feasibility considerations, to arrive at a set of permitted values. Then, we marshalled all of the individual contributing factors  to these convolutive and non-convolutive effects into categories, such as those arising from imperfectly known source characteristics, satellite pointing errors, calibration residuals, PSF modelling errors, detector imperfections (especially arising from radiation damage), and calculated their impact with example numerical values, again weighting these appropriately. We noted that their aggregate contributions must equal or be less than those allocated at the highest level, and again some adjustment and re-balancing may be required. An example of these factors was provided in Table~\ref{tab:aln}.

We then know what is required to achieve the scientific goals of the weak lensing survey.  In order to have made the allocations in the contributions to the overall PSF knowledge budget, we have evaluated what is or may be feasible.  We described briefly in Section~\ref{sec:sims} the ingredients incorporated into the simulations. Because the systematic effects we are controlling have to be known very accurately, a deep understanding of the instrumental effects is required, from the range of variation of the telescope PSF, to the pointing characteristics of the satellite, and to subtle detector effects of various types.  Having made the simulations, we then explained the main steps in the processing of the simulated data. We find that standard processing will be adequate in the flat-fielding and linearity corrections, while most of the other standard data processing procedures (bias subtraction {\it etc.}) contribute second-order effects which don't feed directly into the ellipticity in most cases. The correction for the CCD CTI caused by radiation damage is the principal matter to be addressed at this stage. We noted the algorithm in table 1 of Massey \etal (2010) by which the trailing in the image can be largely corrected by linear combinations of the observed (or, at this stage, simulated) data with copies of these same data passed through a radiation damage model. At the end of this process, the best image data which can be generated using the calibrations and the radiation modelling is available for further analysis.

We then examined whether the performance we can obtain for each constituent contribution remains reasonable. We check  on the effect of the sampling, and conclude that with only three slightly displaced exposures, the slight undersampling at $0.688$ Nyquist does not meet the stringent sampling requirement criteria, but not by a large amount.  While it is not yet quantified how much these sampling criteria can be relaxed without more noticeably impacting the survey's weak lensing systematic error budget, and further work is  needed,  we found in Section~\ref{sec:non-conv} (where the three exposure case including CTI is propagated into the shear power spectrum) that such variation appears to have a limited impact on the Dark Energy FoM.
We continued with the investigation of the level of knowledge that can be reached in the PSF model. The aim here was not to identify the ultimate PSF model to be used, but to show that with an analysis of the modes of PSF variation combined with the Bayesian model fitting, the performance allocated in Equation~\ref{eqn:AMlim}  can be achieved. Further modelling advances will provide additional margin. We generated the eigenmode basis set for the \eu case, over the full field of view, and over a range of optical system characteristics, arising from misalignments and manufacturing errors, finding that the number of modes required is in the range $20-70$. We finally examined how many stars would be required in order to retrieve the PSF from the Bayesian model fitting of the normal modes to the simulated data to the accuracy allocated in Table~\ref{tab:aln}, given a basis set with $40$ components, and the actual pixellised, noisy PSFs. We found that for a reasonable field of view the PSF can be recovered on each field independently, without any reliance on the stability of the optical system from field to field. If such variations can be tracked, additional performance could be achieved.  

We finally examined the residuals in the data caused by the imperfect correction of the CTI caused by radiation damage. We found that the ultimate accuracy of the process is limited by the readout noise of the CCD and detection chain, as this adds uncertainties to the measurement of the charge trails. 
The lack of fidelity of the radiation model, and limited knowledge of the parameters within it contribute to the residuals, but because these are in the frame of the detector, with particular orientations, they can be minimised by iteration of the model.  In addition we found that the effect of the imperfect CTI correction is limited to certain angular scales, of the order those subtended on the sky by the CCD.

We ended with a brief reference to factors affecting the galaxy modelling itself, in particular the effects of imperfections in the linearity correction, of spectral leakage outside of the defined bandpass and of the spatially variable SEDs within the galaxies.


\section*{Acknowledgments}

We thank Rene Laureijs, Pierre Ferruit, Tim Oosterbroek and Ludovic Duvet at ESA for their support. We  thank Elisabetta Semboloni for her comments on the paper. We also thank the referee who made some valuable suggestions which improved the accessibility of this paper.
HH is supported by the Netherlands
Organization for Scientific Research through VIDI grants
and acknowledges support from the Netherlands Research
School for Astronomy (NOVA). TDK was supported by a Royal Astronomical Society 2010 Fellowship and now by a Royal Society University Research Fellowship. RM is supported by a Royal Society University Research Fellowship.
RM and HH also acknowledge
support from ERC International Reintegration Grants. BR acknowledges support from European Research Council in the form of a Starting Grant with number 240672.
JR was supported by JPL, run by Caltech under a contract for NASA.


\end{document}